\documentclass[11pt,a4paper]{article} 
\pdfoutput=1      
\usepackage{jcappub}
\usepackage{bm}
\usepackage{times}
\usepackage{comment}
\usepackage{caption}
\usepackage{subcaption}
\usepackage{amsmath}
\usepackage{ulem}

\usepackage{tikz}
\usepackage{tkz-euclide}
\usepackage{subcaption}
\usepackage{xcolor}

\newcommand{\cc}{\lambda}

\newcommand{\BB}{\mathcal{B}} 
\newcommand{\CC}{\mathcal{C}} 
\newcommand{\VV}{\mathcal{U}} 
\newcommand{\ax}{\alpha_{\chi}}
\newcommand{\px}{\varphi_{\chi}}
\newcommand{\vx}{v_{\chi}}      
\newcommand{\pv}{\Psi}          
\newcommand{\CCG}{\mathcal{G}}  
\newcommand{\LL}{\mathcal{L}}   
\newcommand{\dz}{\delta z}

\newcommand{\ddD}{\delta\mathcal{D}} 

\newcommand{\ttt}{\theta}          
\newcommand{\pp}{\phi}

\newcommand{\up}[1]{{\rm #1}}

\newcommand{\beeq}{\begin{equation}}
\newcommand{\eneq}{\end{equation}}
\newcommand{\bear}{\begin{eqnarray}}
\newcommand{\enar}{\end{eqnarray}}
\newcommand{\nnn}{\nonumber \\}
\newcommand{\RA}{\rightarrow}
\newcommand{\OO}{\mathcal{O}}

\newcommand{\HH}{\mathcal{H}}  
\newcommand{\rbar}{\bar r}

\newcommand{\dnu}{\delta\nu}   
 
\newcommand{\drr}{\delta r}     
\newcommand{\dtt}{\delta\ttt}   
\newcommand{\dpp}{\delta\pp}

\newcommand{\Dquad}{\qquad\qquad}
\newcommand{\rr}{r}
\newcommand{\xx}{x}
\newcommand{\ro}{\tilde r}
\newcommand{\xo}{\tilde x}
\newcommand{\dro}{\de\rr_o}
\newcommand{\gbar}{\bar g}
\newcommand{\al}{\alpha}
\newcommand{\be}{\beta}
\newcommand{\ga}{\gamma}
\newcommand{\de}{\delta}
\newcommand{\ep}{\epsilon}

\newcommand{\ff}{{f_\text{\tiny{K}}}}

\newcommand{\Nn}{N}
\newcommand{\dNn}{\de\Nn}
\newcommand{\bobs}{\bar {\rm o}}
\newcommand{\CK}{\hat k}

\newcommand{\hg}{\hat g}
\newcommand{\NCC}{\mathbb{C}}

\newcommand{\Ga}{\Gamma}
\newcommand{\para}{\parallel}
\newcommand{\GGH}{\widehat{\de\Gamma}{}}
\newcommand{\pa}{\partial}
\newcommand{\TT}{\Theta}
\newcommand{\PP}{\Phi}
\newcommand{\Dcc}{\Delta\cc}
\newcommand{\dDD}{\de\mathcal{D}}
\newcommand{\dD}{\mathcal{D}}
\newcommand{\NP}{n}

\begin{document}

\begin{titlepage}

\setcounter{page}{1} \baselineskip=15.5pt \thispagestyle{empty}

\bigskip

\vspace{1cm}
\begin{center}
{\fontsize{20}{28}\selectfont \bfseries Living in a Non-Flat Universe:\\
Theoretical Formalism}
\end{center}

\vspace{0.2cm}

\begin{center}
{\fontsize{13}{30}\selectfont Sandra Baumgartner$^a$, and
Jaiyul Yoo$^{a,b}$}
\end{center}

\begin{center}
\vskip 8pt
\textsl{$^a$ Center for Theoretical Astrophysics and Cosmology,
Institute for Computational Science}\\
\textsl{University of Z\"urich, Winterthurerstrasse 190,
CH-8057, Z\"urich, Switzerland}

\vskip 7pt

\textsl{$^b$Physics Institute, University of Z\"urich,
Winterthurerstrasse 190, CH-8057, Z\"urich, Switzerland}

\vskip 7pt

\today

\end{center}

\note{sandra.baumgartner@uzh.ch, \hspace{1cm} jyoo@physik.uzh.ch}

\vspace{1.2cm}
\hrule \vspace{0.3cm}
\noindent {\sffamily \bfseries Abstract} \\[0.1cm]
Recent analysis of the Planck measurements opened a possibility that we live in a non-flat universe. Given the renewed interest in non-zero spatial curvature, here we re-visit the light propagation in a non-flat universe and provide the gauge-invariant expressions for the cosmological probes: the luminosity distance, galaxy clustering, weak gravitational lensing, and cosmic microwave background  anisotropies. With the positional dependence of the spatial metric, the light propagation in a non-flat universe is much more complicated than in a flat universe. Accounting for all the relativistic effects and including the vector and tensor contributions, we derive the expressions for the cosmological probes and explicitly verify their gauge invariance. We compare our results to previous work in a non-flat universe, if present, but this work represents the first comprehensive investigation of the cosmological probes in a non-flat universe. Our theoretical formalism in a non-flat universe will play a crucial role in constraining the spatial curvature in the upcoming large-scale surveys.
\vskip 10pt
\hrule

\vspace{0.6cm}
\end{titlepage}

\noindent\hrulefill

\tableofcontents

\noindent\hrulefill

\setcounter{page}{1}
\pagenumbering{arabic}

\section{Introduction}
The spatial curvature~$K$ of the Universe is one of the important cosmological parameters that determine the fate of our Universe. Given the theoretical preference that the inflationary expansion in the early Universe naturally suppresses the curvature parameter~$\Omega_k$, cosmological models with non-zero spatial curvature have been studied far less than those in a flat universe. However, there exist inflationary scenarios with a non-zero spatial curvature \cite{1982PhLB..115..295H, PhysRevD.28.679, 1985PhRvD..32.1899G, 1997PhRvD..55.7461B, 1982Natur.295..304G, 1995PhRvD..52.1837R, 2017PhRvD..96j3534R}, and more importantly, recent Planck analysis of CMB anisotropies allows a small but non-zero value of spatial curvature. Indeed, combining the Planck temperature and polarization power spectra with lensing data gives $\Omega_k=-0.0106\pm0.0065$ \cite{Planck2018CosmParam}. Furthermore, it was shown \cite{2021PhRvD.103d1301H,2020NatAs...4..196D,Di_Valentino_2021,2019Ap&SS.364..134P} (see, however, \cite{2020MNRAS.496L..91E}) that a closed universe can not only explain the anomalous lensing amplitude in the Planck data, but also reduce tensions between CMB measurements and supernova observations at low redshift. A non-zero spatial curvature, on the other hand, gives rise to the heightened discordances with most of the low-redshift cosmological observables, including BAO measurements in galaxy clustering \cite{2021PhRvD.103d1301H,2020NatAs...4..196D,Di_Valentino_2021,2019Ap&SS.364..134P}. It was shown in \cite{2021PDU....3300851V,2021ApJ...908...84V,2021MNRAS.506L...1D} that Cosmic Chronometer measurements can be combined with several other data sets to derive constraints on the curvature parameter. Given the renewed interest in spatial curvature and the recent observational constraints (e.g. \cite{2018ApJ...864...80O, 2021MNRAS.504..300C}), here we revisit the light propagation in a non-flat universe and investigate its impact on the cosmological observables.

In the past, the implication of a non-zero spatial curvature in cosmology has been extensively studied. In a non-flat universe, the eigen-modes of the Helmholtz equation change dramatically, compared to the standard Fourier modes in a flat universe (see e.g., \cite{1967RvMP...39..862H,1982PThPh..68..310T}). In particular, the finite volume of a closed universe is revealed in the discreteness of eigen-modes. The contribution of gravitational waves generated during inflation to the CMB anisotropy in an open Universe was studied in \cite{1997PhRvD..55.7461B}, and it was shown \cite{1985PhRvD..32.1899G, 1982PhLB..115..295H, PhysRevD.28.679} that quantum fluctuations in the inflaton field induce variations in the spatial curvature. Given the large level arm, the effects of a non-zero spatial curvature have been well studied \cite{2014JCAP...09..032L, 1983ApJ...273....2W, 1998ApJ...494..491Z, 1998PhRvD..57.3290H, 2000PhRvD..62d3004C, 2000CQGra..17..871C, 1996astro.ph..4166H, 2002ARA&A..40..171H} for cosmic microwave background (CMB) anisotropies (see, \cite{1996astro.ph..4166H, 2002ARA&A..40..171H} for review). Calculations of the CMB anisotropies in an open universe for both adiabatic and isothermal fluctuations have been carried out in \cite{1983ApJ...273....2W} (see also \cite{2014JCAP...09..032L}). An efficient method to compute the CMB anisotropies was developed in \cite{1998ApJ...494..491Z} by applying the line-of-sight integral method to the Boltzmann equation, and the total angular momentum method was generalized \cite{1998PhRvD..57.3290H} to non-flat FRW universes. \cite{2000PhRvD..62d3004C} examines the generation and propagation of the polarization of the CMB for both flat and non-flat geometries, and \cite{2000CQGra..17..871C} discusses the contribution of gravitational waves to the CMB anisotropy for arbitrary spatial curvature. A method for calculating CMB anisotropies in a curved Friedmann universe was implemented in the Boltzmann code CLASS \cite{2014JCAP...09..032L}.

Relatively less attention was paid in literature to the low-redshift cosmological observables in a non-flat universe such as the luminosity distance, galaxy clustering, and weak gravitational lensing. A pioneering work on the luminosity distance was performed in \cite{1987MNRAS.228..653S}, where a general formula for the magnitude-redshift and distance-redshift relations were derived for a universe with arbitrary constant spatial curvature. The anisotropies of the luminosity distance were also computed in \cite{1989PhRvD..40.2502F, 1999PThPh.101..903S}. For galaxy clustering, it was \cite{2016JCAP...06..013D} that first generalizes the expression for the general relativistic galaxy number counts to spatially curved universes. Finally, a method to determine simultaneously the angular diameter distance and the curvature of the universe using weak-lensing data is described in \cite{2006ApJ...637..598B}. In Section~\ref{comparison}, we will compare our results to those in these early papers in more detail.

Despite the rich history in literature, the light propagation in a non-flat universe has {\it not} been fully understood. For example, a simple prescription for a non-flat universe is to use the formulas derived in a flat universe, but to replace all the background angular diameter distances with those in a non-flat universe. While this method has been widely used in literature, it is not a complete and consistent treatment, but an approximation. In particular, the recent work \cite{2009PhRvD..79b3517Y,PhysRevD.80.083514,PhysRevD.82.083508,2011PhRvD..84f3505B,2011PhRvD..84d3516C,PhysRevD.85.023504}
shows that there exist numerous and subtle relativistic effects associated with the light propagation and observations. Given this development, here we critically re-examine the light propagation in an inhomogeneous universe with a non-zero spatial curvature, closely following the gauge-invariant formalism in \cite{Yoo_2014a,YooGrimmMitsou_2018}, but generalizing it to a non-flat universe. We find that due to the positional dependence of the spatial metric in a non-flat universe, the background solution to the photon geodesic equation is not a solution to the background geodesic equation at a position deviating from a straight path in an inhomogeneous universe, an extra complication, compared to the light propagation in a flat universe. Consequently, the light path in a non-flat universe is much more complicated than previously computed. We then investigate the impact on the cosmological observables.

The organization of this paper is as follows. 
We start by discussing the homogeneous and isotropic, non-flat background universe in Sec.~\ref{sec:NonFlat}. 
Specifically, we introduce the (background) metric and the tetrad vectors to relate physical quantities in the observer rest frame to FRW coordinates in Secs.~\ref{ssec:BGmetric} and \ref{ssec:tetradBG} respectively.
The light propagation in this background universe is discussed in Sec.~\ref{ssec:LightPropagationBG}, and we finish this section with the expression for the comoving angular diameter distance in a non-flat universe in Sec.~\ref{ssec:ComAngDiamDist}.
We then introduce perturbations around this background universe in Sec.~\ref{inhomo}.
The perturbed FRW metric is introduced in Sec.~\ref{sec:metric}, followed by the derivation of the tetrad vectors in an inhomogeneous universe in Sec.~\ref{ssec:tetrad}.
We introduce a conformal transformation in Sec.~\ref{ssec:conformal}, which we use in Sec.~\ref{ssec:GE} to compute the light propagation path.
The source position on the light path is derived in FRW coordinates in Sec.~\ref{ssec:path}. 
In Section~\ref{cosobs}, we proceed by determining expressions for cosmological observables in a non-flat universe, such as the luminosity distance and its fluctuation in Sec.~\ref{sec:LD}, the observed volume in Sec.~\ref{sec:volume} and the observed lensing shear and convergence in Sec.~\ref{sec:WL}. We briefly discuss the collisionless Boltzmann equation in Sec.~\ref{sec:CMB}.
We present useful approximate formulas for the cosmological observables, when the curvature density can be treated as a small perturbation in Sec.~\ref{sec:NearlyFlat}.
Finally, we compare our result to previous work about non-zero curvature in Sec.~\ref{comparison} and finish with a summary and discussion in Sec.~\ref{sec:discussion}.
We explain the geometric relation between the standard coordinate and our coordinate choice in App.~\ref{app:Projection} and we present detailed derivations of the angular solution to the geodesic equation and the angular distortion of the observed source position in App.~\ref{angdetail}.
We summarize our notation convention in Table~\ref{table:notation}.

\begin{table}
\begin{center}
\begin{tabular}{lll}
\hline\hline
Symbols & Definition of the symbols & Equations\\
\hline
$\rr$ & radial coordinate 
& Sec.~\ref{ssec:BGmetric}\\
$\ro$ & comoving angular diameter distance 
& Sec.~\ref{ssec:BGmetric}\\
$\tilde{\chi}$ & comoving radial coordinate 
& Sec.~\ref{ssec:BGmetric}\\
$\omega$, $n^i$ & photon angular frequency \& observed angular direction
& \eqref{kaobs}\\
$\alpha$, $\BB_\alpha$, $\CC_{\alpha\beta}$
& perturbations of the FLRW metric tensor & \eqref{eqn:metric} \\
$\alpha$, $\beta$, $\varphi$, $\gamma$
& scalar metric perturbations & \eqref{eq:decom}\\
$B_\alpha$, $C_\alpha$, $C_{\alpha\beta}$ & vector and tensor metric
perturbations & \eqref{eq:decom}\\
$\xi^\mu$, $T$, $L$, $L^\alpha$
& coordinate transformation \& its decomposition & \eqref{eqn:coordinatetransformation}\\
$\ax$, $\px$, $\pv_\alpha$  & scalar and vector gauge-invariant variables &
\eqref{eqn:GI_scalar_variables}, \eqref{eqn:GI_vector_variable}\\
$u^\mu$, $V_\al$  & four-velocity \& gauge-invariant velocity variable &
\eqref{eqn:4velocity}, \eqref{eqn:defV}\\
$e_a^\mu$  & Local tetrad basis 
($e_0^\mu=u^\mu$, $e_i^\mu$) & Sec.~\ref{ssec:tetrad} \\
$\delta e_i^\eta$, $\delta e_i^\alpha$ & perturbations to the
spatial tetrad vectors & \eqref{eq:tetdef}\\
$\Omega^i$ & 
orientation of the local tetrad basis & \eqref{eqn:asymTet}\\
$\bar\eta_{\bobs}$, $\bar t_{\bobs}$ & age of the (homogeneous) Universe & \eqref{eqn:ObsPosBG}, \eqref{eqn:DeltaEta}\\
$\delta\eta_o$, $\delta x^\alpha_o$ & coordinate lapse and shift 
of the observer position & \eqref{eqn:ObsPos}, \eqref{eqn:DeltaEta}\\
$\hat k^\mu$, $\dnu$, $\dNn^\alpha$ & conformally transformed photon 
wave vector \& its perturbations & \eqref{eqn:RelDiffK}, \eqref{CFWAVE}\\
$\Nn^\al$ & unit directional 3-vector (in a FRW coordinate)  & 
\eqref{choice}\\
$\ttt^i$, $\pp^i$ & unit directions perpendicular to $n^i$ (in observer rest-frame) & 
\eqref{eqn:Definition_ttt_pp}\\
$\TT^\al$, $\PP^\al$ & unit directional 3-vectors perpendicular to $\Nn^\al$ & 
\eqref{eqn:DefTTPP}\\
$\bar\eta_z$, $\lambda_z$ & quantities expressed at the
observed redshift~$z$ & \eqref{eqn:LambdaZ}\\
$\dz$ & perturbation in the observed redshift~$z$ & \eqref{redshift},
\eqref{eqn:dz}\\
$\rbar_z\equiv\rbar_{\cc_z}$ & background radial coordinate at the
observed redshift~$z$ & \eqref{bgrr}\\
$\Delta \lambda_s$ & perturbation in the affine parameter at the source position & \eqref{eqn:LambdaS} \\
$\Delta \eta_s$, $\Delta x^\alpha_s$ & deviation of the source position
from the inferred position &\eqref{eqn:TimeCoordSource}, \eqref{src} \\
$\drr$, $\dtt$, $\dpp$ & geometric decomposition of the spatial deviation
&\eqref{eqn:SourcePos}, \eqref{drrex}, \eqref{tangent} \\
$\widetilde{\drr}$ & deviation of the source angular diameter distance 
&\eqref{eqn:Ro_s}\\
$\delta\tilde{\chi}$ & radial distortion in a tilde coordinate
&\eqref{eqn:DeltaChi}\\
$\dD_L$, $\dD_A$ & luminosity distance \& angular diameter distance
&\eqref{eqn:LumDistAngDistRel}\\
$\dDD$ & distance fluctuation
&\eqref{eqn:DistFlucEquivalence}, \eqref{eqn:deltaD}\\
$g$, $\delta g$ & metric determinant \& its perturbation
&\eqref{eqn:MetricDet}\\
$n^\mu_s$ & photon propagation direction at the source position
&\eqref{eqn:PhotonPropDirSource}\\
$\kappa$ & lensing convergence &\eqref{eqn:StandardConvergence} \\
$\delta V$ & dimensionless volume fluctuation &\eqref{phydV}, \eqref{eqn:dV_GI} \\
$\Delta s^\alpha$ & perturbation to the extended source size
&\eqref{eqn:Dsalpha}\\ 
$n^i_s=(\ttt_s,\pp_s)$ & photon propagation direction
in the source rest-frame  &\eqref{eqn:niSource} \\
$\Delta\ttt$, $\Delta\pp$ & perturbations in the source angle
$(\ttt_s=\ttt+\Delta\ttt$, $\pp_s=\pp+\Delta\pp$) & \eqref{eqn:DThetaSourceRF} \\
$\Delta\ttt^i_s$, $\Delta\pp^i_s$ & perturbations to the source angular
vectors~$\ttt^i_s$ \&~$\pp^i_s$ &\eqref{eqn:SourceDirecVectors}\\
$\Omega^n$, $\Omega^\ttt$, $\Omega^\pp$ & decomposition of the rotation
vector~$\Omega^i$ & \eqref{eqn:DecompRotTetrad} \\
$\hat\kappa$, $\hat\gamma_{1,2}$, $\hat\omega$ &
physical lensing observables (convergence, shear \& rotation) & \eqref{eqn:convergence}, \eqref{eqn:shear}, \eqref{eqn:rotation} \\
\hline\hline
\end{tabular}
\caption{Notation convention used in the paper}
\label{table:notation}
\end{center}
\end{table}

\section{Background Universe with a Non-Zero Spatial Curvature}
\label{sec:NonFlat}
Here we introduce the metric used to model the background
universe with a non-zero
spatial curvature and briefly discuss the light propagation with focus on the
difference compared to the case in a flat universe.

\subsection{Metric convention}
\label{ssec:BGmetric}
For a non-vanishing spatial curvature ($K\neq0$), the Robertson-Walker metric is
\beeq
\label{meme}
ds^2=-a^2d\eta^2+a^2\left[{d\ro^2\over1-K\ro^2}+\ro^2d\Omega^2\right]
=-a^2d\eta^2+a^2\left[d\tilde{\chi}^2+\ro^2d\Omega^2\right]~,
\eneq
where the scale factor~$a(\eta)$ is a function of conformal time~$\eta$
and the comoving angular diameter distance~$\ro$ is related to the
 spatial curvature~$K$ as
\beeq
\ro={1\over\sqrt{K}}\sin\sqrt K\tilde{\chi}~~~\up{for}~K>0~,\Dquad
\ro={1\over\sqrt{-K}}\sinh\sqrt{-K}\tilde{\chi}~~~\up{for}~K<0 ~,
\label{eqn:DefChi}
\eneq
in terms of the comoving radial coordinate~$\tilde{\chi}$ 
\begin{equation}
d\tilde{\chi}^2 = \frac{d\ro^2}{1-K\ro^2}~.
\end{equation}
Mind the notation~$\ro$
for the comoving angular diameter distance. For a positive 
curvature $K>0$, the three-hypersurface is a sphere in a four-dimensional space
with radius $R_K:=1/\sqrt{K}$, in which the comoving angular diameter
 distance~$\ro$ is the radius of lower three dimension of the sphere
\beeq
\ro^2=\xo_1^2+\xo_2^2+\xo_3^2=R_K^2\sin^2\Theta\leq{1\over K}~,
\label{eqn:ro}
\eneq
and the comoving radial coordinate~$\tilde{\chi}=R_K\Theta$ is the length of the arc
in terms of the polar angle~$\Theta$ in 4D. Naturally, the range of~$\ro$
is limited to~$R_K$. A similar parametrization can be obtained for a 
negative curvature with hyperbolic function $(R_K^2=-1/K>0)$.
Given a length unit~$L$, we can re-scale the coordinates 
\beeq
K\RA{K\over|K|L^2}~,\Dquad \ro\RA\sqrt{|K|}\ro L~,\Dquad 
a\RA {a\over L\sqrt{|K|}}~,\Dquad \eta\RA \eta L\sqrt{|K|}~,
\eneq
and the metric remains invariant, while the dimensionful spatial curvature~$K$ is then 
re-scaled as
$K=0,\pm1$ in a new unit system. However, we keep general the spatial curvature~$K$ and
the curvature parameter $\Omega_K=-Kc^2/H_0^2$, where $H_0$ is the Hubble
parameter today.

Another useful representation of a non-flat Universe can be
obtained by a conformal transformation to a untilde coordinate
\beeq
\label{rrro}
\ro=:{\rr\over1+{K\over4}\rr^2}~,\Dquad
1-K\ro^2=\left({1-{K\over4}\rr^2\over1+{K\over4}\rr^2}\right)^2~,
\eneq
and its geometrical relation is
\beeq
\frac12\rr_{\text{\tiny{$\pm$}}}={1\over K\ro}\left[1\pm\sqrt{1-K\ro^2}\right]~,
\label{eqn:relationRr}
\eneq
where $\rr_{\text{\tiny{$\pm$}}}$ is the radius in a 3D plane projected from a non-flat space in 4D.
For $K>0$, the Northern and the Southern hemispheres of a 3D sphere are projected into $\rr_{\text{\tiny{$+$}}}=R_K\sim\infty$ and $\rr_{\text{\tiny{$-$}}}=0\sim R_K$, while a 3D hyperbola is projected into $\rr_{\text{\tiny{$-$}}}=0\sim R_K$.
For both cases, the solid angle in the lower three dimension ($i=1,2,3$) is preserved
\beeq
\label{angconf}
{\xo_i\over \ro}={\xx_i\over\rr}~.
\eneq
The geometric meaning of the projection is discussed in Appendix~\ref{app:Projection}.
The advantage of this parametrization is that the spatial metric becomes
diagonal 
\beeq
\label{metric}
ds^2_3={a^2\over\left(1+{K\over4}\rr^2\right)^2}
\left(d\rr^2+\rr^2d\Omega^2\right)
={a^2\over\left(1+{K\over4}\rr^2\right)^2}
\left(d\xx_1^2+d\xx_2^2+d\xx_3^2\right)~,
\eneq
whereas the original metric in Eq.~\eqref{meme}
is \textit{not} diagonal in a rectangular coordinate.
While the conformal transformation preserves the angles $(\theta,\phi)$ 
in the original metric, it should be noted that the comoving angular
diameter distance is~$\ro$, not~$\rr$, which is then related to the
physical angular diameter distance with $(1+z)$ factor.
As our interest lies in the light propagation in an inhomogeneous universe,
it is computationally convenient to work in a rectangular coordinate.
We will mainly use Eq.~\eqref{metric} for the following computations.

\subsection{Tetrad and photon wave vectors}
\label{ssec:tetradBG}
To relate the physical quantities in the observer rest frame to FRW 
coordinates, we consider four orthonormal
tetrad vectors in the background universe
\beeq
\bar e^\mu_0={1\over a}\left(1,~0\right)~,\Dquad
\bar e^\mu_i={1\over a}\left[0,~\ff(\rr)\de^\al_i\right]~,\Dquad
i=x,y,z~,
\eneq
where we defined a function
\beeq
\ff(\rr):=1+{K\over4}\rr^2~,
\label{eqn:DefF}
\eneq
representing the deviation from a flat space
and the Kronecker delta is related to the three-metric
\beeq
\label{gbar}
ds_3^2=a^2 \gbar_{\al\be}~dx^\al dx^\be~,\Dquad
\gbar_{\al\be}:={\de_{\al\be}\over \ff^2(\rr)}~,
\Dquad \gbar^{\al\be}=\ff^2(\rr)\de^{\al\be}~.
\eneq
We used bars to indicate that they are the quantities in the background
universe, while the same quantities without bars will be those 
in an inhomogeneous universe.
We will use the indices $\mu,\nu,\cdots$ to represent the spacetime 
components, while the indices $\al,\be,\cdots$ represent the spatial
components. The tetrad vectors represent the local time and spatial
directions of the observer rest frame in a FRW coordinate \cite{Mitsou_2020}.

An observer moving with a timelike four-velocity~$\bar u^\mu=\bar e^\mu_0$ at a given spacetime position 
measures the photon angular frequency~$\omega$ and the observed angular 
direction~$(\theta,\phi)$.
The photon wave vector  in the observer rest frame is then written as
\beeq
\label{kaobs}
k^a=\omega(1,-n^i)~,\qquad 1=\de_{ij}n^in^j~,\qquad a\in(0,x,y,z)~, \qquad i,j\in(x,y,z)~,
\eneq
and it can be readily expressed in a FRW coordinate as
\beeq
\bar k^\mu=\bar e^\mu_a k^a
={\omega\over a}\left[1,-\Nn^\al\right]~,\Dquad \Nn^\al:=\ff(\rr)n^i\de_i^\al~,
\eneq
where $n^i=(\sin\ttt\cos\pp, \sin\ttt\sin\pp, \cos\ttt)$ is a unit spatial directional vector in the observer rest frame (or
Minkowski space) and $\Nn^\al$ is a unit 3-vector based on~$\gbar_{\al\be}$.
Note that the wave vector~$k^a$ in the observer rest frame
is independent of FRW coordinates and also independent of
whether the Universe is homogeneous or inhomogeneous.
The null condition for the photon wave vector is trivially satisfied.

\subsection{Light propagation}
\label{ssec:LightPropagationBG}
Now we consider a null path or light propagation in a homogeneous universe. 
The photon
wave vector~$\bar k^\mu$ we derived is one at the observer position, since
it is expressed in terms of the physical frequency and angular direction
measured by the observer in the rest frame. However, due to the symmetry in the
background universe, its parametrization is indeed identical at all positions,
with $\omega$
and~$a$ evaluated at the given spacetime position of a fictitious observer. 
One subtlety we will discuss
later is the
unit directional vector~$n^i$, which is only defined in the observer rest
frame with Minkowski metric and hence is independent of FRW coordinates.
This difference is irrelevant in a homogeneous universe, though.

The light propagation in a homogeneous universe is a straight path, given
the geometry encoded in~$\gbar_{\al\be}$. On a straight path
emanating from the coordinate origin, any observers will record the same
angular directions~$n^i$ in their own rest frame, in which the local
coordinate is aligned with the FRW coordinate.
The geodesic equation for the photon wave 
vector~$\bar k^\mu$ yields the well-known results:
\beeq
\label{sgeo}
\omega\propto{1\over a}~,\Dquad \Nn^{\al\prime}=0~,\Dquad 0=\Nn^\be\Nn^\al
{}_{|\be}~,
\eneq
valid for any spatial curvature, where the prime and the vertical bar 
indicate the time derivative and the covariant derivative with respect 
to~$\eta$ and~$\gbar_{\al\be}$, respectively.
With the metric in Eq.~\eqref{metric}, first 
we can compute the spatial derivative of the function~$\ff$ and the
Christoffel symbol
\beeq
\ff_{,\al}={K\over2}\de_{\al\be}\xx^\be~,\Dquad 
\bar\Gamma^\al_{\be\ga}=-{K\over2\ff}\left(\de^\al_\be \de_{\ga\de}\xx^\de
+\de^\al_\ga \de_{\be\de}\xx^\de-\de_{\be\ga}\xx^\al\right)~,
\label{eqn:ChristoffelSymbolBG}
\eneq
and we can then explicitly verify that our parametrization for~$\Nn^\al$
indeed satisfies the geodesic equation in~\eqref{sgeo}, {\it on the 
condition} 
that the evaluation point~$\xx^\al$ is on a straight photon path, i.e.,
$\xx^\al=\rr ~n^i\de_i^\al$. 
In a flat universe, this condition is not needed, and Eq.~\eqref{sgeo}
is always satisfied with $\Nn^\al\RA n^i\de_i^\al$ (or $\ff\RA1$).
However, since $\Nn^\al$
depends on a position~$\rr$ in a non-flat universe, the evaluation position
must be on a straight path, or the solution of the geodesic equation.
In an inhomogeneous universe, where the photon path deviates from a straight
path, Eq.~\eqref{sgeo} is \textit{not} satisfied on an inhomogeneous photon path, 
though the full geodesic equation
including perturbations to~$\Nn^\al$ is of course satisfied (further details
in Section~\ref{inhomo}).

\subsection{Comoving angular diameter distance}
\label{ssec:ComAngDiamDist}
Without resorting to the photon wave vector, 
a position on the photon path in a homogeneous universe can be readily obtained
by using the null metric $ds^2=0$ and a straight radial path $d\Omega=0$ as
\beeq
d\bar\eta=-{d\bar\rr\over \ff(\bar\rr)}~,
\eneq
where we used the notation~$\bar\eta$ and~$\bar\rr$ to indicate that these
are the solutions to the geodesic equation in a homogeneous universe
(i.e., they are not just coordinates).
By introducing an affine parameter~$\cc$ for the path
\beeq
\label{affine}
{d\bar\eta\over d\cc}\equiv 1~,\Dquad \bar\eta_\cc-\bar\eta_{\bobs}=\cc
=-\int_0^{z_\cc}{dz\over H(z)}~,
\eneq
we can integrate the radial part to obtain the parametrized solution
\beeq
\label{bgrr}
-{d\rbar\over \ff(\rbar)}=d\cc~,\Dquad
\rbar_\cc = \left\{
\begin{array}{lr}
\frac{2}{\sqrt{K}} \tan\left(-\frac{\sqrt{K}}{2}\cc\right)  &\qquad K>0\\
-\cc & \qquad K=0 \\
\frac{2}{\sqrt{|K|}} \tanh\left(-\frac{\sqrt{|K|}}{2}\cc\right) &\qquad K<0
\end{array} \right\}~,
\eneq
where the subscript~$\bobs$ represents the observer position in a 
homogeneous universe and $H(z)$ is the Hubble parameter at redshift~$z$. 
This solution is then related to the comoving 
angular diameter distance~$\ro$ by Eq.~\eqref{rrro} as
\beeq
\label{bgrr2}
\bar{\ro}_\cc= \left\{
\begin{array}{lr}
\frac{1}{\sqrt{K}} \sin\left(-\sqrt{K}\cc\right) & \qquad K>0 \\
-\cc & \qquad K=0 \\
\frac{1}{\sqrt{|K|}} \sinh\left(-\sqrt{|K|}\cc\right) & \qquad K<0 
\end{array} \right\}~.
\eneq
Furthermore, we obtain from Eq.~\eqref{meme}
\begin{equation}
d\bar{\eta}=d\cc=-d\bar{\tilde{\chi}}~, \qquad\qquad
\bar{\tilde{\chi}}=-\cc~,
\end{equation}
for the photon path in a homogeneous universe.
Mind that the physical angular diameter distance is 
\beeq
a\ro={a\rr\over \ff(\rr)}\neq a\rr~.
\eneq
Note that we chose the observer in the background universe to be located at the spatial origin, with conformal time coordinate~$\bar{\eta}_{\bobs}$:
\beeq
\bar{x}_{\bobs}^\mu=(\bar{\eta}_{\bobs}, 0)~, \Dquad
\bar{r}_{\bobs}=0~,\Dquad
\bar{\eta}_{\bobs}=\int_0^\infty \frac{dz}{H(z)}~.
\label{eqn:ObsPosBG}
\eneq

\section{Inhomogeneous Universe with a Non-Zero Spatial Curvature}
\label{inhomo}
Here we compute the light propagation in an inhomogeneous
universe with a non-zero spatial curvature~$K$. The light propagation
in a non-flat universe has been studied in the past, extensively 
in the context of cosmic microwave background \cite{1983ApJ...273....2W,1998ApJ...494..491Z, 1998PhRvD..57.3290H, 1999ApJ...513....1C, 1999PhRvD..59h3506M, 2000PhRvD..62d3004C, 2000GReGr..32.1059C, 2000CQGra..17..871C, 2014JCAP...09..032L, 1996astro.ph..4166H, 2002ARA&A..40..171H}, but also for
luminosity distance \cite{1987MNRAS.228..653S, 1989PhRvD..40.2502F, 1999PThPh.101..903S} and galaxy clustering \cite{2016JCAP...06..013D}, which we will
re-visit in Section~\ref{cosobs}. While relatively simple in a background 
universe with a non-zero curvature,
the light propagation in an inhomogeneous universe involves many subtle
points in the computation of a photon path. As discussed in 
Section~\ref{sec:NonFlat}, the background solution is {\it not} a solution
in an inhomogeneous universe, because it depends on a position, 
which makes the computation substantially more complicated,
compared to the case in a flat universe. We proceed step by step,
paying particular attention to the subtle difference in the light propagation
in a non-flat universe. 

\subsection{Perturbed FRW metric and gauge transformation}
\label{sec:metric}
To account for inhomogeneities in our Universe, we introduce small 
deviations around the background metric 
\beeq
ds^2 = -a^2(1+2\al)d\eta^2 - 2a^2 \mathcal{B}_\alpha d\eta 
d\xx^\alpha + a^2(\bar{g}_{\alpha\beta}+2\mathcal{C}_{\alpha\beta})
d\xx^\alpha d\xx^\beta~,
\label{eqn:metric}
\eneq
where the non-flat nature of 3-hypersurface is encoded in the spatial metric
$\gbar_{\al\be}$ in Eq.~\eqref{gbar} and we assumed a rectangular coordinate
(note that $\gbar_{\al\be}\neq\de_{\al\be}$).
We further decompose the metric tensor into scalar, vector, and tensor, 
according to the symmetry in 3-hypersurface as
\beeq
\label{eq:decom}
\BB_{\alpha}=:\beta_{,\alpha}+B_\alpha~,\Dquad
\CC_{\alpha\beta}=:\varphi~\bar{g}_{\alpha\beta}+\gamma_{,\alpha|\beta}
+C_{(\alpha|\beta)}+C_{\alpha\beta}~,
\eneq
where the comma represents the coordinate derivative and 
the round bracket indicates a symmetrization of the indices 
$C_{(\al|\be)}:=(C_{\al|\be}+C_{\be|\al})/2$. Even in a
rectangular coordinate, the covariant derivative (vertical bar)
is different from a coordinate
derivative due to the non-vanishing spatial curvature~$K$
and the Christoffel symbol~$\bar{\Gamma}^\al_{\be\gamma}$. Two vectors $B_\al$,
$C_\al$ and the tensor~$C_{\al\be}$ are transverse, and the traceless
condition is further imposed for the tensor~$C_{\al\be}$.

The scalar component~$\beta$ in the off-diagonal
metric is trivially obtained by using the transverse condition for the
vector component~$B_\al$ as
\beeq
\nabla^\al B_\al=0~,\Dquad \beta=\Delta^{-1} \nabla^\alpha\BB_\alpha~,  
\eneq
and hence the vector component of the off-diagonal metric is
\beeq
B_\alpha=\BB_\alpha-\nabla_\alpha\Delta^{-1}\nabla^\beta\BB_\beta~,
\eneq
where $\Delta:=\gbar_{\al\be}\nabla^\al\nabla^\be$ is the Laplacian and
$\Delta^{-1}$ is the inverse Laplacian. In the similar way, we derive the two
scalar components in the spatial metric tensor~$\CC_{\al\be}$ as
\bear
\gamma&=&{1\over2}\left(\Delta+{1\over2}\bar R\right)^{-1}
\left(3\Delta^{-1}\nabla^\alpha\nabla^\beta\CC_{\alpha\beta}
-\CC_{\alpha}^{\alpha} \right)~,   \\
\varphi&=&{1\over3}\CC_{\alpha}^{\alpha}-{1\over6}\Delta\left(\Delta
+{1\over2}\bar R\right)^{-1}\left(3\Delta^{-1}\nabla^\alpha\nabla^\beta
\CC_{\alpha\beta}-\CC_{\alpha}^{\alpha} \right)~, 
\enar
and the remaining vector and tensor components are
\bear
C_\alpha&=&2\left(\Delta+{1\over3}\bar R\right)^{-1}\left[\nabla^\beta
\CC_{\alpha\beta}-\nabla_\alpha\Delta^{-1}\nabla^\beta\nabla^\gamma
\CC_{\beta\gamma}\right]~, \\
C_{\alpha\beta}&=&\CC_{\alpha\beta}-{1\over3}\CC^\gamma_\gamma
\gbar_{\alpha\beta}-{1\over2}\left(\nabla_\alpha\nabla_\beta-{1\over3}
\gbar_{\alpha\beta}\Delta\right)\left(\Delta+{1\over2}\bar R\right)^{-1}
\left[3\Delta^{-1}\nabla^\gamma\nabla^\delta\CC_{\gamma\delta}-
\CC^\gamma_\gamma\right] \nonumber\\
&&-2\nabla_{(\alpha}\left(\Delta+{1\over3}\bar R\right)^{-1}\left[
\nabla^\gamma\CC_{\beta)\gamma}-\nabla_{\beta)}\Delta^{-1}\nabla^\gamma
\nabla^\delta\CC_{\gamma\delta}\right]~.
\enar
The presence of the Ricci scalar ($\bar R=6K$) for the 3-hypersurface 
indicates that covariant derivatives are non-commutative:
\beeq
2u_{\al|[\be\ga]}=u_\de\bar R^\de{}_{\al\be\ga}~,\Dquad
\bar R^\al{}_{\be\ga\de}=2K\de^\al_{[\ga}\gbar_{\de]\be}~\qquad
\up{for}~\forall~u^\al~,
\eneq
where the square bracket indicates the anti-symmetrization of the indices
$u_{[\al|\be]}:=(u_{\al|\be}-u_{\be|\al})/2$.
The above SVT decomposition of the metric tensor is unique, except for modes
with infinite wavelength ($k=0$), at which the distinction is ambiguous due to
the inverse Laplacian.

For a general coordinate transformation
\beeq
x^\mu\mapsto\hat{x}^\mu=x^\mu+\xi^\mu~, \Dquad~\xi^\mu=:(T, \LL^\alpha)~,
\Dquad \LL^\alpha=:L^{,\alpha}+L^{\alpha}~,
\label{eqn:coordinatetransformation}
\eneq
the metric tensor transforms as
\beeq
\de_\xi g_{\mu\nu}:=\hat{g}_{\mu\nu}-g_{\mu\nu}=-2\xi_{[\mu;\nu]}~,
\eneq
and the perturbations then gauge transform as
\beeq
\de_\xi\al=-{1\over a}(aT)'~,\Dquad \de_\xi\BB_\al=-T_{,\al}+\LL'_\al~,
\Dquad \de_\xi\CC_{\al\be}=-\HH T\gbar_{\al\be}-\LL_{(\al|\be)}~.
\label{eqn:CoordTransform2}
\eneq
Using the SVT decomposition equations,
we derive the gauge transformation properties
of the metric perturbation variables
\bear
&&
\de_\xi{\beta}=L'-T \, , \Dquad 
\de_\xi{\varphi}=-\HH T \, , \Dquad \de_\xi{\gamma}=-L \, , \\
&&
\de_\xi B_\alpha=L'_\alpha ~,\Dquad
\de_\xi C_\alpha=-L_\alpha \, , \Dquad
\de_\xi C_{\alpha\beta}=0~,
\label{eqn:CoordTransform}
\enar
which are indeed identical to the case in a flat universe as the non-flat
nature is accounted for by~$\gbar_{\al\be}$ and the SVT decomposition equation.

For later use, it is convenient to introduce the scalar shear~$\chi$ and the spatial perturbation variable~$\mathcal{G}_\alpha$ as \cite{1988cpp..conf.....F,Yoo_2014a,YooGrimmMitsou_2018}
\begin{equation}
\chi := a(\beta +\gamma')~, \qquad
\de_\xi \chi = - aT~,\qquad \qquad
\mathcal{G}_\alpha := \gamma_{,\alpha} + C_\alpha~, \qquad 
\de_\xi \mathcal{G}_\alpha = - \mathcal{L}_\alpha~.
\label{eqn:def_G_chi}
\end{equation}
With the scalar shear~$\chi$, we can construct two gauge-invariant potentials
\begin{equation}
\ax:=\alpha-\frac1a~\chi'~,\qquad\qquad
\px:=\varphi-H\chi~,
\label{eqn:GI_scalar_variables}
\end{equation}
corresponding to the Bardeen variables $\alpha_\chi\rightarrow\Phi_A$ and $\varphi_\chi\rightarrow\Phi_H$ in \cite{1980PhRvD..22.1882B}.
A gauge-invariant vector perturbation can be constructed through
\begin{equation}
\Psi_\alpha := B_\alpha + C^{\prime}_\alpha \, ,
\label{eqn:GI_vector_variable}
\end{equation}
which corresponds to $\Psi Q^{(1)}_\alpha$ in \cite{1980PhRvD..22.1882B}.
We parametrize a time-like four-velocity vector as
\beeq
u^\mu={1\over a}\left(1-\al,~\VV^\alpha\right)~,\qquad\qquad
\VV^\alpha:=-U^{,\alpha}+U^{\alpha}~,
\label{eqn:4velocity}
\eneq
where $U^{\alpha}$ is a transverse vector and $U$ is a scalar. 
The gauge transformation of the temporal part is given in Eq.~\eqref{eqn:CoordTransform2}, while the spatial component transforms as 
\beeq
\de_\xi \VV^\alpha=\mathcal{L}^{\alpha\prime}~,\Dquad
\de_\xi U=-L^{\prime}~,\Dquad
\de_\xi U^\alpha=L^{\alpha\prime}
~. 
\eneq
Moreover, we define the scalar velocity potential~$v$ and the corresponding gauge-invariant potential~$\vx$ as
\beeq
v:= U+\beta~,\Dquad
\de_\xi v=-T~,\Dquad
\vx:= v-{1\over a}\chi~.
\eneq
The latter corresponds to the Bardeen variable $\vx\rightarrow v_s^{(0)}$.
Another important gauge-invariant variable
\beeq
\varphi_v := \varphi - \mathcal{H}v~,
\eneq
is the comoving gauge curvature perturbation.
As in \cite{YooGrimmMitsou_2018}, we combine the scalar and the vector perturbation variables to define a gauge-invariant velocity variable:
\begin{equation}
V_\al := -v_{\chi,\al} + U_\alpha - B_\alpha~.
\label{eqn:defV}
\end{equation}

\subsection{Tetrad and photon wave vectors}
\label{ssec:tetrad}
Here we derive the expressions for the tetrad vectors in an inhomogeneous
universe with non-zero spatial curvature, following \cite{YooGrimmMitsou_2018}.
A more general description of the tetrad formalism in cosmology is presented 
in \cite{Mitsou_2020}. 
In an inhomogeneous universe, the observer four-velocity~$u^\mu$ deviates
from the background static motion
\beeq
\label{eq:srcv}
u^\mu:=e_0^\mu=\frac1a\left(1-\al,~\VV^\alpha\right)~,
\Dquad -1=u^\mu u_\mu~,
\eneq
where $\VV^\al$ describes the peculiar motion of the observer and the
observer four-velocity is time-like. The three spatial tetrad vectors in an
inhomogeneous universe can be parametrized as
\beeq
e^\mu_i:=\bar e^\mu_i+\frac1a\left[\delta e_i^\eta~,~\delta e_i^\alpha\right]~,
\label{eq:tetdef}
\eneq
in terms of small deviations from $\bar e_i^\mu$ in the background Universe,
and these perturbations are subject to the normalization condition
$\eta_{ab}=g_{\mu\nu}e^\mu_a e^\nu_b$ with $a,b=0,x,y,z$, which leads to
\beeq
\de e_i^\eta=\ff(\rr)\de^\al_i\left(\VV_\al-\BB_\al\right)~,\Dquad
S_{ij}=\ff^2(\rr)\CC_{\al\be}\de^\al_i\de^\be_j~,
\eneq
where we defined the symmetric and the anti-symmetric part of the spatial
tetrad vector~$\de e_i^\al$ as
\beeq
\frac1a\de e^\al_i:=-\bar e_j^\al\left(S^j{}_i+A^j{}_i\right)~,\Dquad
S_{ij}=S_{(ij)}~,\Dquad A_{ij}=A_{[ij]}~.
\eneq
It is clear from the normalization condition that only the symmetric part
of the spatial tetrad vectors is constrained. 

However, it is crucial
to consider the anti-symmetric part, as the tetrad vectors are four-vectors,
transforming as
\beeq
\de_\xi e^\mu_a=-\pounds_\xi\bar e^\mu_a~,\Dquad
\de_\xi e^\al_i={1\over a}\left(\HH T\ff(\rr)\de^\al_i-{K\over2}
\de_{\be\ga}\xx^\ga\LL^\be\de^\al_i+\LL^\al{}_{,\be}\de^\be_i\right)~,
\eneq
while the symmetric part alone transforms as
\beeq
\de_\xi S_{ij}
=-\ff^2(\rr)\de^\al_i\de^\be_j\left[\HH T\gbar_{\al\be}+\LL_{(\al|\be)}
\right]~.
\eneq
Therefore, the anti-symmetric part of the spatial tetrad vectors should
transform as
\beeq
\de_\xi A_{ji}=-\LL^\al{}_{,\be}\de_{\al[j}\de^\be{}_{i]}~,
\eneq
to make the spatial tetrad vectors a proper four-vector. Given the 
transformation property, we can set the unconstrained anti-symmetric part 
as
\beeq
A_{ji}:=\CCG^\al{}_{,\be}\de_{\al[j}\de^\be{}_{i]}+\ep_{jik}\Omega^k~,
\label{eqn:asymTet}
\eneq
where $\ep_{jik}$ is the Levi-Civita symbol and $\Omega^k$ is gauge-invariant, representing the 
misalignment of the spatial tetrad vectors against the FRW coordinate
(or the misalignment between local and global $x$-$y$-$z$ directions).
The spatial tetrad vector is then
\beeq
\label{srcs}
e_i^\al={\ff(\rr)\over a}\left[\de^\al_i-\de^\be_i\left(\varphi\de^\al_\be
+\CCG^\al{}_{,\be}+C^\al_\be\right)
-\de^{\al j}\ep_{jik}\Omega^k
+{K\over2\ff(\rr)}\de^\al_i\de_{\be\ga}\xx^\ga\CCG^\be\right]~.
\eneq
While the observer velocity is unaffected, the spatial directional vectors
take different form in a universe with non-vanishing spatial curvature at the
linear order in perturbations. 

While the observer in a homogeneous universe is chosen to be at the coordinate 
origin ($x^\al=0$), the observer in an inhomogeneous universe is 
slightly displaced from the coordinate origin by a small perturbation~$\dro$.
With $\ff(\rr_o)=1+\OO(2)$, the spatial tetrad vector at the observer 
position at the linear order in perturbation,
\beeq
\left[ e_i^\al\right]_o
={1\over a}\left[\de^\al_i-\de^\be_i\left(\varphi\de^\al_\be
+\CCG^\al{}_{,\be}+C^\al_\be\right)
-\de^{\al j}\ep_{jik}\Omega^k\right]_{\bobs}~,
\eneq
is identical to one in a universe with a flat space, where the subscript~$\bobs$ indicates
that the quantities in the square bracket are evaluated at the background observer
position~$\bar{x}^\mu_{\bobs}$. Naturally, the
observer at one point has no information about the spatial geometry of the universe, unless
extra information from other spatial position is available.
Having derived
the expression of the local tetrad vectors $[e^\al_i]_o$,
the photon wave vector at the observer position is simply $k^\mu_o=[e^\mu_a]_ok^a$
with the photon wave vector~$k^a$ in Eq.~\eqref{kaobs}
in the observer rest frame, and each component is explicitly
\bear
k^\eta_o &=&
\frac{\omega_o}{a_o}\left[ 1-\al-n^i\de_i^\beta
(\mathcal{U}_\beta-\mathcal{B}_\beta) \right]_{\bobs}~, 
\label{eqn:kobs_0}\\
k^\alpha_o &=&
\frac{\omega_o}{a_o} \left[-n^i\de_i^\al+\mathcal{U}^\alpha+n^i\de_i^\beta 
\left( \delta^\alpha_\beta \varphi+\CCG^\al{}_{,\be}+C^\al_\be\right)
+\de^{\al j}\ep_{jik}n^i\Omega^k\right]_{\bobs}~.
\label{eqn:kobs_alpha}
\enar
It is important to note that the angular direction~$n^i$
 and the photon angular frequency~$\omega_o$ are constants recorded
by the observer in the rest frame, and they are not a function of position.
Last but not least, mind that also the temporal component of the observer coordinate deviates from the one in a homogeneous universe in Eq.~\eqref{eqn:ObsPosBG}:
\beeq
x_o^\mu=:(\bar{\eta}_{\bobs} + \delta\eta_o,~\delta x_o^\alpha)~,
\label{eqn:ObsPos}
\eneq 
where the time lapse~$\delta\eta_o$ and the spatial shift~$\delta x^\al_o$ of the observer
position from $\bar x^\mu_{\bobs}$ are identical to those in a flat universe, as they are determined
by integrating the observer four-velocity~$u^\mu$ along the background trajectory. In our computation of cosmological observables at the linear order, only the time lapse~$\delta\eta_o$ appears in the final expressions:
\beeq
\delta\eta_o=-\int_0^{\bar t_{\bobs}}dt ~\alpha~,\Dquad
\bar t_{\bobs} = \int_0^\infty \frac{dz}{H(z) (1+z)}~.
\label{eqn:DeltaEta}
\eneq
So, note $a_o=a(\eta_o)$ and $\eta_o\neq\bar\eta_{\bobs}$ in Eqs.~\eqref{eqn:kobs_0} and \eqref{eqn:kobs_alpha}.

\subsection{Conformal transformation for photon wave vector}
\label{ssec:conformal}
Since the photon path follows a null path, it is more convenient to compute
the photon path with the conformally transformed metric $\hg_{\mu\nu}$:
\beeq
\hg_{\mu\nu}(x):={1\over a^2}g_{\mu\nu}(\xx)~.
\eneq
The conformal transformation does not affect the null geodesic, hence the null path can be derived via the geodesic equation in the conformally transformed metric. 
Let $\Lambda$ be the affine parameter parametrizing the null path~$x^\mu(\Lambda)$ such that the tangent vector along the photon path is the photon wave vector~$k^\mu$, and let $\lambda$ be the affine parameter parametrizing the null path such that the tangent vector is the conformally transformed wave vector~$\CK^\mu$:
\beeq
k^\mu={dx^\mu\over d\Lambda}~,\Dquad \CK^\mu={dx^\mu\over d\cc}
=\NCC a^2k^\mu~,\Dquad {d\Lambda\over d\cc}=\NCC a^2~.
\label{eqn:RelDiffK}
\eneq
The proportionality constant~$\NCC$ is not constrained through the conformal transformation, and we chose to fix it such that the product (see, \cite{1984ucp..book.....W}),
\beeq
1\equiv(\NCC a\omega)_o~,
\label{eqn:NormalizationC}
\eneq
is unity at the observer position.
Mind that $\NCC$ is a constant, whereas the product $a\omega$ varies with position in an inhomogeneous universe.
With this normalization convention, the conformally transformed wave vector
in the background universe is
\beeq
\hat{\bar k}^\mu_\cc=\left(1,-\Nn^\al\right)_\cc~,\Dquad 
\bar\xx^\al_\cc=\bar\rr_\cc~n^i\de_i^\al~,
\eneq
and it is indeed related to the solution $\bar\rr_\cc$ given in 
Eq.~\eqref{bgrr} in terms of the affine parameter~$\cc$, where we put bars
for~$\bar\rr$ and~$\bar\xx$
to emphasize that they are the solutions in a homogeneous universe.
In an inhomogeneous universe, the conformally transformed wave vector can
be parametrized as
\beeq
\label{CFWAVE}
\CK^\mu_\cc={dx^\mu_\cc\over d\cc}=:
\left(1+\dnu, -\Nn^\alpha - \dNn^\alpha\right)_\cc ~,\Dquad 
\xx^\al_\cc= \bar \xx^\al_\cc+\de \xx^\al_\cc~,
\eneq
where the photon path~$\xx^\al_\cc$ deviates from the straight path 
in the background universe. Regarding the parametrization of $\CK^\mu_\cc$ in 
Eq.~\eqref{CFWAVE}, we emphasize that while the whole photon wave vector is subject
to the null condition $0=\CK_\mu\CK^\mu$ and the geodesic equation
$0=\CK^\nu\CK^\mu{}_{;\nu}$, there exist multiple possibilities for
the separation of~$\CK^\mu$ into the background and perturbation quantities
($\Nn^\al$, $\dNn^\al$). As opposed to the case in a flat universe,
the background direction~$\Nn^\al(\xx_\cc)$ is \textit{position dependent} 
via~$\ff(\rr_\cc)$, and it is a solution to the background
geodesic equation~$\eqref{sgeo}$, only if the 
evaluation position~$\xx_\cc$ is on a straight path~$\bar\xx_\cc$.
This leads to two
possibilities: we could use either $\Nn^\al_\cc:=\Nn^\al(\bar x^\al_\cc)$ or
$\Nn^\al_\cc:=\Nn^\al(x^\al_\cc)$. Either choice results in a different
equation for the evolution of perturbation~$\dNn^\al_\cc$. Since we
prefer to work with local quantities, here we choose to work with the latter:
\beeq
\label{choice}
\Nn^\al_\cc:=\ff(\rr_\cc)n^i\de_i^\al~,
\eneq
where $\rr_\cc$ is the radial coordinate at the photon path~$\xx_\cc$
in an inhomogeneous universe, \textit{not} $\bar\rr_\cc$ on a straight path in a
homogeneous universe.
Of course, the choice is a matter of computational convenience,
and the photon path is independent of any choice. 

With our choice, the directional vector is normalized at a given 
position~$\xx_\cc$:
\beeq
1=\gbar_{\al\be}(\xx_\cc)\Nn^\al_\cc\Nn^\be_\cc~,
\eneq
while the other choice $\Nn^\al(\bar \xx_\cc^\al)$ 
is {\it not} normalized at~$\xx_\cc$ with the metric 
$\gbar_{\al\be}(\xx_\cc)$. However, note that
with our choice the background geodesic equation~\eqref{sgeo} 
is {\it not} satisfied for $\Nn^\al_\cc$ at a position~$\xx_\cc$,
but it can be expanded around~$\bar\xx_\cc$ as
\beeq
\label{Nnint}
\Nn^\al_\cc=\Nn^\al(\bar\xx^\al_\cc)+{K\bar\rr^2_\cc\over2}n^i\de_i^\al\left(
{\drr_\cc \over\bar\rr_\cc}\right)+\OO(2)~,
\eneq
where the first term is the solution to the background
geodesic equation~\eqref{sgeo} and 
\beeq
\label{drrdef}
\drr_\cc:=\de_{i\al}n^i\xx^\al_\cc -\bar\rr_\cc~,
\eneq
is the radial 
distortion in coordinates at a given~$\cc$. Mind that the radial direction
of~$\xx^\al_\cc$ is slightly different from the observed direction~$n^i$
in an inhomogeneous universe. However, at the linear order, the observed
direction~$n^i$ can be used to define the radial distortion~$\drr_\cc$, because
\beeq
\label{rrr}
r_\cc:=\left(\de_{\al\be}\xx^\al_\cc\xx^\be_\cc\right)^{1/2}
=\rbar_\cc\left(1+2\de_{\al\be}
{\bar \xx^\al_\cc\de \xx^\be_\cc\over\rbar_\cc^2}+\OO(2)\right)^{1/2}=
\rbar_\cc+\drr_\cc+\OO(2)~.
\eneq
When
$\Nn^\al_\cc$ is multiplied by a perturbation variable, it can be replaced with
$\Nn^\al(\bar x^\al_\cc)$ at the linear order in perturbations, acting as
an extra advantage for our choice.
In fact, there exists an extra possibility to parametrize the background solution in terms
of the position-dependent angle:
\beeq
n^\al(\xx_\cc):={\xx^\al_\cc\over\rr_\cc}~,
\Dquad \bar\Nn^\al_\cc:=\ff(\rr_\cc)
n^\al_\cc~,
\eneq
with which $\bar\Nn^\al_\cc$ is the solution to Eq.~\eqref{sgeo} at any
position~$\xx_\cc$, because $n^\al$ is the radial direction in FRW coordinates
to the evaluation point~$\xx_\cc$. In this case, however, $n^\al_\cc$ is not the observed angle (mind the difference
$n^\al_\cc\neq n^i\de_i^\al$),
and the use of~$\bar\Nn^\al_\cc$ requires the 
exact knowledge of the photon path~$\xx_\cc$ in an inhomogeneous universe, which we try to obtain by solving the geodesic
equation. With this consideration in mind,
our choice in Eq.~\eqref{choice} results in the boundary conditions
for the conformally transformed wave vector 
\bear
\label{boundary1}
\dnu_o &=&-\left[\al+n^i\de_i^\beta(\mathcal{U}_\beta-\mathcal{B}_\beta)\right]_{\bobs}~,\\
\dNn^\al_o&=&-\left[\mathcal{U}^\alpha+n^i\de_i^\beta 
\left( \delta^\alpha_\beta \varphi+\ga^{,\al}{}_{|\be}
+C^\al{}_{,\be}+C^\al_\be\right)
+\de^{\al j}\ep_{jik}n^i\Omega^k\right]_{\bobs}~,
\label{boundary2}
\enar
where all quantities in the RHS are evaluated at the observer position $\cc=0$,
or at~$\bobs$ up to linear order.

Now we consider the gauge-transformation property of the conformally transformed wave vector~$\CK^\mu_\cc$.
The photon wave vector~$k^\mu$ transforms as a four-vector under the general coordinate transformation in Eq.~\eqref{eqn:coordinatetransformation}
\beeq
\tilde{k}^\mu(\tilde{x}) = \frac{\partial\tilde{x}^\mu}{\partial x^\nu}~k^\nu(x)~,
\eneq
and since it is related to the conformally transformed wave vector~$\CK^\mu$ through Eq.~\eqref{eqn:RelDiffK}, we find
\beeq
\tilde{\CK}^\mu(\tilde{x})
=\tilde{\NCC} a^2(\tilde{\eta})\tilde{k}^\mu(\tilde{x})
=\tilde{\NCC} a^2(\eta)\left[1+2\mathcal{H}T\right]\frac{\partial\tilde{x}^\mu}{\partial x^\nu}~k^\nu(x)
~,
\label{eqn:CTconformalK}
\eneq
at linear order in perturbation.
The normalization condition of the proportionality constant~$\NCC$ in Eq.~\eqref{eqn:NormalizationC} allows us to relate the constant in two different coordinates:
\beeq
\tilde{\NCC}= \NCC\left[1-\mathcal{H}_{\bobs}T_{\bobs}\right]~,
\eneq
where we noted that the frequency~$\omega$ is an observable quantity, independent of coordinate choice.
Equation~\eqref{eqn:CTconformalK} then becomes
\beeq
\tilde{\CK}^\mu(\tilde{x})
=\left[1+2\mathcal{H}T- \mathcal{H}_{\bobs}T_{\bobs}\right]\frac{\partial\tilde{x}^\mu}{\partial x^\nu}~\hat{k}^\nu(x)
~,
\eneq
and the perturbations in the conformally transformed wave vector gauge-transform as
\begin{equation}
\begin{split}
\widetilde{\dnu}(x)&=\dnu + 2\mathcal{H}T - \mathcal{H}_{\bobs}T_{\bobs} + \frac{d}{d\cc}T~, \\\
\widetilde{\dNn}{}^\alpha(x) &= \dNn^\alpha + (2\mathcal{H}T - \mathcal{H}_{\bobs}T_{\bobs})\Nn^\alpha - \frac{d}{d\cc} \mathcal{L}^\alpha - \frac{K\ff}{2}\Nn^\alpha x^\gamma\mathcal{L}_\gamma~,
\end{split}
\label{eq:trwav}
\end{equation}
where $d/d\cc$ is the derivative along the photon path 
\beeq
\label{eq:dlambda}
{d\over d\lambda}\equiv\hat k^\mu{\partial\over\partial x^\mu}=
\left(1+\dnu\right){\partial \over \partial\eta}-\left(\Nn^\alpha+
\dNn^\alpha\right){\partial\over\partial x^\alpha}~,
\eneq
and the function~$\ff(r)$ is related in two coordinates as
\beeq
\ff(\tilde{x})=1+\frac{K}{4}\delta_{\al\be} \tilde{x}^\alpha\tilde{x}^\be 
= \ff(x) + \frac{K}{2}\ff^2(x) x^\alpha\mathcal{L}_\alpha~,
\eneq
and hence
\beeq
\widetilde{\Nn}^\alpha(\tilde{x})\equiv \ff(\tilde{x})n^i\delta_i^\alpha
= \Nn^\alpha(x) + \frac{K}{2}\ff(x)\Nn^\alpha x^\gamma\mathcal{L}_\gamma~.
\eneq
Contrary to $\dnu$, the gauge transformation of $\dNn^\alpha$ is different from that in a flat universe, where the direction vector~$\Nn^\alpha$ is constant and the term proportional to the spatial curvature $K$ vanishes.

\subsection{Solution to the geodesic equation}
\label{ssec:GE}
Having parametrized the conformally transformed wave vector~$\CK^\mu_\cc$ in 
Eqs.~\eqref{CFWAVE} and~\eqref{choice} with the proper boundary condition
in Eqs.~\eqref{boundary1} and~\eqref{boundary2}, we now solve the geodesic equation for the
perturbations~$\dnu_\cc$ and~$\dNn^\al_\cc$. In Section~\ref{ssec:path}
we use these solutions to find the source position~$x^\mu_\cc$ as a function of
the affine parameter~$\cc$.
The null condition for the photon wave vector is
\beeq
\label{null}
0=\hg_{\mu\nu}\CK^\mu\CK^\nu=\dnu -\Nn_\al\dNn^\alpha
+ \al -\BB_\alpha \Nn^\alpha - \CC_{\alpha\beta} \Nn^\alpha  \Nn^\beta~,
\eneq
where all quantities are evaluated at~$\xx_\cc$ and the normalization
$\Nn_\al\Nn^\al=1$ is used (this is valid for our choice only).
The null equation is identical to that in a flat universe, if the
direction vector~$\Nn^\al_\cc$ is replaced with~$n^i\de_i^\al$. Indeed,
this is naturally achieved with our choice for~$\Nn^\al_\cc$ in the limit
$K\RA0$, due to the proper normalization~$\Nn_\al\Nn^\al=1$ 
at a given position.
The geodesic equation for~$\CK^\mu$ is then simply
\beeq
0=\CK^\nu\CK^\mu{}_{;\nu}
=\frac{d}{d\cc}\CK^\mu + \hat{\Gamma}^\mu_{\rho\sigma}\CK^\rho\CK^\sigma~,
\eneq
where $\hat\Gamma^\mu_{\rho\sigma}$ is the Christoffel symbol based on
the conformally transformed metric~$\hg_{\mu\nu}$ and $d/d\cc$ is 
the derivative along the photon path given in Eq.~\eqref{eq:dlambda}.

The temporal component of the geodesic equation can be written as
\beeq
0= \frac{d}{d\cc} \CK^\eta+\hat\Gamma^\eta_{\mu\nu}\CK^\mu\CK^\nu
=\frac{d}{d\cc} \dnu+\delta\hat\Gamma^\eta~,
\label{eqn:GE_temporal_component}
\eneq
where we defined
\bear
\delta\hat\Gamma^\eta&:=& \hat\Gamma^\eta_{\mu\nu}\CK^\mu\CK^\nu=
\al'-2\al_{,\alpha}\Nn^\alpha+\left(\BB_{\alpha|\beta}+\CC_{\alpha\beta}'
\right) \Nn^\alpha \Nn^\beta \\
&=&\frac{d}{d\cc}\left[2\al_\chi -\Psi_{\alpha} \Nn^\alpha  + 2H\chi + 
\frac{d}{d\cc}\left(\frac{\chi}{a}\right)\right] - 
\left( \al_\chi - \varphi_\chi - \Psi_{\alpha}\Nn^\alpha  - 
C_{\alpha\beta}\Nn^\alpha \Nn^\beta \right)'~.\nonumber
\enar
With a non-vanishing spatial curvature, it is important in the derivation
to note the subtle difference
\beeq
0=\Nn^\be\Nn^\al{}_{|\be}+\OO(1)~,\Dquad 
0\neq\Nn^\be\Nn^\al{}_{,\be}=-{d\over d\cc}\Nn^\al+\OO(1)~,
\eneq
even with our choice for~$\Nn^\al_\cc$.
The temporal component of the geodesic equation can be readily solved by
integrating over the affine parameter to yield
\beeq
\label{dnusol}
\dnu_\cc-\dnu_o=
\left[-2\alpha_\chi +\Psi_\parallel - 2H\chi - \frac{d}{d\lambda}
\left(\frac{\chi}{a}\right)\right]_0^\cc
+ \int_0^\cc d\tilde\cc \left( \alpha_\chi - \varphi_\chi - \Psi_\parallel 
- C_\parallel \right)'~,
\eneq
where we introduced the short-hand notation
\beeq
\Psi_\parallel:=\Psi_\alpha\Nn^\alpha~,\Dquad C_\parallel:=C_{\al\be}
\Nn^\al \Nn^\be~.
\eneq
The temporal solution is again identical to one with a vanishing spatial
curvature, if a proper directional vector~$\Nn^\al$ is used.
Next the spatial component of the geodesic equation can be written in
a similar way as
\beeq
\label{SSgeo}
0={d\over d\cc}\CK^\al+\hat\Gamma^\al_{\mu\nu}\CK^\mu\CK^\nu=
\Nn^\be \Nn^\al{}_{|\be}+\dNn^\be \Nn^\al{}_{|\be}-\dNn^{\al\prime}
+\Nn^\be\dNn^\al{}_{|\be}+\widehat{\de\Gamma}^\al~,
\eneq
where we defined 
\bear
\de\hat\Ga^\al&:=&\hat{\Gamma}^\al_{\mu\nu}\CK^\mu\CK^\nu-\hat{\bar\Gamma}^\al
_{\mu\nu}\hat{\bar k}^\mu \hat{\bar k}^\nu=
\hat\Ga^\al_{\eta\eta}-2\hat\Ga^\al_{\eta\be}\Nn^\be
+\hat\Ga^\al_{\be\ga}\CK^\be\CK^\ga-\hat{\bar\Ga}^\al_{\be\ga}\Nn^\be\Nn^\ga\\
&=&
\al^{,\al}-\BB^{\al\prime}-\left(\BB_{\be}{}^{|\al}-\BB^\al{}_{|\be}+2
\CC^{\al\prime}_\be\right)\Nn^\be+\left[2\CC^\al_{(\be|\ga)}-\CC_{\be\ga}{}
^{|\al}\right]\Nn^\be\Nn^\ga+2\hat{\bar\Ga}^\al_{\be\ga}\Nn^\be\dNn^\ga~,
\nonumber\\
\label{GGHDEF}
\widehat{\de\Gamma}{}^\al&:=&\de\hat\Ga^\al
-2\hat{\bar\Ga}^\al_{\be\ga}\Nn^\be\dNn^\ga~,
\enar
and  the terms with the background Christoffel symbol 
in $\hat\Ga^\al_{\mu\nu}\CK^\mu\CK^\nu$
are re-arranged in the geodesic
equation~\eqref{SSgeo} to make the spatial derivative covariant. Mind that only $\widehat{\de\Ga}{}^\al$ is a 3-vector, 
while $\de\hat\Ga^\al$ is not. Once again, the first two terms in
the geodesic equation~\eqref{SSgeo} do not vanish. It proves useful to
re-arrange $\GGH^\al$ as
\bear
\label{GGHCO}
\GGH_\al&=&~\gbar_{\al\be}\GGH^\be
=(\alpha_\chi-\varphi_\chi)_{,\alpha} - \Nn^\beta\Psi_{\beta,\alpha} - \Nn^\beta \Nn^\gamma C_{\beta\gamma,\alpha} - \frac{d}{d\lambda}\left(\Psi_\alpha + 2 \Nn^\beta C_{\alpha\beta}\right) \nnn
&&- 2 \Nn_\alpha\frac{d}{d\lambda} \varphi+\frac{d}{d\lambda}\left[ \frac{1}{\ff^2}\frac{d}{d\lambda}(\ff^2\mathcal{G}_\alpha)\right] + \frac{K\ff}{2}\mathcal{G}_\alpha - K\Nn_\alpha \mathcal{G}_\parallel
\\\
\label{GGHCO2}
&=&~\left(\alpha_\chi-\varphi_\chi- \Psi_\parallel
- C_\parallel\right)_{,\alpha} + \frac{K\bar{r}}{2}\Nn_\alpha
\left(\Psi_\parallel + 2 C_\parallel\right)
- \frac{d}{d\lambda}
\left(\Psi_\alpha + 2 \Nn^\beta C_{\alpha\beta}\right)~ \nnn
&& - 2 \Nn_\alpha\frac{d}{d\lambda} \varphi+ \frac{1}{\ff}\frac{d^2}
{d\lambda^2}(\ff\mathcal{G}_\alpha) + K\left(\mathcal{G}_\alpha -
\Nn_\alpha \mathcal{G}_\parallel\right)~.
\enar
To make further progress, we first define
the radial and the tangential directions. The observer in the rest frame 
measures the observed photon direction~$n^i$ in terms of two angles in the
rest frame and uses it to define the radial direction. Two unit directions
$\ttt^i$ and~$\pp^i$ perpendicular to~$n^i$ 
\beeq
\ttt^i={\pa\over\pa\ttt}~n^i~,\Dquad 
\pp^i={1\over\sin\ttt}{\pa\over\pa\pp}~n^i~,
\label{eqn:Definition_ttt_pp}
\eneq
are then used to define the
observed tangential directions. Apparent from their indices, these unit
directional vectors are defined in the observer rest frame only.
Using these directions, we split the deviation of the photon path
from the background path into the radial and the angular parts:
\beeq
\xx^\al_\cc=:n^i\de_i^\al\left(\rbar_\cc+\drr_\cc\right)+\de\xx^\al_\perp~,
\Dquad \de x^\al_\perp:=\ttt^i\de_i^\al~\rbar_\cc\dtt_\cc
+\pp^i\de_i^\al~\rbar_\cc\sin\ttt~\dpp_\cc~.
\eneq
The decomposition is nothing less than an expression in a spherical
coordinate based on the observed angle $(\ttt,\pp)$ in the rest frame of
the observer, though the photon path~$\xx^\al_\cc$ is a (FRW) coordinate
of the spacetime manifold. Note that $n^i$ is in fact slightly different from
the radial coordinate of~$\xx^\al_\cc$ in an inhomogeneous universe.
However, as shown in Eq.~\eqref{rrr} the radial coordinate 
at the linear order is indeed
\beeq
r_\cc=\rbar_\cc+\drr_\cc+\OO(2)~, \Dquad
\drr_\cc:=\delta_{i\al}n^i x^\al - \rbar_\cc~.
\eneq
We then write three terms with covariant derivative in the geodesic 
equation~\eqref{SSgeo} as
\bear
\Nn^\be \Nn^\al{}_{|\be}&=&-{K\ff\over2}\bigg[n^i\de_i^\al\left(\de_{j\be}n^j
\xx^\be\right)-\xx^\al\bigg]=\frac12K\ff\de\xx_\perp^\al~,\\
\dNn^\be \Nn^\al{}_{|\be}&=&-{K\over2}\bigg[\dNn^\al\left(\de_{j\be}n^j\xx^\be
\right)-x^\al \left(\de_{j\be}n^j\dNn^\be\right)\bigg]=-\frac12K\rbar_\cc
\dNn^\al_\perp~,\\
-\dNn^{\al\prime}+\Nn^\be\dNn^\al{}_{|\be}
&=&-{d\over d\cc}\dNn^\al+\hat{\bar\Gamma}^\al_{\be\ga}\Nn^\be\dNn^\ga
=-{d\over d\cc}\dNn^\al-\frac12K\rbar_\cc\dNn^\al~,
\enar
where we also decomposed
\beeq
\dNn^\al=:\Nn^\al~\dNn_\para+\dNn^\al_\perp~,\Dquad 
\dNn_\para=\Nn_\al\dNn^\al~,\Dquad 0=\Nn_\al\dNn^\al_\perp~.
\eneq
While $x^\al_\cc$ is a coordinate, $\dNn^\al$ is a 3-vector. All these
terms reflect the peculiarities in a non-flat universe that vanish in a flat universe.
So we decomposed the former with~$n^i$ and the
latter with~$\Nn^\al$. At last, 
the spatial geodesic equation~\eqref{SSgeo} becomes
\beeq
\label{SSgeo2}
{d\over d\cc}\dNn^\al=\widehat{\de\Gamma}{}^\al
+\frac12K\left(\ff~\de x^\al_\perp-2\rbar_\cc\dNn^\al_\perp
-\rbar_\cc~\Nn^\al\dNn_\para\right)~.
\eneq
Mind that the derivative of~$\dNn^\al$ with respect to the affine 
parameter~$\cc$
is {\it not} covariant, which is however compensated for by the terms
multiplied with~$K$. The extra terms with~$K$ represent the non-flat
nature of light propagation, which we discuss further below.

Now we are in a position to tackle the spatial geodesic equation. We will
split the geodesic equation into the parallel and the perpendicular
components in terms of the observed angle~$n^i$. Given
that the extra terms with~$K$ in the geodesic equation can be manipulated as
\beeq
\label{trick}
\frac12K\rr={d\ff\over d\rr}={d\ff/d\cc\over d\rr/d\cc} + \mathcal{O}(1)
=-\frac1\ff{d\ff\over d\cc} + \mathcal{O}(1)~,
\eneq
the parallel component of the geodesic equation can be simplified as
\beeq
{d\over d\cc}\left(\Nn_\al\dNn^\al\right)=\Nn_\al\GGH^\al~,
\eneq
where we used
\beeq
\label{trick2}
\Nn_\al=\gbar_{\al\be}\Nn^\be={\de_{\al i}n^i\over \ff}~,\Dquad 
{d\over d\cc}\Nn_\al=-{\Nn_\al\over \ff}{d\ff\over d\cc}+ \mathcal{O}(1)~.
\eneq
Finally, multiplying Eq.~\eqref{GGHCO2} with~$\Nn^\al$  leads to
\beeq
\Nn^\al\GGH_\al=
\Nn^\alpha\frac{\partial}{\partial x^\alpha}\left(\alpha_\chi-\varphi_\chi 
- \Psi_\parallel - C_\parallel\right)- \frac{d}{d\lambda} \left[2\varphi 
+ \Psi_\parallel + 2C_\parallel - 
\frac{d}{d\lambda} \mathcal{G}_\parallel\right]~,
\eneq
where the computation of the term with $\mathcal{G}_\al$ is straightforward:
\beeq
\Nn^\al\frac{1}{\ff}\frac{d^2}{d\lambda^2}(\ff\mathcal{G}_\alpha) = 
n^i\delta_i^\al\frac{d^2}{d\lambda^2}(\ff\mathcal{G}_\alpha)=
\frac{d^2}{d\lambda^2}(\Nn^\al\mathcal{G}_\alpha)~.
\eneq
The parallel component of the geodesic equation can then be readily 
solved by integrating over the affine parameter 
\beeq
(\dNn_\para)_\cc - (\dNn_\para)_o =
-\left[2\varphi + \Psi_\parallel + 2C_\parallel- \frac{d}{d\cc} 
\mathcal{G}_\parallel\right]_{0}^\lambda
+ \int_{0}^\lambda d\tilde\lambda ~\ff(\rbar_{\tilde\cc})
\frac{\partial}{\partial \rbar} 
\left(\alpha_\chi-\varphi_\chi - \Psi_\parallel- C_\parallel\right)~,
\label{eqn:dNparallel}
\eneq
where the gradient in the integration became the derivative along the
observed direction~$\rbar$ in the background universe.
Note that the background photon path $d\rbar=-\ff d\cc$ involves not only the spatial
derivative, but also the time derivative.
The structure of the solution is similar to the case in a flat universe,
if the directional vector~$\Nn^\al$ is replaced with~$n^i\de_i^\al$,
except that the line-of-sight integration includes an extra factor~$\ff(\rr)$, 
reflecting that the geometry is non-flat.

Now we move to the perpendicular component of the geodesic equation.
We find it convenient to define two unit directional 3-vectors perpendicular
to~$\Nn^\al$ in a FRW coordinate:
\beeq
\TT^\al:=\ff(\rr)\ttt^i\delta_i^\alpha ~,\Dquad
\PP^\al:=\ff(\rr)\pp^i\delta_i^\alpha~,
\label{eqn:DefTTPP}
\eneq
where they are normalized in terms of~$\gbar_{\al\be}$. With the same
manipulation in Eqs.~\eqref{trick} and~\eqref{trick2}, the perpendicular
component of the geodesic equation~\eqref{SSgeo2} can be obtained as
\beeq
\label{const}
\frac{d}{d\cc}\left(\TT_\al\dNn^\al\right) =\TT_\al\GGH^\al
+ \frac{1}{2}K\ff(\rr)\TT_\al\de x^\al_\cc
-\frac12K\rbar_\cc \TT_\al\dNn^\alpha  ~.
\eneq
In a flat universe, two extra terms with~$K$ are absent, but
it is clear that in a non-flat universe the light propagation perpendicular
to the observed direction is more involved, as the differential equation
involves~$\TT_\al\de x^\al_\cc$, which we attempt to obtain as a solution.
By taking another derivative of the equation with respect to the affine
parameter and using the definition of~$\CK^\mu$, we arrive at the perpendicular
component of the geodesic equation
\beeq
\frac{d}{d\cc}\left[\frac{1}{\ff}\frac{d}{d\cc}
\left(\TT_\alpha\dNn^\alpha\right)\right] =
 \frac{1}{\ff} \frac{d}{d\cc}\left[\TT_\alpha \GGH^\al\right]~.
\eneq
A formal solution to the geodesic equation can be readily derived as
\beeq
\label{forsol2}
(\TT_\al\dNn^\al)_\cc-(\TT_\al\dNn^\al)_o
= \left[\frac{d}{d\cc}\left(\TT_\alpha\dNn^\al\right)\right]_{\bobs} 
\int_{0}^\cc  d\cc'\ff_{\cc'}
+ \int_{0}^\cc  d\cc' \ff_{\cc'}\int_{0}^{\cc'} 
 d\cc''\frac{1}{\ff_{\cc''}}\frac{d}{d\cc''}
(\TT_\alpha\GGH^\al )_{\cc''}~,
 \eneq
and after some manipulation detailed in Appendix~\ref{angdetail} we obtain
the solution
\bear
\label{polardn}
&&(\TT_\al\dNn^\al)_\cc-(\TT_\al\dNn^\al)_o=
-\frac12K\bar{\rr}_\cc\bigg[\left(\TT_\al\delta\xx^\alpha+\TT^\al\CCG_\alpha
\right)_{\bobs}+\TT^\al\CCG_\al\bigg]
- \int_{0}^{\bar{\rr}_\cc} \frac{d\bar{\rr}'}{\bar{\rr}'}  \TT^\alpha
\left(\Psi_\alpha + 2 \Nn^\beta C_{\alpha\beta} \right)\\
&&\qquad
-\left[\TT^\alpha \Psi_\alpha +2\TT^\alpha \Nn^\beta C_{\alpha\beta}
- \frac{d}{d\cc^{\prime}}\left(\TT^\al\mathcal{G}_\alpha\right)
\right]_{0}^\cc
- \int_{0}^{\bar{\rr}_\cc}  {d\bar{\rr}'\over\rbar'}\left[1+\frac12K
\frac{(\bar{\rr}_\cc-\bar{\rr}')\rbar'}{\ff(\bar{\rr}')}\right]
\frac{\partial}{\partial \theta}\left( \alpha_\chi-\varphi_\chi-
\Psi_\parallel - C_\parallel \right)~,\nonumber
\enar
where $(\TT_\al\dNn^\al)_o$ is given in Eq.~\eqref{bcan} and we replaced the line-of-sight integration over $d\cc$ with the background line-of-sight integration $d\rbar$ as
\beeq
\int_0^\cc d\cc = -\int_0^{\rbar_\cc}\frac{d\rbar}{\ff(\rbar)}+\mathcal{O}(1)~.
\eneq
The most obvious differences to the expression in a flat Universe are two
extra terms proportional to the spatial curvature~$K$.
However, keep in mind that the direction vectors~$\Nn^\alpha$ and $\TT^\alpha$
contain the function~$\ff(r)$ and hence depend on the spatial curvature.
The solution for the azimuthal direction can be readily obtained from
the solution with extra factor~$\sin\ttt$ due to~$\PP^\al$.

Finally, we verify the gauge transformation properties of $\TT_\al\dNn^\al$
in Eq.~\eqref{polardn} as a sanity check. The individual terms 
in Eq.~\eqref{polardn} are already expressed in terms of gauge-invariant
variables, except two terms with $\TT^\al\CCG_\al$
(note that the combination involving $\delta x^\al$ and $\mathcal{G}^\al$ at $\bobs$ is also gauge-invariant). 
The transformation property of $\CCG^\alpha$ is given in Eq.~\eqref{eqn:def_G_chi}, hence Eq.~\eqref{polardn} gauge-transforms as
\beeq
\left. (\TT_\al\widetilde{\dNn^\al})\right|^\cc_0=
\left.\vphantom{\widetilde{\dNn^\al}} (\TT_\al\dNn^\al)\right|^\cc_0
+\frac12K\bar{\rr}_\cc\TT^\al\LL_\alpha
- \left[ \frac{d}{d\cc^{\prime}}\left(\TT^\al\LL_\alpha\right)
\right]_{0}^\cc~.
\label{eqn:GTtang1}
\eneq
The boundary term at the observer position in the LHS is given in Eq.~\eqref{bcan}, and it transforms as
\beeq
(\TT_\al\widetilde{\dNn^\al})_o =(\TT_\al\dNn^\al)_o
-\ttt^i\de_{i\al}\left(\LL^{\alpha\prime}-\Nn^\beta
\LL^{\al}{}_{,\be} \right)_{\bobs}
=(\TT_\al\dNn^\al)_o
-\left[\TT_{\al}\frac{d}{d\cc}\LL^{\alpha}\right]_{\bobs}
~,
\label{eqn:GTtang2}
\eneq
according to the transformation properties derived in Sec.~\ref{sec:metric}.
By noting that 
\beeq
\frac{d}{d\cc}\left(\TT_\al\LL^\al\right) =
\frac{1}{2}K\rbar_\cc \TT_\al\LL^\al + \TT_\al\frac{d}{d\cc}\LL^\al~,
\eneq
and combining Eqs.~\eqref{eqn:GTtang1} and \eqref{eqn:GTtang2}, we find the gauge-transformation of the tangential part of $\dNn^\al$ to be
\beeq
\TT_\al\widetilde{\dNn^\al}=\TT_\al\dNn^\al-\TT_\al{d\over d\cc}\LL^\al~,
\eneq
This is in agreement with the transformation property derived in Eq.~\eqref{eq:trwav}.

\subsection{Source position at the observed redshift}
\label{ssec:path}
\begin{figure}
\centering
\begin{subfigure}{.5\textwidth}
  \centering
  \includegraphics[width=0.95\linewidth]{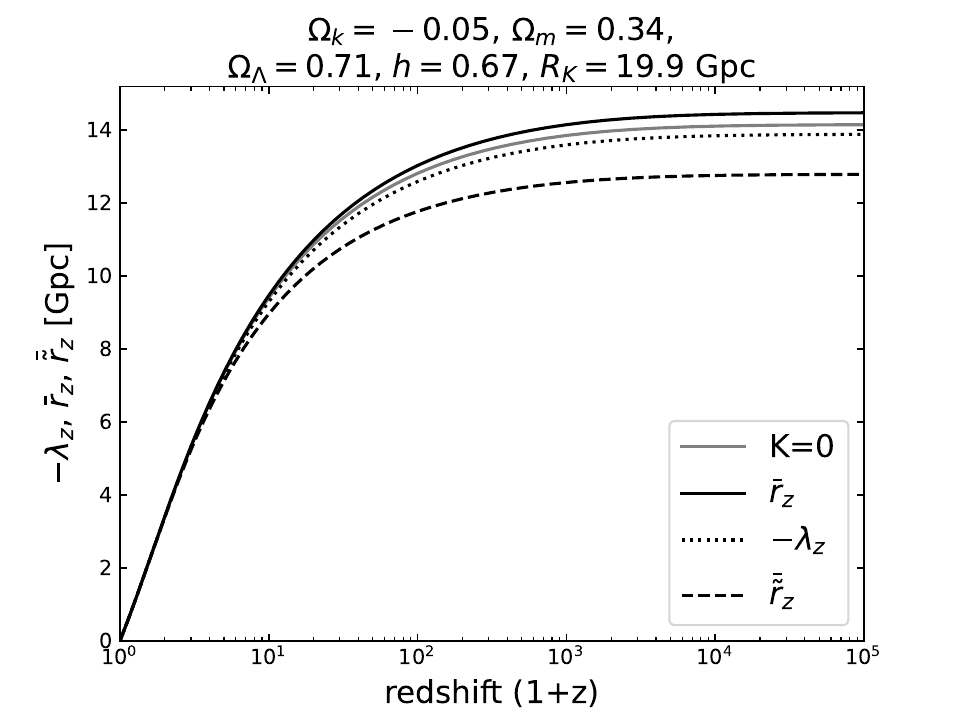}
  \caption{ }
  \label{fig:radius}
\end{subfigure}%
\begin{subfigure}{.5\textwidth}
  \centering
  \includegraphics[width=0.95\linewidth]{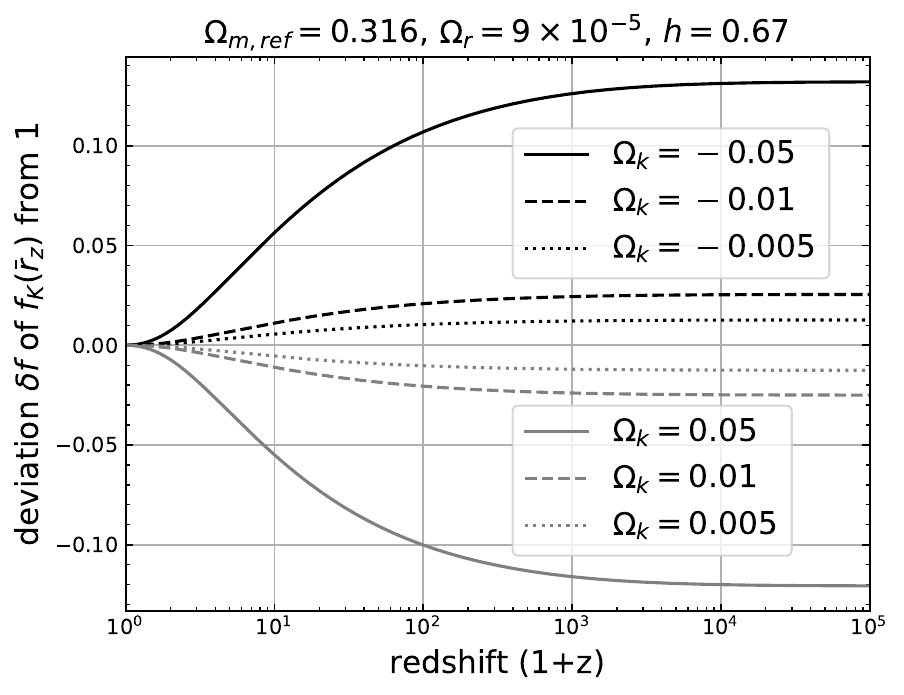}
  \caption{ }
  \label{fig:deltaF}
\end{subfigure}
\caption{(a) The absolute value of the affine parameter~$\cc_z$ (dotted), the background radial coordinate~$\bar{\rr}_z\equiv\bar{\rr}_{\cc_z}$ (solid) and the comoving angular diameter distance~$\bar{\ro}_z\equiv\bar{\ro}_{\cc_z}$ (dashed) as a function of the redshift~$z$ for a closed Universe. The cosmological parameters and the curvature radius~$R_K$ are indicated in the Figure. The grey solid line shows a flat Universe ($\Omega_m=0.316$ and $\Omega_r=9\times10^{-5}$) for comparison, as in that case $\bar{\rr}_z=\bar{\ro}_z=-\cc_z$. 
(b) The deviation~$\delta f \equiv \ff(\bar{\rr}_z)-1 = \frac{1}{4}K\bar{\rr}_z^2$ as a function of the redshift~$z$ for different cosmological models. The values chosen for the curvature density today, $\Omega_k$, are indicated in the Figure. The radiation density today is chosen to be $\Omega_r=9\times10^{-5}$ for all models, while the matter density parameter today is $\Omega_m=\Omega_{m,\text{ref}}-\frac{1}{2}\Omega_k$, and the dark energy density parameter~$\Omega_\Lambda$ today is chosen such that the density parameters add up to unity.}
\label{fig:test}
\end{figure}

So far, we have derived the solution to the geodesic equation in terms of 
affine parameter~$\cc$, and its defining relation in Eq.~\eqref{affine}
yields
\beeq
\bar\eta_\cc-\bar\eta_{\bobs}=\cc~,
\eneq
which is nothing but the temporal solution in a homogeneous universe, while
the spatial position~$\rbar_\cc$ is also given in Eq.~\eqref{bgrr}.
However, neither this affine parameter nor the time coordinate is measurable
in practice. In cosmology, we express the radial position of any light
sources in terms of the observed redshift~$z$, which then acts 
as an observable affine parameter~$\cc_z$, defined as
\beeq
\cc_z:=\bar\eta_z-\bar\eta_{\bobs}=-\int_0^z{dz'\over H(z')}~.
\label{eqn:LambdaZ}
\eneq
Figure~\ref{fig:radius} shows the affine parameter~$\cc_z$ for a slightly closed Universe (dotted line) in comparison to a flat Universe (solid grey line).
The background radial coordinate~$\bar{\rr}_z\equiv\bar{\rr}_{\cc_z}$ (solid line) and the comoving angular diameter distance~$\bar{\ro}_z\equiv\bar{\ro}_{\cc_z}$ (dashed line) are both functions of the redshift due to their dependence on the affine parameter~$\cc_z$ in Eqs.~\eqref{bgrr} and \eqref{bgrr2}. 
In a flat Universe, all three quantities coincide: $\rbar_z=\bar{\ro}_z=-\cc_z$.
Evident in the Figure, the comoving angular diameter distance~$\bar{\ro}_z$ is smaller than $\rbar_z$.
As the denominator in the relation between the two radial coordinates in Eq.~\eqref{rrro} is larger than unity for a closed Universe with $K>0$, it follows that $\bar{\ro}_z\leq\rbar_z$.
The radiation density parameter today is chosen to be $\Omega_r=9\times10^{-5}$, while the values of the remaining density parameters, $\Omega_m$, $\Omega_r$ and $\Omega_\Lambda$, are indicated in the title of the Figure.
The same value for the radiation density parameter was chosen for the flat model, with a matter density parameter of $\Omega_m=0.316$.
Fig.~\ref{fig:deltaF} shows the deviation~$\delta f$ of the function $\ff(\rbar_z)$ in Eq.~\eqref{eqn:DefF} from unity for different values of the curvature density~$\Omega_k$.
The matter density parameter today is chosen to be $\Omega_m=\Omega_{m,\text{ref}}-\frac{1}{2}\Omega_k$, with $\Omega_{m,\text{ref}}=0.316$, while the radiation density is $\Omega_r=9\times10^{-5}$ for all models.
The deviations from unity (i.e. from a flat Universe) are between 12\% and 13\% for $\Omega_k=\pm0.05$ at high redshift $z>100$ and reduce to $\sim$1\% for $\Omega_k=\pm0.005$.
The negative values of $\Omega_k$ are allowed at 1-$\sigma$ confidence level by the Planck measurements \cite{Planck2018CosmParam} in the 1-parameter extensions to the $\Lambda$CDM model for combinations of the Planck power spectra, and the combinations of the Planck power spectra and Planck lensing.
Note that the curve corresponding to a specific value of the curvature density parameter, $\Omega_k=\Omega_k^*$, is not equal to the curve corresponding to $\Omega_k=-\Omega_k^*$ with opposite sign, as the difference in cosmological parameters leads to a difference in $\cc_z$ in Eq.~\eqref{eqn:LambdaZ} and the form of the radius~$\rbar_z$ differs for a closed and an open Universe (see Eq.~\eqref{bgrr}).
Had we plotted the deviation~$\delta f$ as a function of the radius~$\rbar_z$, then the curve corresponding to $K= K^*$ would be equal to the curve corresponding to $K=-K^*$  with opposite sign.
Naturally, the differences in any quantities in the Figure become visible at $z\approx10$, where the characteristic scale is comparable to the curvature radius $\cc_z\approx\rbar_z\sim R_K/2$.
Moreover, due to the scaling of the curvature density, those differences saturate at high redshift.

The observed redshift is simply the ratio of the photon frequency~$\omega_s$ at
emission to one~$\omega_o$ at observation
\beeq
\label{redshift}
1+z:={\omega_s\over\omega_o}={(u_\mu k^\mu)_s\over(u_\mu k^\mu)_o}=:
{1+\dz\over a(\eta_s)}~,
\eneq
where $\eta_s:=\eta_{\cc_s}$ and we defined the perturbation~$\dz$ in the 
observed redshift. 
The perturbation~$\dz$ accounts for all the effects that affect the photon frequency apart from cosmic expansion, such as the peculiar velocity for example.
Multiplying the four-velocity in Eq.~\eqref{eqn:4velocity} with the photon wave vector~$k^\mu$ in Eq.~\eqref{eqn:RelDiffK} leads to
\beeq
u_\mu k^\mu =- \frac{1}{\mathbb{C}a}\left[1+\dnu + \al + \VV_\parallel - \BB_\parallel\right]~,
\eneq
with $\dnu$ given in Eq.~\eqref{dnusol}, so Eq.~\eqref{redshift} yields
\beeq
\dz = -H\chi + (\mathcal{H}_{\bobs}\delta\eta_o+H_{\bobs}\chi_{\bobs}) + \left[V_\parallel - \alpha_\chi+\Psi_\parallel\right]^{\lambda_s}_0 - \int_0^{\bar{\rr}_s} \frac{d\bar{\rr}}{\ff(\bar{\rr})}\left(\alpha_\chi - \varphi_\chi-\Psi_\parallel-C_\parallel\right)'~,
\label{eqn:dz}
\eneq
where we used $a(\eta_o)=1+\mathcal{H}_{\bobs}\delta\eta_o$ and $V_\al$ is the gauge-invariant variable introduced in Eq.~\eqref{eqn:defV}.
The expression for a non-flat Universe is similar,
the only difference being the factor $\frac{1}{\ff}$ in the integrand and $\Nn^\al=\ff(r) n^i\de_i^\al$ for the parallel components.
Since the combination $(\mathcal{H}\delta\eta+H\chi)$ is gauge invariant, the only gauge-dependent term in the expression of $\dz$ is $-H\chi$, and we can define the variable
\beeq
\dz_\chi:=\dz +(H\chi)_s~,
\label{eqn:dz_GI}
\eneq
which is gauge-invariant at linear order.
Furthermore, it is noted that the affine parameter~$\cc_z$ does 
not coincide with the affine parameter~$\cc_s$ at the source position,
except in the background universe. Hence, we define this deviation 
$\Delta \cc_s$ in the affine parameter as
\beeq
\cc_s=:\cc_z + \Dcc_s~,
\label{eqn:LambdaS}
\eneq
and $\Delta\eta_s$ in the corresponding time coordinate as
\beeq
\eta_s=:\bar{\eta}_z +\Delta\eta_s~,\Dquad
\Delta\eta_s = \frac{\delta z}{\mathcal{H}(\bar{\eta}_z)}~,
\label{eqn:TimeCoordSource}
\eneq
where we used Eq.~\eqref{redshift} to express $\Delta\eta_s$ 
in terms of the perturbation~$\dz$ in the redshift.

Having ensured that all the ingredients are in place, we now derive the 
coordinate of a light source by integrating the photon wave 
vector~$\CK^\mu$ over the affine parameter and express it in terms
of the observed position~$\bar x^\mu_z$ and the deviation~$\Delta x^\mu_s$:
\beeq
x^\mu_s=x^\mu_o+\int_0^{\cc_s}d\cc~\CK^\mu_\cc
=:\bar x^\mu_z+\Delta x^\mu_s~,\Dquad \bar x^\mu_z:=\left(\bar\eta_z,~
\rbar_zn^i\de_i^\al\right)~.
\eneq
The time coordinate of the source is then
\beeq
\eta_s=\eta_o+\int_0^{\cc_s}d\cc~\CK^\eta_\cc=\bar\eta_z+
\de\eta_o+\Dcc_s+\int_0^{\cc_z}d\cc~\dnu_\cc\equiv \bar\eta_z+\Delta\eta_s~,
\label{eqn:EtaSource}
\eneq
indeed consistent with expression~$\Delta\eta_s$ in Eq.~\eqref{eqn:TimeCoordSource}. 
A special care is needed for the spatial coordinate of the source,
as a simple integration over the affine parameter yields
\beeq
\label{src}
\xx^\al_s=\xx_o^\al+\int_0^{\cc_s}d\cc~\CK^\al_\cc=\de\xx^\al_o-\int_0^{\cc_s}
d\cc\left(\Nn^\al+\dNn^\al\right)\equiv\rbar_zn^i\de_i^\al+\Delta \xx^\al_s~,
\eneq
in which the integration of~$\Nn^\al(\xx_\cc)$ requires
the correct photon path~$\xx^\mu_\cc$ in an inhomogeneous universe
we try to derive. As noted in Eq.~\eqref{Nnint}, since the integration 
of~$\Nn^\al(\xx_\cc)$ in Eq.~\eqref{src} is
\beeq
-\int_0^{\cc_s}d\cc~N^\al(\xx_\cc)=\left[\rbar_{\cc_s}+
\int_0^{\cc_s}d\cc~{1\over \ff}{d\ff\over d\cc}~\drr_\cc\right]n^i\de_i^\al~,
\eneq
we multiply $n^i\de_{i\al}$ by Eq.~\eqref{src} to obtain equation
for~$\drr_\cc$ and
take the derivative with respect to affine parameter to arrive at
the geodesic equation for the radial distortion
\beeq
{d\over d\cc}\left({\drr_\cc\over \ff}\right)=-\Nn_\al\dNn^\al~.
\eneq
By integrating it again, we obtain the radial distortion at $\cc_s$
\beeq
\label{affinez}
\drr_s=\ff(\rr_s)\dro-\ff(\rr_s)\int_0^{\cc_s}d\cc~
\Nn_\al\dNn^\al~,
\eneq
where $\rr_s=\rr_{\cc_s}$
can be replaced with~$\rbar_z$ as they are multiplied 
by perturbation quantities and the radial distortion at origin is
\beeq
\drr_o=n^i\de_{i\al}\de\xx^\al_o~.
\eneq
The radial distortion of the source position compared to the observed position is then
\beeq
\label{drrDef}
\drr:=n^i\de_{i\al}x^\al_s-\rbar_z=\rbar_{\cc_s}-\rbar_z+\drr_s=
-\Dcc_s \ff(\rbar_z)+\drr_s~,
\eneq
and we obtain the final expression
\bear
\label{drrex}
\drr&=&\ff(\rbar_z)\left[\de\eta_o+\drr_o-{\dz\over\HH_z}+\int_0^{\cc_z}d\cc
\left(\dnu-\Nn_\al\dNn^\al\right)\right] \\
&=&\ff(\rbar_z)\left[\left(\de\eta_o+\chi_{\bobs}\right)+\left(\drr+\Nn^\al\mathcal{G}
_\al\right)_o-{\dz_\chi\over\HH_z}+\int_0^{\rbar_z}{d\rbar\over \ff(\rbar)}
\left(\al_\chi-\varphi_\chi-\Psi_\para-C_\para\right)
-\Nn^\al\mathcal{G}_\al\right] ~, \nonumber
\enar
where we used the null equation~\eqref{null} and manipulated
\beeq
\al-\BB_\al\Nn^\al-\CC_{\al\be}\Nn^\al\Nn^\be=\al_\chi-\varphi_\chi-\Psi_\para
-C_\para+{d\over d\cc}\left({\chi\over a}+\Nn^\al\mathcal{G}_\al\right)~.
\eneq
Note the subtle difference between $\drr$ defined in Eq.~\eqref{drrDef} and $\drr_s\equiv\drr_{\cc_s}$ defined in Eq.~\eqref{drrdef}: While the former represents the radial deviation of the source position~$x^\al_s$ from the observed background position $\bar{x}_z^\al=\rbar_z n^i\de_i^\al$ corresponding to the affine parameter~$\cc_z$, the latter represents the deviation of the source position from the (unobservable) background position $\bar{x}_{\cc_s}^\al:=\rbar_{\cc_s}n^i\de_i^\al$ corresponding to the affine parameter~$\cc_s$ at the source.
In other words, we chose to express the source position in terms of the background position \textit{inferred from the observed redshift}~$z$ instead of the background position corresponding to the source affine parameter~$\cc_s$.
Compared to the expression in a flat universe, the non-flat nature of Eq.~\eqref{drrex} is visible 
with extra factors with~$\ff(\rbar)$. Moreover,
we want to emphasize again that both the radius~$\bar{r}_z$ 
and the direction vector~$\Nn^\alpha$ depend on the spatial curvature.
The radial distortion in Eq.~\eqref{drrex} is already expressed in terms of
gauge-invariant variables, except one gauge-dependent part $\Nn^\al\CCG_\al$,
and hence it is gauge-dependent,
as $\drr$ represents a coordinate position of the source.
The gauge-dependent part makes $\drr$ gauge-transform as
\beeq
\widetilde{\drr}=\drr+\ff\Nn_\al\LL^\al=\drr+n^i\de_{i\al}\LL^\al~,
\label{eqn:GTdeltaR}
\eneq
under a general coordinate transformation in 
Eq.~\eqref{eqn:coordinatetransformation}.
This is exactly as expected for $\tilde{x}^\al_s=x^\al_s+\LL^\al_s$.

The tangential distortion of the source position is then obtained by
multiplying $\ttt^i\de_{i\al}$ or $\pp^i\de_{i\al}$ with Eq.~\eqref{src} as
\beeq
\label{dtt}
\ttt^i\de_{i\al}\xx^\al_s=\ttt^i\de_{i\al}\de\xx^\al_o-\int_0^{\cc_s}
d\cc~\ff\TT_\al\dNn^\al\equiv\rbar_z\dtt~,
\eneq
and noting that the integration is only along the background path 
$-\ff d\cc=d\rbar$, we can  perform the integration by plugging the 
solution in Eq.~\eqref{polardn} to obtain
\bear
\label{tangent}
\rbar_z\dtt&=&\left(1 -  \frac{1}{4}K\bar{\rr}_z^2\right)
\bigg(\TT_{\al}\de\xx^\al +\TT^\al\mathcal{G}_\alpha \bigg)_{\bobs}
+\bar{\rr}_z\left[-\TT_\alpha V^\al+\TT_\al\Nn^\be C^\al_\be
  +\pp^i\Omega^i \right]_{\bobs}    -\ff(\bar{\rr}_z)\TT^\al\mathcal{G}_\alpha
\Dquad\\
&&
-  \int_{0}^{\bar{\rr}_z} d\bar{\rr}~\frac{\bar{\rr}_z}{\bar{\rr}}
~\TT^\alpha\left(\Psi_\alpha + 2\Nn^\beta C_{\alpha\beta}\right) 
- \int_{0}^{\bar{\rr}_z} d\bar{\rr}\left(
\frac{\bar{\rr}_z-\bar{\rr}}{\rbar \ff(\bar{\rr})}\right)
\left(1+\frac{1}{4}K\bar{\rr}_z\rbar\right)
\frac{\partial}{\partial\theta}\left[\alpha_\chi-\varphi_\chi
  - \Psi_\parallel - C_\parallel\right]~,\nonumber
\enar
where the detailed derivation is again given in Appendix~\ref{angdetail}.
The lensing kernel in the integrand can be further manipulated as
\begin{equation}
-\frac{1}{\rbar_z}\int_{0}^{\bar{\rr}_z} d\bar{\rr}\left(
\frac{\bar{\rr}_z-\bar{\rr}}{\rbar \ff(\bar{\rr})}\right)
\left(1+\frac{1}{4}K\bar{\rr}_z\rbar\right) = 
-\int_{0}^{\bar{\tilde{\chi}}_z} d\bar{\tilde{\chi}}\left[\mathcal{S}(\bar{\tilde{\chi}}_z)-\mathcal{S}(\bar{\tilde{\chi}})\right]~,
\end{equation}
where the integrand is
\beeq
\mathcal{S}(\bar{\tilde{\chi}}_z)-\mathcal{S}(\bar{\tilde{\chi}}) = \left\{
\begin{array}{lr}
\sqrt{K}~\frac{\sin\left(\sqrt{K}~(\bar{\tilde{\chi}}_z-\bar{\tilde{\chi}})\right)}{\sin\left(\sqrt{K}~\bar{\tilde{\chi}}_z\right) \sin\left(\sqrt{K}~\bar{\tilde{\chi}}\right)} & \qquad K>0 \\
\frac{1}{\bar{\tilde{\chi}}} - \frac{1}{\bar{\tilde{\chi}}_z} & \qquad K=0 \\
\sqrt{|K|}~\frac{\sinh\left(\sqrt{|K|}~(\bar{\tilde{\chi}}_z-\bar{\tilde{\chi}})\right)}{\sinh\left(\sqrt{|K|}~\bar{\tilde{\chi}}_z\right) \sinh\left(\sqrt{|K|}~\bar{\tilde{\chi}}\right)} & \qquad K<0 
\end{array} \right\}~.
\eneq
Compared to the tangential distortion in a flat universe, the critical
difference lies with extra terms multiplied by the spatial curvature~$K$,
in addition to the usual replacement of the angular vectors. Moreover,
the dependence on~$\ff$ encloses the non-flat nature of the expression.
A similar solution can be obtained for the azimuthal distortion
$\rbar_z\sin\ttt~\dpp$ 
by replacing $\Theta_\al$ in Eq.~\eqref{dtt} with $\Phi_\al$.
The tangential distortion is also gauge-dependent due to the term~$\ff\TT^\al
\CCG^\al$, as it represents a coordinate of the source position.
A direct substitution in Eq.~\eqref{drrex} yields that the tangential distortion also gauge-transforms as
\beeq
\rbar_z\widetilde{\delta \theta}=\rbar_z\dtt+\ttt^i\de_{i\al}
\mathcal{L}^\alpha~, \Dquad
\rbar_z\sin\ttt~\widetilde{\dpp}=\rbar_z\sin\ttt~\dpp
+\pp^i\de_{i\al}\mathcal{L}^\alpha~,
\label{eqn:SourcePositionPerturbationsGT}
\eneq
under a general coordinate transformation, as expected.
To summarize, the position of a source observed at redshift~$z$ and angles~$\ttt$ and $\pp$ is
\beeq
\label{eqn:SourcePos}
\xx^\al_s=\rbar_zn^i\de_i^\al+\Delta \xx^\al_s~,\qquad\qquad
\Delta \xx^\al_s= \drr~n^i\de_i^\al + \rbar_z (\delta\ttt~\ttt^i\de_i^\al + \sin\ttt~\delta\pp~\pp^i\de_i^\al)~,
\eneq
with $\rbar_z$, $\drr$ and $\delta\ttt$ given in Eqs.~\eqref{bgrr}, \eqref{drrex} and \eqref{tangent} respectively.

It proves useful to derive the relation of the source position in our
coordinates to that in the standard coordinates. For the angular position,
since two coordinates are conformally related, the angular position must be
identical:
\beeq
\widetilde{\dtt}(\xo)=\dtt(\xx)~,\Dquad\widetilde{\dpp}(\xo)=\dpp(\xx)~,
\eneq
which is indeed the case by using Eq.~\eqref{angconf} 
\beeq
{\xx^\al\over\rr}=
{\xo^\al\over\ro}=n^i\de_i^\al+\widetilde{\dtt}(\xo)~\ttt^i\de_i^\al+\sin\ttt~
\widetilde{\dpp}(\xo)~\phi^i\de_i^\al+\OO(2)~.
\eneq
The radial position is, however, slightly different in two coordinate systems,
which is already evident in the background relations in Eqs.~\eqref{bgrr} 
and~\eqref{bgrr2}. Using the relation in Eq.~\eqref{rrro}
\beeq
\ro_s=:\bar\ro_z+\widetilde{\drr}\equiv{\rr_s\over \ff(\rr_s)}
= {\rbar_z\over \ff_z} + \left(1-\frac{\ff'}{\ff_z}\rbar_z\right)\frac{\drr}{\ff_z}+\OO(2)
={\rbar_z\over \ff_z}
+\left(1-\frac14K\rbar_z^2\right){\drr\over \ff_z^2}+\OO(2)~,
\label{eqn:Ro_s}
\eneq
we obtain 
\bear
\label{radialdrr}
\widetilde{\drr}(\xo)&=&\left(1-\frac{\ff'}{\ff_z}\rbar_z\right)\frac{\drr(x)}{\ff_z}
~.
\enar
In a tilde coordinate~$\ro$ represents the angular diameter distance in a non-flat universe, rather than the radial distance.
The ``radial'' distortion in a tilde coordinate is
\beeq
\delta \tilde{\chi}=\frac{\drr}{\ff}
= \left(\de\eta_o+\chi_{\bobs}\right)+\left(\drr+\Nn^\al\mathcal{G}
_\al\right)_o -{\dz_\chi\over\HH_z}+\int_0^{\rbar_z}{d\rbar\over \ff(\rbar)}
\left(\al_\chi-\varphi_\chi-\Psi_\para-C_\para\right)
-\Nn^\al\mathcal{G}_\al
~.
\label{eqn:DeltaChi}
\eneq
The time coordinate remains unaffected by two coordinate systems.

\section{Cosmological Observables in a Non-Flat Universe}
\label{cosobs}
Having computed the source position in a non-flat universe, 
we are now ready to derive gauge-invariant expressions 
for the cosmological observables 
at a position specified by the observed redshift~$z$ with 
two observed angles~$\theta$  and~$\phi$.
We start by deriving a unit physical area~$dA_\text{phy}$ 
in Section~\ref{sec:LD} for the luminosity distance and a unit physical 
volume~$dV_\text{phy}$ in Section~\ref{sec:volume} for galaxy clustering.
We then derive the distortion matrix in Section~\ref{sec:WL} 
for weak gravitational lensing in a non-flat universe, and 
we finish in Section~\ref{sec:CMB} by discussing the impact on 
cosmic microwave background anisotropies.

\subsection{Luminosity distance: Physical area}
\label{sec:LD}
The relativistic effects in the luminosity distance have been extensively discussed in literature \cite{2005PhRvD..71f3537B,2006PhRvD..73l3526H,2006PhRvD..73b3523B,2006PhRvL..96s1302B,2012MNRAS.426.1121C,2013PhRvL.111i1302F,2014CQGra..31t2001U,2014PhRvL.112v1301B,2015MNRAS.450..883K,2017JCAP...03..062F}.
The fluctuation in the luminosity distance in a non-flat universe was first
derived in \cite{1987MNRAS.228..653S} by using the optical scalar equation 
(see also, \cite{1989PhRvD..40.2502F, 1999PThPh.101..903S}). Drawing on our calculations of the source position in 
Section~\ref{inhomo}, we derive the luminosity distance by computing a unit
physical area. In particular, we are interested in the physical area and
shape perpendicular to the light propagation in the rest frame of the source.
This area is observed in the sky of the observer subtended by the observed
solid angle $(d\ttt,d\pp)$, and the shape of this area is related to the
weak lensing observables in Sec.~\ref{sec:WL}.
Since the
luminosity distance~$\dD_L$ is related to the angular diameter distance~$\dD_A$
at the observed redshift~$z$ as
\beeq
\dD_L=(1+z)^2\dD_A~,
\label{eqn:LumDistAngDistRel}
\eneq
the fluctuations of the luminosity distance and the angular diameter distance
are identical
\beeq
\dDD:=\ddD_L=\ddD_A~,
\label{eqn:DistFlucEquivalence}
\eneq
where we defined the fluctuations via
\beeq
\dD_L=:\bar\dD_L(z)(1+\dDD_L)~,\Dquad\dD_A=:\bar\dD_A(z)(1+\dDD_A)~,
\label{eqn:DistancesBGPert}
\eneq
and the background distances are
\beeq
\bar\dD_L(z)=(1+z)^2\bar\dD_A(z)~,\Dquad \bar\dD_A(z)={\bar{\ro}(z)\over1+z}
={\bar\rr_z\over \ff(\bar\rr_z)(1+z)}~.
\label{eqn:BGDistances}
\eneq

We derive the fluctuation in the angular diameter distance 
by considering a unit physical area covered by the solid angle \cite{1972gcpa.book.....W,2017JCAP...04..045B,2016JCAP...09..046Y}
\beeq
\label{phydA}  
dA_\up{phy}=\sqrt{-g}~\epsilon_{\mu\nu\sigma\rho}~u^\mu_s\NP^\nu_s
\frac{\partial \xx^\sigma_s}{\partial\theta}\frac{\partial \xx^\rho_s}
     {\partial\phi}~d\theta~d\phi=\dD_A^2(z)d\Omega~,
\eneq
where the Levi-Civita symbol is normalized as $\epsilon_{\eta xyz}=1$ and
the metric determinant is
\beeq
g:=\up{det}~g_{\mu\nu}=-\frac{a^8}{\ff^6(r)}(1+2~\delta g)~,\Dquad
\delta g := \al + \CC^\alpha_{\alpha}=\al+\varphi+\Delta\ga~.
\label{eqn:MetricDet}
\eneq
The covariant expression represents the physical area covered
by the photon path at the source position over the observed solid angle
$d\Omega=\sin\ttt ~d\ttt d\pp$, and the Levi-Civita symbol imposes that the
area is perpendicular to the time-direction~$u^\mu_s$ of the source and the
photon propagation direction~$\NP^\mu_s$ at the source position. 
The photon propagation direction is naturally related to the photon wave vector as:
\beeq
n^\mu_s=-\frac{1}{\omega}k^\mu + u^\mu
=\frac{1}{a}\Big[\VV_\parallel-\BB_\parallel,~\Nn^\alpha + \dNn^\alpha + \VV^\alpha - \Nn^\alpha(\dnu + \al + \VV_\parallel-\BB_\parallel)
\Big]~,
\label{eqn:PhotonPropDirSource}
\eneq
where the photon wave vector~$k^\mu$ and the four-velocity~$u^\mu$ are given in Eqs.~\eqref{eqn:RelDiffK} and \eqref{eqn:4velocity} respectively, and the photon frequency is related to the latter through
\beeq
\omega=-u_\mu k^\mu = - \frac{1}{\mathbb{C}a^2}u_\mu \hat{k}^\mu 
~.
\eneq
A few subtle points are in order:
two angular derivatives in Eq.~\eqref{phydA} are 
with respect to the observed angles in the observer rest frame, not with
respect to FRW coordinates, such that only $n^i\de_i^\al$, $\ttt^i\de_i^\al$, and
$\pp^i\de_i^\al$ are affected, while other quantities are affected via its
positional dependence on~$\bar\xx^\al_z$. Moreover, those derivatives are
partial derivatives with respect to observed angle, 
while the other observables such as the observed redshift are fixed.

To the linear order in perturbations, the covariant expression 
in Eq.~\eqref{phydA} can be greatly simplified, first 
by considering the metric determinant
\beeq
\label{det}
\sqrt{-g}={a^4_s\over \ff^3(\rr_s)}(1+\de g)={1\over(1+z)^4\ff^3_z}\left(1+4~\dz
-3{\ff'\over \ff}\drr+\de g\right)~,
\eneq
where $\ff_z:=\ff(\bar\rr_z)$ and we expanded $a(\eta_s)$ in terms of the redshift by
using Eq.~\eqref{redshift} and 
$\ff(\rr_s)$ by 
\beeq
\ff(\rr_s)=\ff(\bar{\rr}_z+\drr) 
= \ff_z + \left.\ff'(r)\right|_{r=\bar{\rr}_z}\drr
= \ff_z + \frac{K}{2}\bar{\rr}_z \drr
~.
\label{eqn:ExpansionF}
\eneq
Second, we explicitly write the 
Levi-Civita terms
\beeq
\label{LEVI}
\epsilon_{\mu\nu\sigma\rho}u^\mu_s\NP^\nu_s \frac{\partial \xx^\sigma}
 {\partial\theta}\frac{\partial \xx^\rho}{\partial\phi}=
 \epsilon_{\eta\alpha\beta\gamma}{\NP^\al_s\over a} 
 \frac{\partial\xx^\beta}{\partial\theta}\frac{\partial \xx^\gamma}
 {\partial\phi} -
 \epsilon_{\eta\alpha\beta\gamma}{1\over a}\al~ \bar\NP^\al_s
 \frac{\partial \bar\xx^\beta}{\partial\theta}\frac{\partial\bar\xx^\gamma}
 {\partial\phi} +
 \epsilon_{\alpha\nu\sigma\rho}{1\over a}~\VV^\al \bar\NP^\nu_s\frac{\partial
\bar\xx^\sigma}
  {\partial\theta}\frac{\partial \bar\xx^\rho}{\partial\phi} ~,
\eneq
where we omitted the subscript~$s$ and the third term is already at
 the second order
in perturbation because $\bar\NP^\eta_s=0$ and $\bar\eta_s$ is independent of
angle. Multiplied by the metric determinant, the second term becomes
\beeq
\sqrt{-{\bar g}}\times(-1)
\epsilon_{\eta\alpha\beta\gamma}\frac{1}{a}\al~\bar\NP^\alpha_s \frac{\partial\bar
\xx^\beta}{\partial\theta}\frac{\partial\bar\xx^\gamma}{\partial\phi} d\ttt
d\pp=-\al~\bar\dD_A^2(z)d\Omega~,
\eneq
which is then canceled by the same contribution with~$\al$ in~$\de g$.
The first term in Eq.~\eqref{LEVI}
contains the background and the perturbations as
\beeq
 \epsilon_{\eta\alpha\beta\gamma}{\ff n^i\de_i^\al\over a^2} 
 \frac{\partial\bar\xx^\beta}{\partial\theta}\frac{\partial\bar \xx^\gamma}
 {\partial\phi} +
\epsilon_{\eta\alpha\beta\gamma}{\de\NP^\al_s\over a} 
 \frac{\partial\bar\xx^\beta}{\partial\theta}\frac{\partial\bar \xx^\gamma}
 {\partial\phi} +
\epsilon_{\eta\alpha\beta\gamma}{\bar\NP^\al_s\over a} 
 \frac{\partial\Delta\xx^\beta}{\partial\theta}\frac{\partial\bar \xx^\gamma}
 {\partial\phi} +\epsilon_{\eta\alpha\beta\gamma}{\bar\NP^\al_s\over a} 
 \frac{\partial\bar\xx^\beta}{\partial\theta}\frac{\partial\Delta \xx^\gamma}
 {\partial\phi} ~,
\eneq
where~$\ff$ and~$a$ in the first background terms need to be further expanded
again by using Eqs.~\eqref{eqn:ExpansionF} and~Eq.~\eqref{redshift}, which reduces the numerical
factor~4 for~$\dz$ in Eq.~\eqref{det} to~2 and the factor~$-3$ for~$\drr$ 
to~$-2$. The next term is 
\beeq
\epsilon_{\eta\alpha\beta\gamma}{\de\NP^\al_s\over a} 
 \frac{\partial\bar\xx^\beta}{\partial\theta}\frac{\partial\bar \xx^\gamma}
 {\partial\phi} =n^i\de_{i\al}~\de\NP^\al_s{\rbar_z^2\sin\ttt\over a}
\RA-\CC_\parallel{\ff\rbar_z^2\sin\ttt\over a^2}~.
\eneq
Regarding two terms involving~$\Delta\xx^\al$, we first compute 
by using the expression for $\Delta\xx^\al$ in Eq.~\eqref{eqn:SourcePos}, 
\bear
\label{DDx1}
{\pa\over\pa\ttt}\Delta\xx^\al&=&n^i\de_i^\al\left({\pa\over\pa\ttt}\drr
-\rbar_z\dtt\right)
+\rbar_z\ttt^i\de_i^\al\left({\drr\over\rbar_z}+{\pa\over\pa\ttt}\dtt\right)
+\rbar_z\sin\ttt~\pp^i\de_i^\al\left(\cot\ttt~\dpp+{\pa\over\pa\ttt}
\dpp\right)~,~~\qquad\\
\label{DDx2}
{\pa\over\pa\pp}\Delta\xx^\al&=&n^i\de_i^\al\left({\pa\over\pa\pp}\drr
-\rbar_z\sin^2\ttt~\dpp\right)+\rbar_z\ttt^i\de_i^\al\left({\pa\over\pa\pp}\dtt
-\sin\ttt\cos\ttt~\dpp\right)\\
&&+\rbar_z\sin\ttt~\pp^i\de_i^\al\left({\drr\over\rbar_z}+\cot\ttt~
\dtt+{\pa\over\pa\pp}\dpp\right)~,\nonumber
\enar
and readily derive 
\bear
\epsilon_{\eta\alpha\beta\gamma}{\bar\NP^\al_s\over a} 
 \frac{\partial\Delta\xx^\beta}{\partial\theta}\frac{\partial\bar \xx^\gamma}
 {\partial\phi} &=&{\ff\over a^2}\rbar_z^2\sin\ttt\left({\drr\over\rbar_z}+
{\pa\over\pa\ttt}\dtt\right)~,\\
\epsilon_{\eta\alpha\beta\gamma}{\bar\NP^\al_s\over a} 
 \frac{\partial\bar\xx^\beta}{\partial\theta}\frac{\partial\Delta \xx^\gamma}
 {\partial\phi} &=&{\ff\over a^2}\rbar_z^2\sin\ttt\left({\drr\over\rbar_z}
+\cot\ttt~\dtt+{\pa\over\pa\pp}\dpp\right)~.
\enar
The sum of these two terms is proportional to the standard expression of the gravitational lensing convergence~$\kappa$ defined as:
\beeq
\kappa:=-\frac{1}{2}\left(\cot\ttt~\dtt + \frac{\partial}{\partial\ttt}\dtt + \frac{\partial}{\partial\pp}\dpp\right)~.
\label{eqn:StandardConvergence}
\eneq
Evident from Eq.\eqref{eqn:SourcePositionPerturbationsGT} and stated in \cite{YooGrimmMitsou_2018}, this expression is gauge-dependent, hence $\kappa$ cannot be an observable. 
In Section~\ref{sec:WL} we show that the gauge-invariant expression for the observed convergence~$\hat{\kappa}$ is identical to the fluctuation~$\dDD$ up to a minus sign: $\hat{\kappa}=-\dDD$.
Finally, adding all terms together and taking the square-root,
we arrive at the luminosity distance fluctuation
\beeq
\dDD = \dz -\kappa + \left(\frac{1}{\bar\rr_z} - \frac{\ff'}{\ff}
\right)\drr
+\frac{1}{2}\left(\CC^\alpha_{\alpha}-\CC_\parallel\right) ~, 
\label{LD}
\eneq
where the radial distortion term is indeed
\beeq
{\drr\over\rbar_z}-{\ff'\over \ff}\drr=\left(1-{K\rbar_z^2\over 2\ff}\right)
{\drr\over\rbar_z}=\left(1-\frac14K\rbar_z^2\right){\drr\over \ff_z\rbar_z}~.
\eneq
In a closed universe with $K>0$, the term in the brackets vanishes for $\rbar_z=2/\sqrt{K}$.
However, notice that this corresponds to a (background) comoving angular diameter distance of $\bar\ro_z=1/\sqrt{K}=R_K$, i.e. when we are back at the origin.
One might be puzzled by the difference in 
the expression for the luminosity distance fluctuation in Eq.~\eqref{LD},
compared to the expression in the flat universe, because the difference
is hidden in the expression for~$\dz$, $\drr$, and~$\kappa$, while the
source position is parametrized in an identical way. 
Noting that the angular diameter distance in the background is $a\bar\ro$, 
not $a\bar\rr$, we can readily verify that this is indeed the case 
thanks to Eq.~\eqref{radialdrr}:
\beeq
\left(1-\frac{1}{4}K\bar{\rr}^2_z\right)\frac{\drr}{\ff_z\bar\rr_z}
={\widetilde{\drr}(\xo)\over\bar\ro_z}~.
\label{eqn:deltaRrelation}
\eneq
In other words, the expression for~$\dDD$ is identical in a
parametrized form as in Eq.~\eqref{LD} with Eq.~\eqref{eqn:deltaRrelation}, but individual components such
as~$\dz$, $\kappa$, and~$\drr$ are different in a non-flat universe, compared
to those in a flat universe.
Note that $\delta\ro/\bar{\ro}$ in a tilde coordinate represents the fluctuation in the angular diameter distance, rather than the fluctuation in the radial position.

To verify the sanity of Eq.~\eqref{LD}, we check if the expression for the luminosity distance fluctuation in Eq.~\eqref{LD} is gauge-invariant.
Although the individual components are gauge-dependent, the whole combination should be gauge-invariant.
As mentioned previously, the standard expression of the lensing convergence~$\kappa$ in Eq.~\eqref{eqn:StandardConvergence} is gauge dependent, and it transforms as
\beeq
\tilde{\kappa}=\kappa+\frac{\ff_z}{\rbar_z}\left(
\Nn_\al
- \frac{1}{2}\TT_\al\frac{\partial}{\partial\ttt}
-\frac{1}{2}\PP_\al\frac{1}{\sin\ttt} \frac{\partial}{\partial\pp}
\right)\mathcal{L}^\alpha~,
\label{eqn:GTkappa}
\eneq
where we used Eq.~\eqref{eqn:SourcePositionPerturbationsGT} and $\Nn_\al=\frac{1}{\ff}\de_{i\al}n^i$. 
The gauge transformation of the perturbation in the redshift is derived from Eq.~\eqref{eqn:dz} as
\beeq
\widetilde{\dz}=\dz + \mathcal{H}T~,
\label{eqn:GTdz}
\eneq
while the transformation of~$\drr$ is given in Eq.~\eqref{eqn:GTdeltaR}.
The term $\CC^\alpha_{\alpha}-\CC_\parallel$ is
\bear
\CC^\alpha_{\alpha}-\CC_\parallel 
&=& 2\varphi - C_\parallel + \mathcal{G}^\alpha_{~~|\alpha}-\Nn^\alpha\Nn^\beta\mathcal{G}_{\alpha|\beta}\nnn
&=& 2\varphi - C_\parallel +\frac{\ff_z}{\rbar_z}\left(\TT_\alpha \frac{\partial}{\partial\ttt} + \PP_\alpha \frac{1}{\sin\ttt} \frac{\partial}{\partial\pp}\right)\mathcal{G}^\alpha - K\rbar_z \mathcal{G}_\parallel
~,
\enar
and it gauge-transforms as
\beeq
\widetilde{\CC}^\alpha_{\alpha}-\widetilde{\CC}_\parallel =
\CC^\alpha_{\alpha}-\CC_\parallel 
- 2\mathcal{H}T -\frac{\ff_z}{\rbar_z}\left(\TT_\alpha \frac{\partial}{\partial\ttt} + \PP_\alpha \frac{1}{\sin\ttt} \frac{\partial}{\partial\pp}\right)\mathcal{L}^\alpha + K\rbar_z \mathcal{L}_\parallel
~,
\eneq
where we used the decomposition of the metric tensor introduced in Sec.~\ref{sec:metric}.
The angular derivatives cancel with the ones in the gauge transformation of $\kappa$, and the term proportional to $\mathcal{H}T$ cancels with the gauge transformation of $\dz$.
The remaining term is proportional to $\mathcal{L}_\parallel$, and it cancels with the corresponding terms in the transformation of $\kappa$ and $\drr$.
Therefore, the expression for~$\delta\mathcal{D}$ is gauge-invariant and we can express it in terms of gauge-invariant variables
\begin{equation}
\delta \mathcal{D} = \delta z_\chi + \varphi_\chi - \frac{1}{2} C_\parallel + \left(\frac{1}{\bar{r}_z}-\frac{\ff'}{\ff}\right)\delta\mathcal{R} -\mathcal{K}~.
\label{eqn:deltaD}
\end{equation}
where we defined the two gauge-invariant variables
\bear
\delta\mathcal{R} &:=& \drr + (\ff\Nn_{\alpha}\mathcal{G}^\alpha)_s=\drr + \Nn_{\alpha o}\mathcal{G}^\alpha_s~, \nnn
\mathcal{K} &:=& \kappa + \frac{\ff_z}{\rbar_z}\left(\Nn_\alpha - \frac{1}{2}\TT_\alpha\frac{\partial}{\partial\ttt}- \frac{1}{2}\PP_\alpha\frac{1}{\sin\ttt}\frac{\partial}{\partial\pp}\right)\mathcal{G}^\alpha_s~.
\label{eqn:GIradius&kappa}
\enar

\subsection{Galaxy clustering: Physical volume}
\label{sec:volume}
Now we are interested in a unit physical volume at
the observed redshift that subtends by the observed angle. 
The fluctuation in the physical volume is the dominant contribution to galaxy clustering. 
The relativistic effects in galaxy clustering have been extensively studied \cite{2009PhRvD..79b3517Y,PhysRevD.80.083514,PhysRevD.82.083508,2011PhRvD..84f3505B,2011PhRvD..84d3516C,PhysRevD.85.023504,Yoo_2014b,2014JCAP...12..017D,2014JCAP...09..037B,2016JCAP...01..016D,2017JCAP...03..034U,2018JCAP...07..050K}.
While relatively
less attention was paid in literature 
to galaxy clustering in a non-flat universe,
\cite{2016JCAP...06..013D} derived the physical volume in a non-flat universe, accounting
for the relativistic effects. Their results are different from ours presented
 in this subsection, and we discuss the difference in Section~\ref{comparison}.

In a redshift survey, galaxies are measured in terms of
observed redshift and angle, and their number density $n_g^\up{obs}$
is estimated by counting
the number~$dN_g$ of galaxies in a small volume characterized by redshift 
interval~$dz$ and solid angle~$d\Omega$. Of course, the observed number~$dN_g$
is the product of the physical number density~$n_g^\up{phy}$ of the source
galaxies and the physical volume~$dV_\up{phy}$ occupied by the source galaxies
that is observed at the observed redshift and angle. The discrepancy
between the physical and the observed quantities gives rise to all
the fluctuations in galaxy clustering (see, e.g., \cite{2009PhRvD..79b3517Y}).
The physical volume~$dV_\text{phy}$ in the rest-frame of the source 
enclosed within the observed solid angle~$d\Omega$ over the observed redshift 
interval~$dz$ can be obtained 
\cite{PhysRevD.80.083514,PhysRevD.82.083508}
in a similar way to Eq.~\eqref{phydA} as
\beeq
\label{phydV}
dV_\text{phy}=\sqrt{-g}~\epsilon_{\mu\nu\sigma\rho}~u^\mu_s ~
\frac{\partial\xx^\nu_s}{\partial z} \frac{\partial \xx^\sigma_s}
{\partial\theta}\frac{\partial \xx^\rho_s}{\partial\phi}~dz~d\theta~d\phi
=:dV_\up{obs}(1+\de V)~,
\eneq
where we defined the dimensionless fluctuation~$\de V$ in volume, 
compared to the volume $dV_\up{obs}$ in a homogeneous universe 
\beeq
dV_\up{obs}:={\bar\ro_z^2~d\Omega~dz\over H_z(1+z)^3}
={\rbar_z^2~d\Omega~dz\over H_z\ff_z^2(1+z)^3}~
\label{eqn:obsVol}
\eneq
(mind the difference between~$\bar\ro_z$ and~$\rbar_z$).
The physical volume~$dV_\text{phy}$ deviates from the volume~$dV_\up{obs}$ inferred by the observer due to the inhomogeneity of the universe.
Compared to the physical area in Eq.~\eqref{phydA}, the redshift interval
is an additional dimension, covering the line-of-sight direction in the source
frame, in addition to the physical area. Again, the partial
derivatives in the covariant expression fix the other observables 
among $(z,\ttt,\pp)$ in the derivative. Moreover, the redshift
derivative in Eq.~\eqref{phydV} is the line-of-sight derivative,
which affects the quantities via its radial positional dependence 
on~$\bar\xx^\al_z$, but is independent of observed angles. Specifically,
the derivative with respect to the observed redshift can be expressed
 by using Eqs.~\eqref{affine} and~\eqref{bgrr} as
\beeq
{\pa\over \pa z}={d\cc_z\over dz}{d\over d\cc_z}=-\frac1{H_z}\left({\pa\over
\pa\bar\eta_z}-\ff_z{\pa\over\pa \rbar_z}\right)~.
\label{eqn:partialZ}
\eneq

To the linear order in perturbations, the covariant expression in
Eq.~\eqref{phydV} can be simplified in a similar way, first by explicitly
writing the Levi-Civita terms
\beeq
\label{LEVI2}
\epsilon_{\mu\nu\sigma\rho}u^\mu~ \frac{\partial\xx^\nu}{\partial z} 
\frac{\partial\xx^\sigma}{\partial\theta}\frac{\partial\xx^\rho}
{\partial\phi} =
\epsilon_{\eta\alpha\beta\gamma}\frac{1}{a} \frac{\partial\xx^\alpha}
{\partial z} \frac{\partial\xx^\beta}{\partial\theta}\frac{\partial\xx^\gamma}
{\partial\phi}
-\epsilon_{\eta\alpha\beta\gamma}\frac{\al}{a} \frac{\partial\bar\xx^\alpha}
{\partial z} \frac{\partial\bar\xx^\beta}{\partial\theta}
\frac{\partial\bar\xx^\gamma}{\partial\phi}
+\epsilon_{\alpha\nu\sigma\rho} \frac{\VV^\alpha}{a}
\frac{\partial\bar\xx^\nu}{\partial z} \frac{\partial\bar\xx^\sigma}
{\partial\theta}\frac{\partial\bar\xx^\rho}{\partial\phi} ~,
\eneq
where we omitted the subscript~$s$. Since the time coordinate~$\bar\eta_z$
of the source in the background depends only on the redshift, the last term
is non-vanishing only for the combination
\beeq
\sqrt{-\bar g}\times
\epsilon_{\alpha\eta\beta\ga} \frac{\VV^\alpha}{a}
\frac{\partial\bar\xx^\eta}{\partial z} \frac{\partial\bar\xx^\be}
{\partial\theta}\frac{\partial\bar\xx^\ga}{\partial\phi} ={a^4\over \ff_z^3}
\times \VV^\al n^i\de_{i\al}{\rbar_z^2\sin\ttt\over aH_z}
={\VV_\para~\rbar_z^2\sin\ttt\over H_z\ff_z^2(1+z)^3}~,
\eneq
where we defined the line-of-sight velocity component
\beeq
\VV_\para:=\Nn_\al\VV^\al~.
\eneq
The second term with the metric determinant~$\sqrt{-g}$ yields
\beeq
\sqrt{-\bar g}\times(-1)
\epsilon_{\eta\alpha\beta\gamma}\frac{\al}{a} \frac{\partial\bar\xx^\alpha}
{\partial z} \frac{\partial\bar\xx^\beta}{\partial\theta}
\frac{\partial\bar\xx^\gamma}{\partial\phi}=-{\al~\rbar_z^2\sin\ttt\over
H_z\ff_z^2(1+z)^3}~,
\eneq
which is again canceled by the same contribution with~$\al$ in~$\de g$.
The first term in Eq.~\eqref{LEVI2}
contains the background and the perturbation terms as
\beeq
\epsilon_{\eta\alpha\beta\gamma}\frac{1}{a} \frac{\partial\bar\xx^\alpha}
{\partial z} \frac{\partial\bar\xx^\beta}{\partial\theta}
\frac{\partial\bar\xx^\gamma}{\partial\phi}+
\epsilon_{\eta\alpha\beta\gamma}\frac{1}{a} \frac{\partial\Delta\xx^\alpha}
{\partial z} \frac{\partial\bar\xx^\beta}{\partial\theta}
\frac{\partial\bar\xx^\gamma}
{\partial\phi}+
\epsilon_{\eta\alpha\beta\gamma}\frac{1}{a} \frac{\partial\bar\xx^\alpha}
{\partial z} \frac{\partial\Delta\xx^\beta}{\partial\theta
}\frac{\partial\bar\xx^\gamma}{\partial\phi}+
\epsilon_{\eta\alpha\beta\gamma}\frac{1}{a} \frac{\partial\bar\xx^\alpha}
{\partial z} \frac{\partial\bar\xx^\beta}{\partial\theta}
\frac{\partial\Delta\xx^\gamma}{\partial\phi}~,
\eneq
where the scale factor~$a$ in the first term needs to be further expanded
by using Eq.~\eqref{redshift}, which yields the numerical factor~3
for~$\dz$ in combination with $\sqrt{-g}$ in Eq.~\eqref{det}, i.e.,
\beeq
\sqrt{-g}\times
\epsilon_{\eta\alpha\beta\gamma}\frac{1}{a} \frac{\partial\bar\xx^\alpha}
{\partial z} \frac{\partial\bar\xx^\beta}{\partial\theta}
\frac{\partial\bar\xx^\gamma}{\partial\phi}
=\left(1+3~\dz-3~{\ff'\over \ff}\drr
+\de g\right){\rbar_z^2\sin\ttt\over H_z\ff_z^2(1+z)^3}~.
\eneq
The remaining
three terms can be computed by using Eqs.~\eqref{DDx1} and~\eqref{DDx2},
in addition to 
\beeq
{\pa\over\pa z}\Delta\xx^\al=n^i\de_i^\al~{\pa\over\pa z}\drr
+\rbar_z\ttt^i\de_i^\al\left({\pa\ln\rbar_z\over\pa z}+
{\pa\over\pa z}\right)\dtt+\rbar_z\sin\ttt~\pp^i\de_i^\al\left(
{\pa\ln\rbar_z\over\pa z}+{\pa\over\pa z}\right)\dpp~,
\eneq
as
\bear
\sqrt{-\bar g}\times
\epsilon_{\eta\alpha\beta\gamma}\frac{1}{a} \frac{\partial\Delta\xx^\alpha}
{\partial z} \frac{\partial\bar\xx^\beta}{\partial\theta}
\frac{\partial\bar\xx^\gamma}{\partial\phi}&=&
{\rbar_z^2\sin\ttt\over H_z\ff_z^2(1+z)^3}\left({H_z\over \ff_z}{\pa\over\pa z}
\drr\right)~,\\
\sqrt{-\bar g}\times
\epsilon_{\eta\alpha\beta\gamma}\frac{1}{a} \frac{\partial\bar\xx^\alpha}
{\partial z} \frac{\partial\Delta\xx^\beta}{\partial\theta
}\frac{\partial\bar\xx^\gamma}{\partial\phi}&=&
{\rbar_z^2\sin\ttt\over H_z\ff_z^2(1+z)^3}
\left({\drr\over\rbar_z}+{\pa\over\pa\ttt}\dtt\right)~,\\
\sqrt{-\bar g}\times
\epsilon_{\eta\alpha\beta\gamma}\frac{1}{a} \frac{\partial\bar\xx^\alpha}
{\partial z} \frac{\partial\bar\xx^\beta}{\partial\theta}
\frac{\partial\Delta\xx^\gamma}{\partial\phi}&=&
{\rbar_z^2 \sin\ttt\over H_z\ff_z^2(1+z)^3}\left({\drr\over\rbar_z}+\cot\ttt~\dtt
+{\pa\over\pa\pp}\dpp\right)~.
\enar
Combining all the terms, we arrive at the dimensionless volume
fluctuation 
\beeq
\de V=3~\dz+\CC^\al_\al+\VV_\para
+2{\drr\over \rbar_z}- 3 \frac{\ff'}{\ff}\drr
+{H_z\over \ff_z}{\pa\over\pa z}\drr-2~\kappa~.
\label{eqn:dV}
\eneq
The three terms proportional to the radial distortion~$\drr$ can be rewritten in terms of $\widetilde{\drr}(\tilde{x})$ and $\delta\tilde{\chi}$ as
\beeq
H_z{\pa\over\pa z}\delta\tilde{\chi} + 2~{\widetilde{\drr}\over\bar\ro_z}
~.
\eneq
The expression in a non-flat universe is similar to that in a flat universe, but the difference between the angular diameter distance and the radial distortions is manifest.

Despite the gauge dependence of the individual terms in Eq.~\eqref{eqn:dV}, their combination, i.e. the volume fluctuation~$\de V$, is gauge-invariant, which we verify as follows.
The term $\CC^\alpha_{\alpha}$ can be rewritten as
\beeq
\CC^\alpha_{\alpha} = 3\varphi + \mathcal{G}^\alpha_{~~|\alpha}
= 3\varphi + \ff_z\left[\Nn_\al \frac{\partial}{\partial\rbar} + \frac{1}{\rbar_z}\left(\TT_\alpha \frac{\partial}{\partial\ttt} + \PP_\alpha \frac{1}{\sin\ttt} \frac{\partial}{\partial\pp}\right)\right]\mathcal{G}^\al - \frac{3}{2}K\rbar_z\mathcal{G}_\parallel
~,
\eneq
and it gauge-transforms as
\beeq
\widetilde{\CC}^\alpha_{\alpha} = \CC^\alpha_{\alpha} - 3\mathcal{H}T - \ff_z\left[\Nn_\al \frac{\partial}{\partial\rbar} + \frac{1}{\rbar_z}\left(\TT_\alpha \frac{\partial}{\partial\ttt} + \PP_\alpha \frac{1}{\sin\ttt} \frac{\partial}{\partial\pp}\right)\right]\mathcal{L}^\al + \frac{3}{2}K\rbar_z\mathcal{L}_\parallel
~.
\eneq
The first term cancels the gauge-dependent part of the redshift distortion~$\dz$ in Eq.~\eqref{eqn:GTdz}, and the angular derivatives cancel the angular derivatives of the gauge-dependent part in the lensing convergence~$\kappa$ in Eq.~\eqref{eqn:GTkappa}.
The remaining terms in the transformation of $\CC^\alpha_{\alpha}$ and $\kappa$ cancel with the transformation of the source velocity
\beeq
\tilde{\VV}^\alpha=\VV^\alpha + \mathcal{L}^{\al\prime}~,
\eneq
and the transformation of the radial distortion~$\drr$ in Eq.~\eqref{eqn:GTdeltaR}, for which we use Eq.~\eqref{eqn:partialZ} to find
\beeq
\frac{\partial}{\partial z}(\ff_z\mathcal{L}_\parallel)=-\frac{1}{H_z}\left[\ff_z \mathcal{L}'_\parallel - \ff_z^2 \Nn_\al \frac{\partial}{\partial \rbar_z} \mathcal{L}^\al\right]~.
\eneq
Therefore, we can express the volume fluctuation entirely in terms of gauge-invariant variables:
\beeq
\de V=3\delta z_\chi + 3\varphi_\chi + V_\parallel + \Psi_\parallel +   \left(\frac{2}{\bar{r}_z}-K\bar{r}_z\right)\frac{\delta \mathcal{R}}{\ff(\bar{r}_z)} + \frac{H(z)}{\ff(\bar{r}_z)}\frac{\partial}{\partial z}\delta \mathcal{R} -2\mathcal{K} 
~,
\label{eqn:dV_GI}
\eneq
where $\dz_\chi$, $\delta\mathcal{R}$ and $\mathcal{K}$ were introduced in Eqs.~\eqref{eqn:dz_GI} and \eqref{eqn:GIradius&kappa} respectively.

Having computed the dimensionless volume fluctuation $\de V$, we are now in position to relate the observed number density~$n_g^\up{obs}$ of galaxies with the physical number density~$n_g$ of galaxies at the source position:
\beeq
dN_g^\up{obs} = n_g^\up{obs}~dV_\up{obs} = n_g~dV_\text{phy}~, \qquad\qquad
n_g^\up{obs}=n_g(1+\de V)~,
\eneq
where $dN_g^\up{obs}$ is the observed number of galaxies counted by the observer within the observed redshift and solid angle.
In addition to the volume fluctuation~$\de V$, the observed galaxy number density is also affected by the physical number density~$n_g$ expressed in terms of the observed position~$\bar{x}^\mu_z$.
The volume fluctuation includes the redshift-space distortion and the lensing effects, and the fluctuation in $n_g$ includes the matter density fluctuation and the magnification bias, in addition to numerous relativistic effects (see, e.g., \cite{2018JCAP...10..024S,2020JCAP...11..064G}).

\subsection{Weak gravitational lensing: Observed shear and convergence}
\label{sec:WL}
Since gravitational lensing is a relativistic phenomenon, the relativistic effects are naturally part of the dominant contributions, and many groups have studied the additional relativistic contributions to the weak lensing observables \cite{1967ApJ...150..737G,1991ApJ...380....1M,1992ApJ...388..272K,1998MNRAS.301.1064K,2002ApJ...574...19C,2005PhRvD..72j3004D,2008PhRvD..78l3530B,2010PhRvD..81h3002B,2012PhRvD..86h3527S,2013JCAP...08..051Y,2014PhRvD..89l3006S,2015JCAP...09..033C,2018JCAP...07..067G,2021PhRvD.104h3548G}.
Similar to the case in galaxy clustering, relatively less
attention has been paid to weak gravitational
lensing in a non-flat universe (see, however,
\cite{2006ApJ...637..598B}). Naturally, the primary difference in a non-flat universe 
lies in the angular diameter distance, and the lensing kernel depends on
the background angular diameter distances.
This simple prescription of using the angular diameter distance in a
non-flat universe but the same theoretical description in a flat universe
has been mostly used to compute the gravitational lensing
observables in a non-flat universe. It is already evident that the light
propagation in a non-flat universe is substantially more complicated.
Here we derive the gravitational lensing observables in a non-flat universe.
This subsection is a generalization of the fully gauge-invariant formalism 
 \cite{YooGrimmMitsou_2018} of cosmological weak lensing to a non-flat 
Universe. In particular, we closely follow the derivations described 
in Section~5.1 of \cite{YooGrimmMitsou_2018}, 
accounting for the difference in a non-flat universe.

The FRW coordinates~$x^\mu_s$ of a point-like source observed at redshift~$z$ and angular direction~$n^i$ in the observer rest-frame were derived in Sec.~\ref{ssec:path}. 
Here we consider an extended source at observed redshift~$z$ that appears subtended by the infinitesimal angular size $(d \ttt, d \pp)$ in the sky. 
The source size in FRW coordinates is then
\bear
dx^\alpha_s&=& \frac{\partial x^\alpha_s}{\partial\ttt} d\ttt + \frac{\partial x^\alpha_s}{\partial\pp} d\pp 
= -\bar{\rr}_z n^i\de_i^\alpha\left[d\ttt\left(\dtt - \frac{1}{\bar{\rr}_z}\frac{\partial}{\partial\ttt}\drr\right) + d\pp\left(\sin^2\ttt~\dpp - \frac{1}{\bar{\rr}_z}\frac{\partial}{\partial\pp}\drr\right) \right]\nnn
&&+ \bar{\rr}_z \ttt^i\de_i^\alpha\left[d\ttt\left(1+ \frac{\drr}{\bar{\rr}_z} + \frac{\partial}{\partial\ttt}\dtt\right) + d\pp\left(\frac{\partial}{\partial\pp}\dtt - \sin\ttt\cos\ttt~\dpp\right) \right]\nnn
&&+ \bar{\rr}_z \sin\ttt~\pp^i\de_i^\alpha\left[d\ttt\left(\cot\ttt+  \frac{\partial}{\partial\ttt}\right)\dpp + d\pp\left(1+ \frac{\drr}{\bar{\rr}_z}  + \frac{\partial}{\partial\pp}\dpp + \cot\ttt~\dtt\right) \right]~,
\label{eqn:SourceSizeFRW}
\enar
and we define the perturbation~$\Delta s^\alpha$ as
\beeq
dx^\alpha_s:=\bar{\rr}_z(dn^\alpha+\Delta s^\alpha)~,\qquad\qquad
dn^\alpha=\ttt^i\de_i^\alpha d\ttt + \sin\ttt~\pp^i\de_i^\alpha d\pp~.
\label{eqn:Dsalpha}
\eneq
The expression for the perturbation~$\Delta s^\alpha$ is identical in parametrized form to the expression in a flat Universe, derived in Eq.~(5.6) in \citep{YooGrimmMitsou_2018}.
However, the individual components such as the background radial coordinate~$\bar{\rr}_z$ at observed redshift~$z$ and the perturbations~$\delta\rr$, $\delta\ttt$ and $\delta\pp$ in the source position do of course differ from the expressions in a flat Universe, as discussed in Sec.~\ref{ssec:path}.
In computing the extended source size, we need to consider that the source size is indeed described by 4D $dx^\mu_s$, instead of 3D $dx^\al_s$ in Eq.~\eqref{eqn:SourceSizeFRW}.
However, we will only need the spatial component for our purposes at the linear order.

Next, we compute the photon propagation direction~$n^i_s$ in the source rest frame, in which the physical shape of the observed source galaxies is defined.
The temporal and the spatial components of the photon propagation direction~$n^\mu_s$ in FRW coordinates are already introduced in Eq.~\eqref{eqn:PhotonPropDirSource} as
\bear
n^\eta_s &=& \frac{1}{a}\left\{\VV_\parallel - \BB_\parallel\right\}~, \nnn
n^\alpha_s&=&\frac{1}{a}\left\{
\Nn^\alpha+\Nn^\alpha(C_\parallel-\varphi_\chi-H\chi) - 2\Nn^\gamma C^\alpha_\gamma
+ (\TT^\alpha\TT_\gamma + \PP^\alpha\PP_\gamma)V^\gamma - \Nn^\gamma\mathcal{G}^\alpha_{~,\gamma} \phantom{\frac{K}{2}}\right.\nnn
&&+ \frac{K\bar{\rr}_z}{2}\Nn^\alpha\mathcal{G}_\parallel 
+ \ff(\bar{\rr}_z)\left[-(\TT^\alpha\TT_\gamma + \PP^\alpha\PP_\gamma)V^\gamma + \Nn^\gamma C^\alpha_\gamma - \Nn^\alpha C_\parallel - \epsilon_{ij}^\alpha n^i\Omega^j \right]_{\bobs} \nnn
&&-\frac{K\bar{\rr}_z}{2} \ff(\bar{\rr}_z)\left[\delta x^\alpha + \mathcal{G}^\alpha - \Nn^\alpha(\delta x_\parallel + \mathcal{G}_\parallel)\right]_{\bobs} \nnn
&&- \ff(\bar{\rr}_z)\int_0^{\bar{\rr}_z}\frac{d\bar{\rr}}{\bar{\rr}\ff(\bar{\rr})}\left(1+ \frac{K}{2}\frac{\bar{\rr}(\bar{\rr}_z - \bar{\rr})}{\ff(\bar{\rr})}\right) \left[\TT^\alpha\frac{\partial}{\partial \ttt}+ \frac{1}{\sin\ttt}\PP^\alpha\frac{\partial}{\partial\pp}\right] \left[\alpha_\chi - \varphi_\chi - \Psi_\parallel - C_\parallel\right] \nnn
&&\left.-\ff(\bar{\rr}_z)\int_0^{\bar{\rr}_z}\frac{d\bar{\rr}}{\ff(\bar{\rr})}\frac{1}{\bar{\rr}}\left[\TT^\alpha\TT_\gamma + \PP^\alpha\PP_\gamma\right]\left(\Psi^\gamma + 2\Nn^\epsilon C^\gamma_\epsilon\right)
\right\}~,
\enar
where we used the expressions for $\dnu$ and $\dNn^\al$ derived in Sec.~\ref{ssec:GE}.
With a tetrad vector~$\left[e_a^\mu\right]_s$ at the source position, we can project the photon propagation direction~$n^\mu_s$ to obtain the one $n^i_s$ in the rest frame of the source:
\bear
n_{i\,s}&=&g_{\mu\nu}~e_i^\mu n^\nu_s = g_{\alpha\beta}~e_i^\alpha n^\beta_s + \mathcal{O}(2)~,
\enar
and the direction~$n^i_s$ measured by an observer at the source position is 
\bear
n^i_s&=&\delta^{ij} n_{j\,s} = n^i + \left[ (\ttt^i\TT_\gamma + \pp^i\PP_\gamma)V^\gamma - \ff^2 n^j\delta_j^\alpha \delta^{i\gamma}C_{\alpha\gamma} + n^i C_\parallel + \epsilon^i{}_{jk} n^j \Omega^k\right]_{\bobs}^s \nnn
&&-\frac{K\bar{\rr}_z}{2}\left[\delta^i_\alpha(\delta x^\alpha + \mathcal{G}^\alpha) -n^i (\delta x_\parallel + \mathcal{G}_\parallel) \right]_{\bobs}
- \int_0^{\bar{\rr}_z}\frac{d\bar{\rr}}{\bar{\rr}}\left[\ttt^i\TT_\gamma + \pp^i\PP_\gamma\right] \left(\Psi^\gamma+2\Nn^\beta C^\gamma_\beta\right)\nnn
&&-\int_0^{\bar{\rr}_z} \frac{d\bar{\rr}}{\bar{\rr}}  \left[1+ \frac{K}{2}\frac{\bar{\rr}(\bar{\rr}_z - \bar{\rr})}{\ff(\bar{\rr})}\right]\left[\ttt^i\frac{\partial}{\partial\ttt} + \frac{1}{\sin\ttt}\pp^i \frac{\partial}{\partial\pp}\right]\left[\alpha_\chi - \varphi_\chi - \Psi_\parallel - C_\parallel\right]~.
\enar
This corresponds to Eq.~(5.10) in \cite{YooGrimmMitsou_2018}, where
\bear
- \left[\ttt^i\TT_\gamma + \pp^i\PP_\gamma\right] \left(\Psi^\gamma + 2\Nn^\beta C^\gamma_\beta\right) 
- \left[\ttt^i\frac{\partial}{\partial\ttt} + \frac{1}{\sin\ttt}\pp^i \frac{\partial}{\partial\pp}\right]\left[\alpha_\chi - \varphi_\chi - \Psi_\parallel - C_\parallel\right]  \nnn
=-\hat{\nabla}^i(\alpha_\chi-\varphi_\chi)+\Nn^\alpha\hat{\nabla}^i\Psi_\alpha + \Nn^\alpha\Nn^\beta\hat{\nabla}^iC_{\al\be} ~.\qquad
\enar
Compared to the case in a flat universe, two extra contributions are present in proportion to $K$.
By introducing the perturbations~$\Delta\ttt:=\ttt_s-\ttt$ and $\Delta\pp:=\pp_s-\pp$ to account for the deviation of the two angles~$\ttt_s$ and $\pp_s$ observed in the source rest-frame from the two angles~$\ttt$ and $\pp$ observed in the observer rest-frame, the light propagation direction in the source rest-frame can be written as
\beeq
n^i_s=
\begin{pmatrix}
\sin\ttt_s \cos\pp_s\\
\sin\ttt_s \sin\pp_s\\
\cos\ttt_s
\end{pmatrix}
= n^i + \ttt^i \Delta\ttt + \sin\ttt~\pp^i\Delta\pp+\mathcal{O}(2)~.
\label{eqn:niSource}
\eneq
The perturbations in angles are therefore given by
\bear
\label{eqn:DThetaSourceRF}
\Delta\theta&=&\ttt_i n^i_s=\left[ \TT_\gamma V^\gamma - \Nn^\alpha \TT^{\gamma}C_{\alpha\gamma} - \pp_k \Omega^k\right]_{\bobs}^s 
-\frac{K\bar{\rr}_z}{2}\ttt_i \delta^i_\alpha(\delta x^\alpha + \mathcal{G}^\alpha)_{\bobs}\\
&&- \int_0^{\bar{\rr}_z}\frac{d\bar{\rr}}{\bar{\rr}}\TT_\gamma  \left(\Psi^\gamma+2\Nn^\beta C^\gamma_\beta\right)
-\int_0^{\bar{\rr}_z} \frac{d\bar{\rr}}{\bar{\rr}} \left[1 + \frac{K}{2}\frac{\bar{\rr}(\bar{\rr}_z - \bar{\rr})}{\ff(\bar{\rr})}\right]\frac{\partial}{\partial\ttt} \left[\alpha_\chi - \varphi_\chi - \Psi_\parallel - C_\parallel\right]~,\nonumber
\enar
and equivalently for $\Delta\pp=\frac{1}{\sin\ttt}\pp_i n^i_s$.
The corresponding expression in a flat Universe was derived in Eq.~(5.16) in \cite{YooGrimmMitsou_2018}.
As in Eq.~\eqref{eqn:Definition_ttt_pp}, we form an orthonormal basis by introducing the two direction vectors~$\ttt^i_s$ and $\pp^i_s$ perpendicular to $n^i_s$:
\bear
\ttt^i_s&=&\frac{\partial}{\partial\ttt_s}n^i_s = \ttt^i-n^i\Delta\ttt + \cos\ttt~\pp^i\Delta\pp+\mathcal{O}(2)=:\ttt^i + \Delta\ttt^i_s~,\nnn
\pp^i_s&=&\frac{1}{\sin\ttt_s}\frac{\partial}{\partial\pp_s}n^i_s=\pp^i-(\sin\ttt~n^i+\cos\ttt~\ttt^i )\Delta\pp+\mathcal{O}(2)=:\pp^i + \Delta\pp^i_s~.
\label{eqn:SourceDirecVectors}
\enar
The physical shape in the source rest-frame that appears in the observer sky extended by $(d\ttt, d\pp)$ can be determined by projecting $dx^\alpha_s$ in Eq.~\eqref{eqn:SourceSizeFRW} along the two orthogonal directions~$\ttt^i_s$ and $\pp_s^i$ in the plane perpendicular to the light propagation direction~$n^i_s$:
\bear
\label{eqn:dLttt1}
dL_{\ttt_s} &=& g_{\mu\nu}~e_i^\mu dx^\nu_s \ttt^i_s = g_{\alpha\beta}~e_i^\alpha dx^\beta_s \ttt^i_s + \mathcal{O}(2) \nnn
&=&a(\eta_s)\ff(\rr_s)\bar{\rr}_z\left[\frac{1}{\ff^2}d\ttt 
+ \bar{g}_{\alpha\beta}\ttt^i\de_i^\alpha\Delta s^\beta 
+ 2\ttt^i\de_i^\alpha \CC_{\alpha\beta} dn^\beta
+ \bar{g}_{\alpha\beta}dn^\beta \left(\delta_i^\alpha\Delta\ttt^i_s \phantom{\frac{K}{2}} \right.\right.\nnn
&&\left.\left.\qquad\qquad\qquad- \ttt^i\de_i^\gamma\left(\delta^\alpha_\gamma\varphi + \mathcal{G}^\alpha_{~,\gamma} + C^\alpha_\gamma - \frac{K\ff}{2}\delta^\alpha_\gamma x^\delta\mathcal{G}_\delta\right) - \delta^{\alpha j}\epsilon_{jik}\ttt^i\Omega^k\right)
\right]~,\\
dL_{\pp_s} &=& g_{\mu\nu}~e_i^\mu dx^\nu_s \pp^i_s = g_{\alpha\beta}~e_i^\alpha dx^\beta_s \pp^i_s + \mathcal{O}(2) \nnn
&=&a(\eta_s)\ff(\rr_s)\bar{\rr}_z\left[\frac{1}{\ff^2}\sin\ttt~d\pp 
+ \bar{g}_{\alpha\beta}\pp^i\de_i^\alpha\Delta s^\beta 
+ 2\pp^i\de_i^\alpha \CC_{\alpha\beta} dn^\beta
+ \bar{g}_{\alpha\beta}dn^\beta \left(\delta_i^\alpha\Delta\pp^i_s \phantom{\frac{K}{2}} \right.\right.\nnn
&&\left.\left.\qquad\qquad\qquad- \pp^i\de_i^\gamma\left(\delta^\alpha_\gamma\varphi + \mathcal{G}^\alpha_{~,\gamma} + C^\alpha_\gamma - \frac{K\ff}{2}\delta^\alpha_\gamma x^\delta\mathcal{G}_\delta\right) - \delta^{\alpha j}\epsilon_{jik}\pp^i\Omega^k\right)
\right]~.
\label{eqn:dLppp1}
\enar
This equation corresponds to Eq.~(5.18) in \cite{YooGrimmMitsou_2018}.
Apart from the appearance of some factors of the function~$\ff$ in our expression, they have the same parametrized form, as the terms in the second line of Eqs.~\eqref{eqn:dLttt1} and \eqref{eqn:dLppp1} are simply $\frac{1}{\ff}\delta e_i^\alpha \ttt^i$ or $\frac{1}{\ff}\delta e_i^\alpha \pp^i$ respectively. 
For the terms proportional to $\Delta s^\beta$, we notice that $\bar{g}_{\alpha\beta}\ttt^i\de_i^\alpha=\frac{1}{\ff^2}\delta_{\alpha\beta}\ttt^i\de_i^\alpha=\frac{1}{\ff^2}\delta_{i\beta}\ttt^i$, hence
\bear
\bar{g}_{\alpha\beta}\ttt^i\de_i^\alpha\Delta s^\beta &=& \frac{1}{\ff^2}\left[d\ttt\left(\frac{\drr}{\bar{\rr}_z} + \frac{\partial}{\partial\ttt}\dtt\right) + d\pp\left(\frac{\partial}{\partial\pp}\dtt - \sin\ttt\cos\ttt~\dpp\right)\right]~, \nnn
\bar{g}_{\alpha\beta}\pp^i\de_i^\alpha\Delta s^\beta &=& \frac{1}{\ff^2}\left[d\ttt\left(\cos\ttt+ \sin\ttt \frac{\partial}{\partial\ttt}\right)\dpp + \sin\ttt~d\pp\left(\frac{\drr}{\bar{\rr}_z}  + \frac{\partial}{\partial\pp}\dpp + \cot\ttt~\dtt\right)\right]~.
\enar
Apart from the overall factor of $\ff^{-2}$, our expressions coincide with the parametrization in Eqs.~(5.19) and (5.20) in \cite{YooGrimmMitsou_2018} derived for a flat universe.
The next terms in Eqs.~\eqref{eqn:dLttt1} and \eqref{eqn:dLppp1} are proportional to the perturbation~$\CC_{\alpha\beta}$ in the spatial part of the metric, which are decomposed according to Eq.~\eqref{eq:decom}:
\bear
2\ttt^i\de_i^\alpha \CC_{\alpha\beta} dn^\beta &=& \frac{2}{\ff^2}\varphi~d\ttt + 2\ttt^i\de_i^\alpha\mathcal{G}_{(\alpha|\beta)}dn^\beta + 2C_{\alpha\beta}\ttt^i\de_i^\alpha dn^\beta~, \nnn
2\pp^i\de_i^\alpha \CC_{\alpha\beta} dn^\beta &=&\frac{2}{\ff^2}\varphi\sin\ttt~d\pp + 2\pp^i\de_i^\alpha\mathcal{G}_{(\alpha|\beta)}dn^\beta + 2C_{\alpha\beta}\pp^i\de_i^\alpha dn^\beta~,
\enar
where we used that $\gamma_{,\alpha|\beta}=\gamma_{,\beta|\alpha}$.
Finally, the last terms in Eqs.~\eqref{eqn:dLttt1} and \eqref{eqn:dLppp1} are
\bear
&&\bar{g}_{\alpha\beta}dn^\beta \left(\delta_i^\alpha\Delta\ttt^i_s - \ttt^i\de_i^\gamma\left(\delta^\alpha_\gamma\varphi + \mathcal{G}^\alpha_{~,\gamma} + C^\alpha_\gamma - \frac{K\ff}{2}\delta^\alpha_\gamma x^\delta\mathcal{G}_\delta\right) - \delta^{\alpha j}\epsilon_{jik}\ttt^i\Omega^k\right) \nnn
&&\quad=\frac{1}{\ff^2}\left[ d\ttt\left(\frac{K\ff}{2}x^\delta \mathcal{G}_\delta - \varphi\right) + \sin\ttt~d\pp\left(\cos\ttt~\Delta\pp - \epsilon_{jik}\pp^j\ttt^i\Omega^k\right) \right]
-\bar{g}_{\alpha\beta}dn^\beta\ttt^i\de_i^\gamma(\mathcal{G}^\alpha_{~,\gamma}+C^\alpha_\gamma)~, \nnn
&&\bar{g}_{\alpha\beta}dn^\beta \left(\delta_i^\alpha\Delta\pp^i_s - \pp^i\de_i^\gamma\left(\delta^\alpha_\gamma\varphi + \mathcal{G}^\alpha_{~,\gamma} + C^\alpha_\gamma - \frac{K\ff}{2}\delta^\alpha_\gamma x^\delta\mathcal{G}_\delta\right) - \delta^{\alpha j}\epsilon_{jik}\pp^i\Omega^k\right) \\
&&\quad=\frac{1}{\ff^2}\left[ -d\ttt\left(\cos\ttt \Delta\pp + \epsilon_{jik}\ttt^j\pp^i\Omega^k\right) + \sin\ttt~d\pp\left(\frac{K\ff}{2}x^\delta \mathcal{G}_\delta -\varphi\right) \right]
-\bar{g}_{\alpha\beta}dn^\beta\pp^i\de_i^\gamma(\mathcal{G}^\alpha_{~,\gamma}+C^\alpha_\gamma)~.\nonumber
\enar
Combining all these terms leads to
\bear
\frac{dL_{\ttt_s}}{a(\eta_s)\ff(\rr_s)\bar{\rr}_z} &=&\frac{1}{\ff^2}d\ttt \left[1
+ \frac{\drr}{\bar{\rr}_z} + \frac{\partial}{\partial\ttt}\dtt
+\varphi
+\frac{K\ff}{2}x^\delta \mathcal{G}_\delta
\right]\nnn
&&+\frac{1}{\ff^2}\sin\ttt~d\pp \left[ \frac{1}{\sin\ttt}\frac{\partial}{\partial\pp}\dtt + \cos\ttt(\Delta\pp-\dpp) + \Omega^n
\right]\nnn
&&+ \left(2\ttt^i\de_i^\alpha \mathcal{G}_{(\alpha|\beta)} 
-\bar{g}_{\alpha\beta}\ttt^i\de_i^\gamma\mathcal{G}^\alpha_{~,\gamma}\right)dn^\beta
+ C_{\alpha\beta} \ttt^i\de_i^\alpha dn^\beta~,\nnn
\frac{dL_{\pp_s}}{a(\eta_s)\ff(\rr_s)\bar{\rr}_z} &=&\frac{1}{\ff^2}d\ttt\left[\cos\ttt~\dpp+ \sin\ttt \frac{\partial}{\partial\ttt}\dpp
-\cos\ttt \Delta\pp - \Omega^n
\right] \nnn
&&+\frac{1}{\ff^2}\sin\ttt~d\pp \left[1+ \frac{\drr}{\bar{\rr}_z}  + \frac{\partial}{\partial\pp}\dpp + \cot\ttt~\dtt
+ \varphi +\frac{K\ff}{2}x^\delta \mathcal{G}_\delta 
\right] \nnn
&&+\left(2\pp^i\de_i^\alpha\mathcal{G}_{(\alpha|\beta)} -\bar{g}_{\alpha\beta}\pp^i\de_i^\gamma\mathcal{G}^\alpha_{~,\gamma}\right)dn^\beta + C_{\alpha\beta}\pp^i\de_i^\alpha dn^\beta~,
\label{eqn:derivationL1}
\enar
where we decomposed the rotation~$\Omega^i$ of the local tetrad basis as
\beeq
\Omega^i =: n^i \Omega^n + \ttt^i \Omega^\ttt + \pp^i \Omega^\pp~,
\label{eqn:DecompRotTetrad}
\eneq
hence $\epsilon_{jik}\ttt^j\pp^i\Omega^k=-\epsilon_{jik}\pp^j\ttt^i\Omega^k=\Omega^n$.
We first have a closer look on the two terms proportional to the perturbation~$\mathcal{G}_\alpha$.
The covariant derivative with respect to the background three-metric~$\bar{g}_{\alpha\beta}$ is related to the coordinate derivative through
\beeq
\mathcal{G}_{\alpha|\beta}=\mathcal{G}_{\alpha,\beta} - \bar{\Gamma}^\gamma_{\beta\alpha}\mathcal{G}_\gamma = \mathcal{G}_{\alpha,\beta}  + {K\over2\ff}\left( \de_{\alpha\de}\xx^\de\mathcal{G}_\be + \de_{\be\de}\xx^\de\mathcal{G}_\alpha- \de_{\be\alpha}\xx^\gamma\mathcal{G}_\gamma\right)~,
\eneq
and since
\beeq
\bar{g}_{\alpha\beta} \mathcal{G}^\alpha_{~,\gamma} = \mathcal{G}_{\beta,\gamma} + {K\over \ff}\de_{\gamma\de}\xx^\de\mathcal{G}_\be~,
\eneq
the part proportional to $\mathcal{G}_\alpha$ becomes
\bear
2\ttt^i\de_i^\alpha\mathcal{G}_{(\alpha|\beta)} -\bar{g}_{\alpha\beta}\ttt^i\de_i^\gamma\mathcal{G}^\alpha_{~,\gamma} &=&  \ttt^i\de_i^\alpha\mathcal{G}_{\alpha,\beta} 
+  {K\over \ff}\left( \de_{\be\de}\xx^\de\ttt^i\de_i^\alpha\mathcal{G}_\alpha- \ttt^i\de_{i\be}\xx^\gamma\mathcal{G}_\gamma\right)\nnn
&=&  \ttt^i\de_i^\alpha\mathcal{G}_{\alpha,\beta} 
+  {K\bar{\rr}\over \ff^2}\left( n^i\de_{i\be}\TT^\alpha\mathcal{G}_\alpha- \ttt^i\de_{i\be}\mathcal{G}_\parallel\right)~, 
\enar
where we used that $x^\alpha=\bar{\rr}n^i\de_i^\alpha+ \mathcal{O}(1)$ along the photon propagation path.
Multiplying this with $dn^\beta$ leads to
\bear
(2\ttt^i\de_i^\alpha \mathcal{G}_{(\alpha|\beta)}  -\bar{g}_{\alpha\beta}\ttt^i\de_i^\gamma\mathcal{G}^\alpha_{~,\gamma})dn^\beta &=& \frac{1}{\ff^2}\left[(d\ttt~\TT^\beta  + d\pp~\sin\ttt~\PP^\beta )\TT^\alpha\mathcal{G}_{\alpha,\beta} - K\bar{\rr}\mathcal{G}_\parallel d\ttt\right]~,\qquad\quad \nnn
(2\pp^i\de_i^\alpha \mathcal{G}_{(\alpha|\beta)}  -\bar{g}_{\alpha\beta}\pp^i\de_i^\gamma\mathcal{G}^\alpha_{~,\gamma})dn^\beta &=& \frac{1}{\ff^2}\left[(d\ttt~\TT^\beta  + d\pp~\sin\ttt~\PP^\beta )\PP^\alpha\mathcal{G}_{\alpha,\beta} - K\bar{\rr}\sin\ttt~\mathcal{G}_\parallel d\pp\right]~.\qquad\quad 
\enar
The tensor terms in Eq.~\eqref{eqn:derivationL1} are
\bear
C_{\alpha\beta} \ttt^i\de_i^\alpha dn^\beta = \frac{1}{\ff^2}
(C_{\TT\TT}~d\ttt + C_{\TT\PP}\sin\ttt~d\pp)
~,\nnn
C_{\alpha\beta} \pp^i\de_i^\alpha dn^\beta = \frac{1}{\ff^2} (C_{\PP\TT}~d\ttt + C_{\PP\PP}\sin\ttt~d\pp)~,
\enar
where we introduced the notation
\beeq
C_{\TT\TT}=C_{\alpha\beta}\TT^\alpha\TT^\beta~, \qquad\quad
C_{\PP\PP}=C_{\alpha\beta}\PP^\alpha\PP^\beta~,\qquad\quad
C_{\PP\TT}=C_{\alpha\beta}\PP^\alpha\TT^\beta=C_{\TT\PP}~.
\eneq
The physical shape of the source in Eq.~\eqref{eqn:derivationL1} can be rewritten as
\bear
dL_{\ttt_s} &=&\frac{a(\eta_s)\bar{\rr}_z}{\ff(\rr_s)}\left\{d\ttt \left[1
+ \frac{\drr}{\bar{\rr}_z} + \frac{\partial}{\partial\ttt}\dtt
+\varphi
-\frac{K\bar{\rr}_z}{2} \mathcal{G}_\parallel
+ \TT^\alpha\TT^\beta\mathcal{G}_{\alpha,\beta} + C_{\TT\TT}
\right]\right.\nnn
&&\left.+\sin\ttt~d\pp \left[ \frac{1}{\sin\ttt}\frac{\partial}{\partial\pp}\dtt + \cos\ttt(\Delta\pp-\dpp) - \epsilon_{jik}\pp^j\ttt^i\Omega^k + \TT^\alpha\PP^\beta\mathcal{G}_{\alpha,\beta} + C_{\TT\PP}
\right]\right\}~,\nnn
dL_{\pp_s} &=&\frac{a(\eta_s)\bar{\rr}_z}{\ff(\rr_s)}\left\{d\ttt\left[\cos\ttt~\dpp+ \sin\ttt \frac{\partial}{\partial\ttt}\dpp
-\cos\ttt \Delta\pp - \epsilon_{jik}\ttt^j\pp^i\Omega^k
+ \PP^\alpha\TT^\beta\mathcal{G}_{\alpha,\beta} + C_{\PP\TT}
\right]\right. \nnn
&&\left.+\sin\ttt~d\pp \left[1+ \frac{\drr}{\bar{\rr}_z}  + \frac{\partial}{\partial\pp}\dpp + \cot\ttt~\dtt
+ \varphi -\frac{K\bar{\rr}}{2} \mathcal{G}_\parallel
+ \PP^\alpha\PP^\beta\mathcal{G}_{\alpha,\beta} + C_{\PP\PP}
\right]\right\}~.
\label{eqn:derivationL2}
\enar
This corresponds to Eq.~(5.23) in \cite{YooGrimmMitsou_2018} in a flat Universe.
Using Eq.~\eqref{redshift} and expanding the function $\ff(\rr_s)=\ff(\bar{\rr}_z+\delta\rr)$ around the background radial coordinate~$\bar{\rr}_z$ leads to
\beeq
\frac{a(\eta_s)\bar{\rr}_z}{\ff(\rr_s)} =  \frac{\bar{\rr}_z}{(1+z)\ff(\bar{\rr}_z)}\left[1+\dz-\frac{K}{2\ff(\bar{\rr}_z)}\bar{\rr}_z \delta\rr\right]
= \bar\dD_A(z)\left[1+\dz-\frac{K}{2\ff(\bar{\rr}_z)}\bar{\rr}_z \delta\rr\right]~,
\eneq
at linear order in perturbation theory, where $\bar\dD_A(z)$ is the angular diameter distance in the background, given in Eq.~\eqref{eqn:BGDistances}.
So the physical shape in the source rest-frame $(dL_{\ttt_s}, dL_{\pp_s} )$ in Eq.~\eqref{eqn:derivationL2} can be expressed at the observed redshift as 
\beeq
\begin{pmatrix}
dL_{\ttt_s} \\
dL_{\pp_s} 
\end{pmatrix}
=: \bar{\mathcal{D}}_A(z)
 \begin{pmatrix}
\hat{\mathbb{D}}_{11}  & \hat{\mathbb{D}}_{12} \\
\hat{\mathbb{D}}_{21} & \hat{\mathbb{D}}_{22}
\end{pmatrix}
 \begin{pmatrix}
d{\ttt} \\
\sin\ttt~d{\pp} 
\end{pmatrix}~,
\eneq
in terms of the observed source size~$(d\ttt, d\pp)$ in the sky and the distortion matrix elements
\bear
\hat{\mathbb{D}}_{11} &=&\left( 1 + \frac{\partial}{\partial\ttt}\dtt\right)+\delta z + \left(1-\frac{1}{4}K\bar{\rr}_z^2\right)\frac{\delta\rr}{\bar{\rr}_z \ff(\bar{\rr}_z)} + \varphi + \TT^\alpha\TT^\beta \mathcal{G}_{\alpha,\beta} - \frac{K\bar{\rr}_z}{2} \mathcal{G}_\parallel + C_{\TT\TT}~, \nnn
\hat{\mathbb{D}}_{22} &=& \left(1 + \cot\ttt~\dtt + \frac{\partial}{\partial\pp}\dpp\right)+\delta z + \left(1-\frac{1}{4}K\bar{\rr}_z^2\right)\frac{\delta\rr}{\bar{\rr}_z \ff(\bar{\rr}_z)} + \varphi + \PP^\alpha\PP^\beta \mathcal{G}_{\alpha,\beta} - \frac{K\bar{\rr}_z}{2} \mathcal{G}_\parallel + C_{\PP\PP}~, \nnn
\hat{\mathbb{D}}_{12} &=&\left( \frac{1}{\sin\ttt}\frac{\partial}{\partial\pp}\dtt - \cos\ttt~\dpp\right)   + \TT^\alpha\PP^\beta \mathcal{G}_{\alpha,\beta} + C_{\TT\PP} + \Omega^n + \cos\ttt~\Delta\pp~,\nnn
\hat{\mathbb{D}}_{21} &=&\left( \cos\ttt + \sin\ttt\frac{\partial}{\partial\ttt}\right)\dpp + \PP^\alpha\TT^\beta \mathcal{G}_{\alpha,\beta} + C_{\PP\TT} - \Omega^n - \cos\ttt~\Delta\pp~.
\label{eqn:DistortionMatrixElements}
\enar
These matrix elements correspond to Eqs.~(5.26)$-$(5.29) in \cite{YooGrimmMitsou_2018}.
Mind that the term proportional to the radial distortion~$\drr$ expressed in terms of the coordinates where the radius in the background corresponds to the comoving angular diameter distance~$\ro$ is just $\widetilde{\drr}(\tilde{x})/\bar{\ro}_z$, as evident from Eq.~\eqref{eqn:deltaRrelation}.
The appearance of the term proportional to $K\delta\rr$ emerges from the expansion of the function~$\ff(\rr_s)$ around $\bar{\rr}_z$. 
These terms do not appear in $\hat{\mathbb{D}}_{12}$ and $\hat{\mathbb{D}}_{21}$ as we work at linear order in perturbation theory.
Note that
\bear
\TT^\alpha\TT^\beta\mathcal{G}_{\alpha,\beta} - \frac{K\bar{\rr}_z}{2} \mathcal{G}_\parallel &=& \TT^\alpha\TT^\beta\mathcal{G}_{\alpha|\beta} ~, \qquad\qquad
\TT^\alpha\PP^\beta\mathcal{G}_{\alpha,\beta} = \TT^\alpha\PP^\beta\mathcal{G}_{\alpha|\beta}~,\nnn
\PP^\alpha\PP^\beta\mathcal{G}_{\alpha,\beta} - \frac{K\bar{\rr}_z}{2} \mathcal{G}_\parallel &=& \PP^\alpha\PP^\beta\mathcal{G}_{\alpha|\beta} ~, \qquad\qquad
\PP^\alpha\TT^\beta\mathcal{G}_{\alpha,\beta} = \PP^\alpha\TT^\beta\mathcal{G}_{\alpha|\beta}~,
\enar
and
\beeq
\frac{\partial}{\partial x^\alpha}=\frac{\partial}{\partial \bar{x}^\alpha} + \mathcal{O}(1)
= n^i\de_{i\alpha}\frac{\partial}{\partial \bar{\rr}} + \frac{1}{\bar{\rr}}\ttt^i\de_{i\alpha}\frac{\partial}{\partial \ttt} + \frac{1}{\bar{\rr}\sin\ttt}\pp^i\de_{i\alpha}\frac{\partial}{\partial \pp} + \mathcal{O}(1)~,
\eneq
hence
\bear
\TT^\alpha\TT^\beta\mathcal{G}_{\alpha,\beta} &=& \ff^2(\bar{\rr}) \ttt^i\de_i^\alpha\ttt^j\de_j^\beta\mathcal{G}_{\alpha,\beta}
= \frac{\ff^2(\bar{\rr})}{\bar{\rr}} \ttt^i\de_i^\alpha\frac{\partial}{\partial\ttt}\mathcal{G}_\alpha~,\nnn
\PP^\alpha\PP^\beta\mathcal{G}_{\alpha,\beta} &=& \ff^2(\bar{\rr}) \pp^i\de_i^\alpha\pp^j\de_j^\beta\mathcal{G}_{\alpha,\beta}
= \frac{\ff^2(\bar{\rr})}{\bar{\rr}\sin\ttt} \pp^i\de_i^\alpha\frac{\partial}{\partial\pp}
\mathcal{G}_\alpha~,\nnn
\TT^\alpha\PP^\beta\mathcal{G}_{\alpha,\beta} &=& \ff^2(\bar{\rr}) \ttt^i\de_i^\alpha\pp^j\de_j^\beta\mathcal{G}_{\alpha,\beta}
= \frac{\ff^2(\bar{\rr})}{\bar{\rr}\sin\ttt} \ttt^i\de_i^\alpha\frac{\partial}{\partial\pp}
\mathcal{G}_\alpha~,\nnn
\PP^\alpha\TT^\beta\mathcal{G}_{\alpha,\beta} &=& \ff^2(\bar{\rr})\pp^i\de_i^\alpha\ttt^j\de_j^\beta\mathcal{G}_{\alpha,\beta}
= \frac{\ff^2(\bar{\rr})}{\bar{\rr}}
\pp^i\de_i^\alpha\frac{\partial}{\partial\ttt} \mathcal{G}_\alpha~.
\label{eqn:identities1}
\enar

Having derived the relation of the physical shape of the source and the observed angular shape, now we are in a position to derive the weak lensing observables.
First, the convergence~$\hat{\kappa}$ of the distortion matrix~$\hat{\mathbb{D}}$ is
\bear
\hat{\kappa}&\equiv& 1-\frac{1}{2}\text{Tr}(\hat{\mathbb{D}}) \nnn
&=&\kappa-\delta z - \left(1-\frac{1}{4}K\bar{\rr}_z^2\right)\frac{\delta\rr}{\bar{\rr}_z \ff_z} -\varphi
-\frac{1}{2}\left[ C_{\TT\TT} + C_{\PP\PP} + 
\left( \TT^\alpha\TT^\beta + \PP^\alpha\PP^\beta \right)\mathcal{G}_{\alpha|\beta}
\right]~,
\label{eqn:convergence}
\enar
where
\bear
\left( \TT^\alpha\TT^\beta + \PP^\alpha\PP^\beta \right)\mathcal{G}_{\alpha|\beta}
=\left( \TT^\alpha\TT^\beta + \PP^\alpha\PP^\beta \right)\mathcal{G}_{\alpha,\beta}
- K\bar{\rr}_z \mathcal{G}_\parallel
 ~.
\enar
Contrary to the standard expression~$\kappa$ in Eq.~\eqref{eqn:StandardConvergence}, the expression for the lensing convergence~$\hat{\kappa}$ is gauge invariant, as the additional terms in Eq.~\eqref{eqn:convergence} exactly cancel the gauge-dependent terms of $\kappa$.
Comparing the distance fluctuation~$\delta \mathcal{D}$ in Eq.~\eqref{LD} with the convergence~$\hat{\kappa}$ in Eq.~\eqref{eqn:convergence} and noting that 
\beeq
\CC^\alpha_\alpha-\CC_\parallel = 2\varphi 
+(\TT^\alpha\TT^\beta+\PP^\alpha\PP^\beta)\mathcal{G}_{\alpha,\beta}
- K\bar{\rr}\mathcal{G}_\parallel - C_\parallel~,
\eneq
and
\beeq
C_\parallel = - C_{\TT\TT} - C_{\PP\PP}~,
\eneq
as the tensor~$C_{\alpha\beta}$ is traceless, we find that the two quantities are equivalent up to sign, 
\beeq
\hat{\kappa}=-\delta \mathcal{D}~,
\label{eqn:RelKappadD}
\eneq
at linear order in perturbation theory.
This relation has already been pointed out in \cite{YooGrimmMitsou_2018} in their Eq.~(5.41), and since we proved in Sec.~\ref{sec:LD} the gauge invariance of the distance fluctuation, it follows that the convergence~$\hat{\kappa}$ is indeed gauge invariant.

Next, the two gravitational lensing shear components of the distortion matrix~$\hat{\mathbb{D}}$ are
\bear
2\hat{\gamma}_1 &\equiv& \hat{\mathbb{D}}_{22} - \hat{\mathbb{D}}_{11} 
=\cot\ttt~\dtt + \frac{\partial}{\partial\pp}\dpp - \frac{\partial}{\partial\ttt}\dtt + (\PP^\alpha\PP^\beta-\TT^\alpha\TT^\beta)\mathcal{G}_{\alpha|\beta} + C_{\PP\PP} -  C_{\TT\TT}~,~~~\nnn
-2\hat{\gamma}_2 &\equiv& \hat{\mathbb{D}}_{12} + \hat{\mathbb{D}}_{21} 
=\frac{1}{\sin\ttt} \frac{\partial}{\partial\pp}\dtt +\sin\ttt \frac{\partial}{\partial\ttt}\dpp + (\TT^\alpha\PP^\beta+\PP^\alpha\TT^\beta)\mathcal{G}_{\alpha|\beta}  + C_{\TT\PP} + C_{\PP\TT}~.
\label{eqn:shear}
\enar
We can verify their gauge-invariance by using the gauge transformation properties of the two angular distortions~$\dtt$ and $\dpp$ in Eq.~\eqref{eqn:SourcePositionPerturbationsGT} and the identities in Eq.~\eqref{eqn:identities1}.
The shear components derived in a flat Universe in \cite{YooGrimmMitsou_2018} in their Eqs.~(5.42)-(5.45) have the same parametrized form as our result, but the individual components such as $\dtt$ and $\dpp$ are different and depend on the value of the curvature~$K$.
Mind that there is also a factor of the function~$\ff_z$ in the definition of $\TT^\al$ and $\PP^\al$.
The terms proportional to the metric perturbation variables~$\mathcal{G}_\alpha$ and $C_{\al\beta}$ are coming from the change of the FRW frame to the rest-frame of the source \cite{YooGrimmMitsou_2018}.

Finally, the rotation~$\hat{\omega}$ of the distortion matrix~$\hat{\mathbb{D}}$ is
\bear
\label{eqn:rotation}
2\hat{\omega}&\equiv& \hat{\mathbb{D}}_{21} - \hat{\mathbb{D}}_{12}\\
&=&\left( \sin\ttt\frac{\partial}{\partial\ttt}\dpp - \frac{1}{\sin\ttt}\frac{\partial}{\partial\pp}\dtt + 2\cos\ttt~\dpp\right)
+  (\PP^\alpha\TT^\beta -  \TT^\alpha\PP^\beta )\mathcal{G}_{\alpha,\beta}
- 2\Omega^n - 2\cos\ttt~\Delta\pp
~.\nonumber
\enar
Since
\beeq
\dtt\ni\pp_i \Omega^i_o~, \qquad\qquad
\dpp\ni-\frac{1}{\sin\ttt}\ttt_i \Omega^i_o~, \qquad\qquad
\Delta\pp\ni\frac{1}{\sin\ttt}\ttt_i (\Omega^i_s - \Omega^i_o)~,
\eneq
the rotation~$\hat{\omega}$ is related to the rotation~$\Omega^i$ of the spatial tetrad vectors against the FRW coordinate as
\beeq
2\hat{\omega}\ni 2(\Omega_o^n-\Omega_s^n) + 2\cot\ttt~(\Omega^\ttt_o-\Omega^\ttt_s)~.
\eneq
The rotation can be measured once the local coordinates of the source and the observer are synchronized. 
The only physical way for the synchronization is to parallel transport the observer basis~$e_i^\mu$ to the source position, which completely determines $\Omega_s^i$ in terms of $\Omega_o^i$. 
It was shown \cite{YooGrimmMitsou_2018} that the extra contributions in Eq.~\eqref{eqn:rotation} are exactly what cancel the contributions from the parallel transport of $e_i^\mu$ in $\Omega_s^i$, in a flat universe.
In other words, while it appears that the local basis vectors rotate as they are parallel transported in a FRW coordinate, there exists no rotation in the rest frame at the linear order against its local coordinates. The rotation in a FRW coordinate without accounting for $\Omega_o^i$ and $\Omega_s^i$ is a coordinate artifact.
We believe that the rotation~$\hat{\omega}$ would be zero at the linear order in a non-flat universe, which will be proved in a later work.

\subsection{Cosmic microwave background anisotropies}
\label{sec:CMB}
\label{sec:CMB}
With the large angular diameter distance to the last
scattering surface, the impact of a non-zero spatial curvature on
cosmic microwave background anisotropies has been extensively studied
\cite{1983ApJ...273....2W, 1998ApJ...494..491Z, 1998PhRvD..57.3290H, 1999ApJ...513....1C, 1999PhRvD..59h3506M, 2000PhRvD..62d3004C, 2000GReGr..32.1059C, 2000CQGra..17..871C, 2014JCAP...09..032L, 1996astro.ph..4166H, 2002ARA&A..40..171H} in literature. Given our computation of the light propagation
in a non-flat universe, here we revisit the issue and discuss the impact
on CMB.

With conservation of the phase space density~${\cal F}$ for photon 
distributions, the propagation of CMB photons is described by the Boltzmann 
equation (see, e.g., \cite{1995ApJ...455....7M}).
Free electrons scatter off CMB photons, giving rise to
the correction to the phase-space conservation, but the collisions 
are local events and well described by QFT in Minkowski metric, which is
independent of whether the universe is flat or non-flat. However,
the collisionless Boltzmann equation involves the light propagation over
cosmological scales, which is naturally affected by the curvature of
space. At the linear order in perturbations, the propagation of CMB photons
is on a straight background path, as we are interested in the linear-order
perturbation~$f$ 
of the phase-space distribution function. Hence the only part that is affected
by the spatial curvature in the collisionless Boltzmann equation 
\beeq
0=f'+\hat q^\al{\pa f\over\pa \xx^\al}+
{d\bar f\over d\ln q}{d\ln q\over d\eta}~,\Dquad q:=a\omega~,
\label{eqn:BoltzmannEqn}
\eneq
is the change of the comoving momentum~$q$ in a non-flat universe~$\frac{d\ln q}{d\eta}$,
where ${\cal F}:=\bar f+f$ and $\hat q^\al$ is a unit directional vector
of the photon propagation.

Here we investigate if the change of the comoving momentum in a non-flat 
universe is different from that in a flat universe.
First, the photon wave vector~$k^\mu$ along a photon path can be expressed
by using the tetrad expressions in Eqs.~\eqref{eq:srcv} and~\eqref{srcs} as
\bear
\label{k0}
k^\eta &=&\frac{q}{a^2}\left[ 1-\al-\Nn^\beta (\VV_\beta-\BB_\beta) 
\right]~, \\
\label{ka}
k^\alpha &=&\frac{q}{a^2} \left[-\Nn^\alpha+\VV^\alpha+\Nn^\beta
\left( \delta^\alpha_\beta \varphi +\CCG^\alpha{}_{,\beta}+C^\alpha_\beta 
\right)+ \epsilon^\alpha{}_{\be k}\Nn^\be\Omega^k
 - \frac{K}{2\ff}\Nn^\alpha \de_{\be\ga}x^\de \CCG^\ga\right]~,
\enar
where the photon wave vector is evaluated at the propagation position.
Note that the photon propagation vector~$\Nn^\al$ here in this subsection 
represents 
\beeq
\Nn^\al=\ff n^i\de_i^\al~,
\eneq
with the observed photon propagation direction~$n^i$ in the rest frame 
of ``fictitious observers'' along the photon propagation direction, not
the observer at the origin (or us at the Earth). Since we will deal with
only the background photon path, this subtle distinction vanishes, 
and~$\Nn^\al$ here in this subsection is equivalent to one used
throughout the paper. Similarly, the photon angular frequency~$\omega$
in the comoving momentum $q=a\omega$ is also one observed in the rest frame
of fictitious observers, not one we measure in our rest frame. But at
the background $\omega\propto1/a$ and the comoving momentum is constant.
Second, consider the time derivative of the expression for~$k^\eta$ 
in Eq.~\eqref{k0}
\beeq
\frac{d}{d\eta}k^\eta= \left(\frac{d\ln q}{d\eta} -2\HH \right)k^\eta
- \frac{q}{a^2}\frac{d}{d\eta}\left[ \al+\Nn^\beta (\VV_\beta-\BB_\beta) 
\right]~,
\eneq
and the change of the comoving momentum is then
\beeq
{d\ln q\over d\eta}=2\HH+{d\over d\eta}\ln k^\eta+ 
\frac{d}{d\eta}\left[ \al+\Nn^\beta (\VV_\beta-\BB_\beta)\right]~,
\label{eqn:GE1}
\eneq
where we used $\bar k^\eta=q/a^2$ to simplify the equation. Given that 
\beeq
{d\over d\eta}={d\cc\over d\eta}{d\over d\cc}=
\frac{\partial}{\partial \eta} - \Nn^\alpha\frac{\partial}{\partial x^\alpha}
+\OO(1)~,
\eneq
the last term can be readily computed as
\beeq
\frac{d}{d\eta}\left[ \al+\Nn^\beta (\VV_\beta-\BB_\beta) \right] = 
\al'-\al_{,\para} + \left(\VV_\parallel -\BB_\parallel\right)'
-(\VV_\al-\BB_\al)_{|\beta} \Nn^\alpha\Nn^\beta ~,
\eneq
and our task is to compute the middle term~$d\ln k^\eta / d\eta$ using the geodesic equation.

Since we want the time derivative of the physical photon wave vector~$k^\eta$
rather than $\CK^\eta$, we use the physical affine parameter~$\Lambda$ in 
Eq.~\eqref{eqn:RelDiffK} to derive
\beeq
\frac{d}{d\eta}\ln k^\eta = \frac{d\Lambda}{d\eta}\frac{d}{d\Lambda}\ln k^\eta 
= \frac{1}{k^\eta}\frac{d}{d\Lambda}\ln k^\eta~,
\eneq
and the temporal component of the geodesic equation yields
\bear
\frac{d}{d\Lambda}k^\eta &=& -\Gamma^\eta_{\mu\nu}k^\mu k^\nu 
=-\Ga^\eta_{\eta\eta}(k^\eta)^2-2\Ga^\eta_{\eta\al}\bar k^\eta \bar k^\al
-\Ga^\eta_{\al\be}k^\al k^\be \\
&=& \frac{q^2}{a^4}\left[-2\HH + 4\HH(\al+\VV_\parallel-\BB_\parallel) -
\al' + 2\al_{,\parallel}-
 \varphi'- (\BB_{\alpha|\beta}+\CCG'_{\alpha|\beta})
\Nn^\alpha\Nn^\beta -C^\prime_\parallel \right]~.\nonumber
\enar
Hence, we derive
\beeq
{d\over d\eta}\ln k^\eta=-2\HH-
\al' + 2\al_{,\parallel}- \varphi'- (\BB_{\alpha|\beta}+\CCG'_{\alpha|\beta})
\Nn^\alpha\Nn^\beta -C^\prime_\parallel ~,
\eneq
and the change of the comoving momentum in a non-flat universe is finally
obtained as
\beeq
{d\ln q\over d\eta}= \al_{,\parallel}- \varphi'
 + \left(\VV_\parallel -\BB_\parallel\right)'
- \left(\VV_\al+\CCG'_\al\right)_{|\be}\Nn^\alpha\Nn^\beta 
-C^\prime_\parallel ~.
\label{eqn:dlnqdeta}
\eneq
The change of the comoving momentum is zero in the background, and in an
inhomogeneous universe, it varies exactly the same way in a flat universe, once
the photon propagation direction is properly replaced by~$\Nn^\al$.
Therefore, there are
no further corrections in the Boltzmann equation in a non-flat universe.
The angular multipole decomposition of the phase-space distribution 
for the observed CMB anisotropies is of course unaffected by the spatial
curvature.
Eq.~\eqref{eqn:dlnqdeta} can be rewritten in terms of gauge-invariant variables as
\beeq
{d\ln q\over d\eta}= {\al_\chi}_{,\parallel}- \varphi_\chi'
 + V_\parallel'
- \left(V_\al+\Psi_\al\right)_{|\be}\Nn^\alpha\Nn^\beta 
-C^\prime_\parallel  - \frac{d}{d\cc}(H\chi)~,
\eneq
and the last term with $H\chi$ is gauge-dependent, which is exactly what is needed to cancel the gauge-dependent part of the distribution function in the Boltzmann equation~\eqref{eqn:BoltzmannEqn} \cite{Mitsou_2020,2021PhRvD.103f3516B}:
\beeq
\de_\xi f=-\mathcal{H}T  \frac{d \bar{f}}{d\ln q}~.
\eneq

\section{Expressions in a nearly flat universe}
\label{sec:NearlyFlat}
In this Section we expand the expressions for the cosmological observables in a real universe with non-zero spatial curvature around a fiducial cosmology without spatial curvature ($\Omega_k=0$) and treat the curvature density~$\Omega_k$ in a real universe as a perturbation. 
In these two models, the cosmological parameters are different, and the perturbation evolutions are different, while the observed quantities such as the observed redshift are identical.
Given the tight constraints on the spatial curvature $|\Omega_k|\lesssim 0.02$ (1-$\sigma$ constraint from CMB alone \cite{Planck2018CosmParam}), we can use the perturbation variables in a flat fiducial cosmology, where the evolution is simpler to compute.
The formulas in this Section are, however, only an approximation to the exact expressions in a non-flat universe, which are valid only when $|\Omega_k|\ll1$ is as small as other perturbations.
In both models, we keep the same energy densities $\rho_r$, $\rho_m$, $\rho_\Lambda$ of radiation, matter, and a cosmological constant.
The Hubble parameter in a real universe with non-zero spatial curvature is
\begin{equation}
H^2=\frac{8\pi G}{3}(\rho_r + \rho_m + \rho_\Lambda) - \frac{K}{a^2} = 
H_0^2\left[\frac{\Omega_r}{a^4} + \frac{\Omega_m}{a^3} + \Omega_\Lambda + \frac{\Omega_k}{a^2}\right]~,
\end{equation}
where the density parameters are 
\begin{equation}
\Omega_{r,m,\Lambda}
 = \frac{8\pi G}{3H_0^2}~\bar{\rho}_{r,m,\Lambda}~,\qquad\qquad
 \Omega_k=-\frac{K}{H_0^2}~.
\end{equation}
In the fiducial model the Hubble parameter is then
\begin{equation}
(H^\text{fid})^2=\frac{8\pi G}{3}(\rho_r + \rho_m + \rho_\Lambda) = 
(H_0^\text{fid})^2\left[\frac{\Omega_r^\text{fid}}{a_\text{fid}^4} + \frac{\Omega_m^\text{fid}}{a_\text{fid}^3} + \Omega_\Lambda^\text{fid} \right]~,
\end{equation}
where the energy densities~$\rho_r(a)$, $\rho_m(a)$, $\rho_\Lambda$ are identical to those in a non-flat universe and we labeled quantities in the fiducial flat model with the superscript ``$\text{fid}$''.
Note that the presence of spatial curvature changes $H_0$ in a real universe and hence the density parameters:
\begin{equation}
H_0=H_0^\text{fid}\left(1+\frac{1}{2}\Omega_k\right) + \mathcal{O}(\Omega_k^2)~, \qquad
\Omega_i = \frac{8\pi G}{3H_0^2} \bar{\rho}_{i,0}= \Omega_i^\text{fid} \left(\frac{H_0^\text{fid}}{H_0}\right)^2
= \Omega_i^\text{fid} \left(1-\Omega_k\right) + \mathcal{O}(\Omega_k^2)~.
\label{eqn:NearlyFlatOmegai}
\end{equation}
In consequence, the Hubble parameters are related to each other as
\begin{equation}
H(a)=H^\text{fid}(a) \left[1 + \frac{1}{2}\frac{\Omega_k}{a^2} \left(\frac{H_0^\text{fid}}{H^\text{fid}}\right)^2\right] + \mathcal{O}(\Omega_k^2)~,
\end{equation}
and note that the functions describing the evolution of the scale factor and the conformal time coordinate for the observer and the observed redshift are different in the two models, i.e.
\begin{equation}
a(\eta) \neq a^\text{fid}(\eta)~,\qquad \qquad
\bar{\eta}_o \neq \bar{\eta}^\text{fid}_o~,
\end{equation}
while the value of the observed redshift itself is the same.
The observed redshift~$z$ relates the scale factor~$a$ in the non-flat model to the one in the fiducial flat model:
\begin{equation}
1+z = \frac{a(\bar{\eta}_o)}{a(\bar{\eta}_z)} = \frac{a^\text{fid}(\bar{\eta}^\text{fid}_o)}{a^\text{fid}(\bar{\eta}^\text{fid}_z)}~,\qquad \qquad
\bar{\eta}_z \neq \bar{\eta}^\text{fid}_z~.
\end{equation}
but we are free to choose the scale factor today to be unity in both models:
\begin{equation}
a(\bar{\eta}_o)=1=a^\text{fid}(\bar{\eta}^\text{fid}_o)~.
\end{equation}

First, we work out the relation for the background quantities in the two models.
We derive the affine parameter at the observed redshift 
\begin{equation}
\cc_z= \cc_z^\text{fid} + \frac{1}{2}\Omega_k (H_0^\text{fid})^2 \int_0^z \frac{(1+z')^2 dz'}{\left[H^\text{fid}(z')\right]^3} + \mathcal{O}(\Omega_k^2)~,
\end{equation}
and by using Eq.~\eqref{bgrr}, the radial coordinate~$\rbar_z$ in the background is related to $\rbar_z^\text{fid}$ as
\begin{equation}
\rbar_z=\rbar_z^\text{fid} \left[ 1 - \Omega_k(H_0^\text{fid})^2\left\{\frac{1}{12}(\rbar_z^\text{fid})^2 + \frac{1}{2\rbar_z^\text{fid}}\int_0^z \frac{(1+z')^2 dz'}{\left[H^\text{fid}(z')\right]^3} \right\}\right] + \mathcal{O}(\Omega_k^2)~,
\end{equation}
where the radius~$\rbar_z^\text{fid}$ in the flat fiducial model coincides with the comoving angular diameter distance:
\begin{equation}
\rbar_z^\text{fid}=-\cc_z^\text{fid}=\bar{\ro}_z^\text{fid}~.
\end{equation}
The background angular diameter distance in Eq.~\eqref{eqn:BGDistances} in a non-flat universe contains an additional factor arising from the function $\ff(\rbar_z)$, leading to
\begin{equation}
\bar\dD_A(z) = \bar\dD_A^\text{fid}(z) \left[1 + \Omega_k(H_0^\text{fid})^2\left\{\frac{1}{6} (\rbar_z^\text{fid})^2 - \frac{1}{2\rbar_z^\text{fid}}\int_0^z \frac{(1+z')^2 dz'}{\left[H^\text{fid}(z')\right]^3}
\right\}\right] + \mathcal{O}(\Omega_k^2)~,
\label{eqn:dDA_fid}
\end{equation}
and the observed volume in Eq.~\eqref{eqn:obsVol} in a homogeneous universe is
\begin{equation}
dV_\up{obs} = dV_\up{obs}^\text{fid} \left[ 1 + \Omega_k(H_0^\text{fid})^2\left\{ \frac{1}{3}(\rbar_z^\text{fid})^2 - \frac{1}{2}\left(\frac{1+z}{H^\text{fid}(z)}\right)^2 - \frac{1}{\rbar_z^\text{fid}}\int_0^z \frac{(1+z')^2 dz'}{\left[H^\text{fid}(z')\right]^3}\right\}\right] + \mathcal{O}(\Omega_k^2)~.
\label{eqn:dVobs_fid}
\end{equation}
Note that the function $\ff$ evaluated at the background radius~$\rbar_z$ is
\begin{equation}
\ff(\rbar_z)=1+\frac{K}{4}(\rbar_z^\text{fid})^2 + \mathcal{O}(\Omega_k^2)~,
\end{equation}
and the direction vector~$\Nn^\alpha_z$ becomes
\begin{equation}
\Nn^\alpha_z= n^i\delta_i^\alpha \ff(\rbar_z) = \Nn^\alpha_\text{fid} \left[1 + \frac{K}{4} (\rbar_z^\text{fid})^2\right] + \mathcal{O}(\Omega_k^2)~,
\end{equation}
where $\Nn^\alpha_\text{fid}\equiv n^i\delta_i^\alpha$ is constant as we have $\ff^\text{fid}(r)=1$ for all radii~$r$.

Next we discuss the relation between the linear-order perturbation variables
in two cosmological models. Since we assumed that the physical quantities are
identical in both models such as the matter density, we also assume that the
initial conditions are identical in both models, for example, the spectral
index~$n_s$, the fluctuation amplitude~$A_s$. Furthermore, since the evolution
of the perturbation variables depend on the background quantities such as the
Hubble parameter~$H(a)$ and their difference is only at the linear order,
the perturbation variables in two models can be assumed to evolve in the same
way. 
Therefore, each perturbation variable in the non-flat model coincides with the corresponding perturbation variable in the fiducial model, for example:
\begin{equation}
\dz = \dz^\text{fid}+\mathcal{O}(2)~, \qquad
\drr = \drr^\text{fid}+\mathcal{O}(2)= \widetilde{\drr}^\text{fid}+\mathcal{O}(2)~, \qquad
\dtt = \dtt^\text{fid}+\mathcal{O}(2)~,
\end{equation}
where the three quantities are defined at the real source position, $x_s$ and $x_s^\text{fid}$, but their difference is already at the linear order.
By combining this with Eqs.~\eqref{phydV} and \eqref{eqn:dVobs_fid}, we obtain the physical volume element
\begin{equation}
dV_\text{phy}
= dV_\text{phy}^\text{fid} \left[ 1 + \Omega_k(H_0^\text{fid})^2\left\{ \frac{1}{3}(\rbar_z^\text{fid})^2 - \frac{1}{2}\left(\frac{1+z}{H^\text{fid}(z)}\right)^2 - \frac{1}{\rbar_z^\text{fid}}\int_0^z \frac{(1+z')^2 dz'}{\left[H^\text{fid}(z')\right]^3}\right\}\right] + \mathcal{O}(\Omega_k^2) ~,
\end{equation}
and by using Eq.~\eqref{eqn:DistancesBGPert} and Eq.~\eqref{eqn:dDA_fid}, we obtain the angular diameter distance
\begin{equation}
\dD_A
= \dD_A^\text{fid} \left[1 + \Omega_k(H_0^\text{fid})^2\left\{\frac{1}{6} (\rbar_z^\text{fid})^2 - \frac{1}{2\rbar_z^\text{fid}}\int_0^z \frac{(1+z')^2 dz'}{\left[H^\text{fid}(z')\right]^3}\right\}
\right] + \mathcal{O}(\Omega_k^2)~.
\end{equation}
Since the perturbation variables in the non-flat model coincide with the corresponding perturbation variables in the fiducial model, it is evident that the lensing convergence and the shear are
\begin{equation}
\hat{\kappa} = \hat{\kappa}^\text{fid} +\mathcal{O}(2)~, \qquad\qquad
\hat{\gamma}_{1,2} = \hat{\gamma}_{1,2}^\text{fid} +\mathcal{O}(2)~,
\end{equation}
hence at linear order nothing changes in the lensing observables that have no non-vanishing background quantities.
Finally, we emphasize that the function~$\ff$ evaluated at the observer position is unity up to third order:
\begin{equation}
\ff(\rr_o)=\ff(\delta \rr_o)= 1 + \frac{K}{4}\delta\rr_o^2 = 1 + \mathcal{O}(3)~.
\end{equation}

\section{Comparison to previous work}
\label{comparison}
In this Section we compare our results to previous work in literature on the cosmological observables in a non-flat universe.

\subsection{Sasaki 1987}
The pioneering paper \cite{1987MNRAS.228..653S} presents the first and complete derivation of the luminosity distance with general spatial curvature~$K$.
The derivation is based on the optical scalar equation, while in our work we derived the luminosity distance by computing the unit physical area in Sec.~\ref{sec:LD}.
While both approaches are consistent, the expression for the luminosity distance derived in \cite{1987MNRAS.228..653S} is not directly applicable for a universe with non-zero spatial curvature. 
The final expression in \cite{1987MNRAS.228..653S} is stated in their Eq.~(3.18)
\begin{equation}
\begin{split}
&\mathcal{D}_L(\cc_s) \propto \exp\left[ -\frac{1}{2} \int_0^{\cc_s}  \frac{d\cc}{s(\cc)} 
\int_\cc^{\cc_s}  d\cc'~s(\cc')4K \frac{d}{d\cc'}(\gamma_i\delta x^i)
\right]~, \\\
&s(\cc):=\sinh^2\left[\sqrt{-K}(\cc-\cc_s-\Delta\cc_s)\right]~,
\end{split}
\label{eqn:Sasaki1}
\end{equation}
where $\cc_s$ is the conformal affine parameter at the source, $\gamma^i$ is the spatial component of the conformally transformed wave vector in the background, $\delta x^i(\cc)$ is the perturbation along the null geodesic induced by the presence of the inhomogeneities, and $\Delta\cc_s$ is the infinitesimal affine distance corresponding to the source radius. 
The integrand of the solution in Eq.~\eqref{eqn:Sasaki1} in the exponential contains the spatial curvature~$K$ and the distortion~$\delta x^i$.
Since the latter is not solved for, the integral cannot be computed and the expression for the luminosity distance is applicable only if $K=0$.
We faced a similar complication in our approach: the perpendicular component of the geodesic equation in Eq.~\eqref{const} involves a term proportional to the spatial curvature~$K$ multiplied by the fluctuation in the position~$\delta x^\al_\cc$.
We resolved this issue by taking another derivative of the whole equation with respect to the affine parameter.
This allowed us to derive the solution to the geodesic equation and subsequently to find an expression for the source position in Sec.~\ref{ssec:path}.
Similar steps need to be taken from Eq.~(3.18) in \cite{1987MNRAS.228..653S} to reach the expression for the luminosity distance applicable for a universe with non-zero spatial curvature.

\subsection{Di Dio et al. 2016}
The source position and the volume fluctuation were derived for a non-flat universe in Newtonian gauge in the comprehensive paper \cite{2016JCAP...06..013D}. 
They computed the galaxy number counts and modified the CLASS code accordingly, such that it includes relativistic galaxy number counts in spatially curved geometries.
Continuing the computation, they derived an expression for the angular power spectrum.
Vector and tensor perturbations were neglected throughout the whole computation, only scalar perturbations were considered.
They work in a spherical coordinate, in which the radial coordinate (henceforth labeled as $r^\text{\tiny{D}}$) is equivalent to our comoving radial coordinate~$\tilde{\chi}$ in Eq.~\eqref{eqn:DefChi} and it is related to our radial coordinates~$\rr$ and $\ro$ through
\begin{equation}
\frac{\rr}{\ff(\rr)} = \ro = S_k(r^\text{\tiny{D}}) = \left\{
\begin{array}{lr}
\frac{1}{\sqrt{K}} \sin\left(\sqrt{K}~r^\text{\tiny{D}}\right)  &\qquad K>0\\
r^\text{\tiny{D}} & \qquad K=0 \\
\frac{1}{\sqrt{|K|}} \sinh\left(\sqrt{|K|}~r^\text{\tiny{D}}\right) &\qquad K<0
\end{array} \right.~.
\label{eqn:ComparisonRadius}
\end{equation}
The calculations proceed in a similar way, but there exist differences in the key equations. 
First of all, the contributions at the observer position are all neglected in \cite{2016JCAP...06..013D}. 
The distortion in the observed redshift in our Eq.~\eqref{eqn:dz} can be expressed in terms of their notation convention in Newtonian gauge as
\begin{equation}
\dz = \mathcal{H}_o \delta\eta_o - \hat{n}^\text{\tiny{D}}\cdot \vec{v}_s - \Psi_s - \int_{\bar{\tau}_s}^{\bar{\tau}_o} d\tau (\dot{\Psi} + \dot{\Phi}) + \hat{n}^\text{\tiny{D}}\cdot \vec{v}_o + \Psi_o~,
\label{eqn:ComparisonDZ}
\end{equation}
where $\tau$ is the conformal time, a dot represents the conformal time derivative, $\Psi$ and $\Phi$ are the two Bardeen potentials and $-\hat{n}^\text{\tiny{D}}$ is the direction of observation. The velocity $\vec{v}$ in their notation is related to our velocity variable as $v^\alpha = \VV^\alpha/\ff$.
Note that they adopted a different convention to express the observed redshift~$z$, and their distortion in the observed redshift is related to $\dz$ in our Eq.~\eqref{eqn:dz} through
\begin{equation}
z =: \bar{z}^\text{\tiny{D}}+\dz^\text{\tiny{D}}~, \qquad\qquad
\dz = \mathcal{H}_o \delta\eta_o + \frac{\dz^\text{\tiny{D}}}{1 + \bar{z}^\text{\tiny{D}}}~.
\end{equation}
Comparing Eq.~\eqref{eqn:ComparisonDZ} to their Eq.~(2.14) for $\dz^\text{\tiny{D}}$, we notice that the only differences are the three contributions in Eq.~\eqref{eqn:ComparisonDZ} at the observer position.
Next, we consider the source position. 
Expressing our radial distortion~$\drr$ in Eq.~\eqref{drrex} in terms of the notation convention in \cite{2016JCAP...06..013D} leads to
\begin{equation}
\frac{\drr}{\ff(\rbar_z)} = \delta\eta_o + \drr_o - \frac{\dz}{\mathcal{H}(z)} + \int_{\bar{\tau}_z}^{\bar{\tau}_o} d\tau (\Psi + \Phi)~,
\label{eqn:ComparisonDr}
\end{equation}
where the radial coordinate~$r^\text{\tiny{D}}$ is related to our radial coordinate~$\rr$ through Eq.~\eqref{eqn:ComparisonRadius}
\begin{equation}
\delta r^\text{\tiny{D}} = \frac{\drr}{\ff(\rbar_z)}~.
\end{equation}
The radial distortion~$\delta r^\text{\tiny{D}}$ in their Eq.~(2.33) coincides with the last term in Eq.~\eqref{eqn:ComparisonDr}, but the other three terms are not present.
The lack of the~$\delta z$ contribution is likely due to the different definition.
For the angular part of the source position, our Eq.~\eqref{tangent} becomes
\bear
\dtt&=&\left(\frac{1}{\rbar_z} -  \frac{1}{4}K\bar{\rr}_z\right)
(\TT_{\al}\de\xx^\al)_{\bobs}
-\left(\TT_\alpha v^\al\right)_{\bobs} 
- \int_{0}^{\bar{\rr}_z} d\bar{\rr}\left(
\frac{\bar{\rr}_z-\bar{\rr}}{\rbar_z\rbar \ff(\bar{\rr})}\right)
\left(1+\frac{1}{4}K\bar{\rr}_z\rbar\right)
\frac{\partial}{\partial\theta}\left[\Psi+\Phi\right]~,\nonumber
\enar
in their notation convention and it can be manipulated as
\begin{equation}
-\int_{0}^{\bar{\rr}_z} d\bar{\rr}\left(
\frac{\bar{\rr}_z-\bar{\rr}}{\rbar_z\rbar \ff(\bar{\rr})}\right)
\left(1+\frac{1}{4}K\bar{\rr}_z\rbar\right)
\frac{\partial}{\partial\theta}\left[\Psi+\Phi\right] 
=
-\int_{\bar{\tau}_o}^{\bar{\tau}_z} d\bar{\tau}~\frac{\ff^2(\rbar)}{\rbar^2} \int_{\bar{\tau}_o}^{\bar{\tau}} d\bar{\tau}'~\frac{\partial}{\partial\theta}\left[\Psi+\Phi\right]~, 
\label{eqn:ComparisonDT}
\end{equation}
which is exactly the expression for $\dtt$ found in their Eq.~(2.41).
Finally, we compare the fluctuation~$\delta V$ in the physical volume.
It turns out that their expression for the volume fluctuation in their Eq.~(2.48) is equivalent to our results for scalar fluctuations in the Newtonian gauge, ignoring the contributions at the observer position.

\section{Summary and Discussion}
\label{sec:discussion}
We have developed a gauge-invariant theoretical framework to compute the light propagation in a non-flat universe and to study the effects of a non-zero curvature on the cosmological observables. Compared to the light propagation in a flat universe, its background solution depends on the position, and it is valid only on a straight path in a background universe. In an inhomogeneous universe, the photon path deviates from a straight path, and the background solution in a non-flat universe is {\it not} a solution on a real photon path that deviates from a straight path. This subtlety results in differences in the geodesic equation for perturbations and its solution in a non-flat universe, compared to the cases in a flat universe. Table~\ref{table:summary} provides the references to the key equations and our findings are summarized as follows.
\begin{itemize}
\item The position $x_s^\mu$ of a light source can be expressed in terms of the observed position $\bar x_z^\mu=(\bar \eta_z, \bar x^\al_z)$ in a homogeneous universe described by the observed redshift~$z$ and observed angle~$(\theta,\phi)$ and the deviation~$\Delta x^\mu_s$ from the observed position. The spatial deviation can be decomposed into the radial distortion $\delta \tilde{\chi}$ in Eq.~\eqref{eqn:DeltaChi} and the angular distortion $(\delta\theta, \delta\phi)$
in Eq.~\eqref{tangent}. 
In a non-flat universe, $\widetilde{\drr}$ ($\neq\delta\tilde{\chi}$) represents the distortion in the angular diameter distance, rather than the radial distortion.
Compared to such distortions in a flat universe, their expressions take a similar form, where the nature of a non-flat universe is encoded in the function~$\ff(r)=1+Kr^2/4$.
\item Using the most general metric representation including the vector and tensor contributions, we have derived all the expressions for the cosmological observables without choosing a gauge condition and demonstrated that those expressions are indeed gauge-invariant.
\item In Section~\ref{cosobs} we have derived the expressions for the luminosity distance, the physical volume and the lensing observables. 
These expressions in a non-flat universe take the same form in a flat universe presented in \cite{YooGrimmMitsou_2018}, when expressed in terms of the distortions ($\delta \tilde{\chi}$, $\widetilde{\drr}$, $\delta\theta$, $\dpp$, $\delta z$) in the source position. However, we emphasize that the individual components of course are different from those in a flat universe.
\item The linear-order Boltzmann equation describes the evolution of the CMB anisotropies along the background photon path. Consequently, it is only the change of the photon comoving momentum that matters in the Boltzmann equation. We showed that the change of the comoving momentum in a non-flat universe is identical to the case in a flat universe, when the propagation vectors are properly normalized. Consequently, there is no further correction in the Boltzmann equation in a non-flat universe.
\item In Section \ref{sec:NearlyFlat}, we expanded the expressions for the cosmological observables in a real universe with non-zero spatial curvature around a fiducial flat cosmology by treating the curvature density~$\Omega_k$ as a perturbation. 
These expressions in a flat universe are an approximation to the expressions in a real non-flat universe, but they are simpler to compute, though the validity of those observables are limited to the case when $\Omega_k$ is as small as the other perturbations.
\item We have compared our calculations to the previous work in literature. The pioneering work \cite{1987MNRAS.228..653S} derived an expression for the luminosity distance, accounting for all the relativistic effects, but in a non-flat universe the final expression is not complete, as it depends on the deviation of the source position from the background path. Further steps need to be taken to reach the final expression we obtained in this work. Regarding galaxy clustering in a non-flat universe, the fluctuation in the physical volume was derived in the comprehensive work \cite{2016JCAP...06..013D},
and its impact on future surveys was quantified. While they only considered the scalar perturbations and ignored all the contributions at the observer position, their expression is consistent with our calculation.
\end{itemize}
We have shown that the light propagation in a non-flat universe is much more
involved than in a flat universe. While we have developed the theoretical descriptions of the cosmological observables in a non-flat universe, we have not yet quantified their impact on cosmological observations. Numerical computation of
the cosmological observables in a non-flat universe will be performed in a future work to quantify the viability of a non-flat universe. In literature, a simple prescription has been widely used for computing galaxy clustering or the weak
lensing observables, in which the same expressions in a flat universe are used, while the background angular diameter distance is replaced with one in a non-flat universe. It is clear in Section~\ref{sec:NearlyFlat} that this simple prescription is incorrect,
but it remains to be quantified how much systematic errors arise by adopting this simple prescription.
Given the current observational constraint $|\Omega_k|\lesssim 0.02$ (1-$\sigma$ constraint from CMB alone, $|\Omega_k|\lesssim 0.003$ in combination with BAO data \cite{Planck2018CosmParam}), the spatial curvature is unlikely to be a dominant component. However, the detection of its small but non-zero value in future surveys would have far-reaching consequences on our understanding of the Universe. It is well known that primary CMB anisotropies are degenerate in terms of spatial curvature, which is broken only when extra measurements at low redshift are combined. Future galaxy surveys will measure galaxy clustering and weak lensing signals with high precision, providing a great opportunity to constrain the spatial curvature of the Universe. Our accurate descriptions of the cosmological observables in a non-flat universe will play a key role in the upcoming era of precision cosmology.

\begin{table}
\begin{center}
\begin{tabular}{lll}
\hline\hline
Symbols & Definition of the symbols & Equations\\
\hline
$\dz$ & perturbation in the redshift
& \eqref{eqn:dz}\\
$\drr$ & radial distortion of the source position
& \eqref{drrex}\\
$\dtt$, $\dpp$ & angular distortions of the source position 
& \eqref{tangent}\\
$\delta \mathcal{D}$ & luminosity distance fluctuation (dimensionless)
& \eqref{eqn:deltaD}, \eqref{LD}\\
$\delta V$ & volume fluctuation (dimensionless)
& \eqref{eqn:dV_GI}, \eqref{eqn:dV}\\
$\hat{\kappa}$ & observable lensing convergence 
& \eqref{eqn:convergence}, \eqref{eqn:RelKappadD}\\
$\hat{\gamma}_{1,2}$ & observable lensing shear components
& \eqref{eqn:shear}\\
\hline\hline
\end{tabular}
\caption{Summary of the main results}
\label{table:summary}
\end{center}
\end{table}

\acknowledgments
We acknowledge support
by the Swiss National Science Foundation (Grant No. SNF PP00P2$\_$176996).
S.B. and J.Y. are further supported by a Consolidator Grant of
the European Research Council (ERC-2015-CoG Grant No. 680886).

\appendix
\section{Coordinates in a non-flat universe}
\label{app:Projection}
\subsection{Stereographic projection for $K>0$}
\label{app:StereographicProjection}
\begin{figure}
\centering
\begin{subfigure}{0.3\textwidth}
\begin{tikzpicture} 
every edge quotes/.append style = {anchor=south, sloped}
                        ]
\coordinate (O) at (0,0);
\coordinate (N) at (0,1.5);
\coordinate (X) at (2.5,0);
\coordinate (S) at (0,-1.5);
\coordinate (XN) at (-2,0);
\coordinate (Q) at (1.2,-0.9);
\coordinate (Qp) at (0.6,0);
\coordinate (Bp) at (2.2,0);
\coordinate (B) at (1.6,0.5385);
\draw[-, color=black] (XN) - -(X);
\draw[dashed, color=black] (S) - -(N);
\draw[dashed, color=black] (Q) - -(N);
\draw[dashed, color=black] (Bp) - -(N);
\draw (O) circle (1.5);
\draw (N) node[anchor=south, above]{$\tilde{N}$};
\draw (S) node[anchor=north east, below]{$\tilde{S}$};
\draw (O) node[anchor=north east, below left]{$\hat{S}$};
\draw (Q) node[anchor=north west, below ]{$\tilde{Q}$};
\draw (Qp) node[anchor=north east, below ]{$\hat{Q}$};
\draw (B) node[anchor=south west, above ]{$\tilde{B}$};
\draw (Bp) node[anchor=north east, below ]{$\hat{B}$};
\filldraw[black] (N) circle (1pt);
\filldraw[black] (S) circle (1pt);
\filldraw[black] (O) circle (1pt);
\end{tikzpicture}
\subcaption{ }
\label{subfig:StereographicProjection1}
\end{subfigure}
\qquad
\begin{subfigure}{0.3\textwidth}
\begin{tikzpicture} 
every edge quotes/.append style = {anchor=south, sloped}
                        ]
\coordinate (O) at (0,0);
\coordinate (N) at (0,1.5);
\coordinate (X) at (3.2,0);
\coordinate (XN) at (-3.2,0);
\coordinate (S) at (0,-1.5);
\coordinate (Sp) at (intersection of XN--X and N--S);
\coordinate (Q) at (1.2,-0.9);
\coordinate (Qp) at (intersection of XN--X and N--Q);
\coordinate (QN) at (-1.2,-0.9);
\coordinate (QNp) at (intersection of XN--X and N--QN);
\coordinate (W) at (1,-1.118);
\coordinate (Wp) at (intersection of XN--X and N--W);
\coordinate (WN) at (-1,-1.118);
\coordinate (WNp) at (intersection of XN--X and N--WN);
\coordinate (E) at (0.8,-1.269);
\coordinate (Ep) at (intersection of XN--X and N--E);
\coordinate (EN) at (-0.8,-1.269);
\coordinate (ENp) at (intersection of XN--X and N--EN);
\coordinate (R) at (0.6,-1.375);
\coordinate (Rp) at (intersection of XN--X and N--R);
\coordinate (RN) at (-0.6,-1.375);
\coordinate (RNp) at (intersection of XN--X and N--RN);
\coordinate (T) at (0.4,-1.446);
\coordinate (Tp) at (intersection of XN--X and N--T);
\coordinate (TN) at (-0.4,-1.446);
\coordinate (TNp) at (intersection of XN--X and N--TN);
\coordinate (Z) at (0.2,-1.487);
\coordinate (Zp) at (intersection of XN--X and N--Z);
\coordinate (ZN) at (-0.2,-1.487);
\coordinate (ZNp) at (intersection of XN--X and N--ZN);
\coordinate (U) at (1.38,-0.588);
\coordinate (Up) at (intersection of XN--X and N--U);
\coordinate (UN) at (-1.38,-0.588);
\coordinate (UNp) at (intersection of XN--X and N--UN);
\coordinate (A) at (1.47,-0.298);
\coordinate (Ap) at (intersection of XN--X and N--A);
\coordinate (AN) at (-1.47,-0.298);
\coordinate (ANp) at (intersection of XN--X and N--AN);
\coordinate (D) at (1.47,0.298);
\coordinate (Dp) at (intersection of XN--X and N--D);
\coordinate (DN) at (-1.47,0.298);
\coordinate (DNp) at (intersection of XN--X and N--DN);
\coordinate (F) at (1.38,0.588);
\coordinate (Fp) at (intersection of XN--X and N--F);
\coordinate (FN) at (-1.38,0.588);
\coordinate (FNp) at (intersection of XN--X and N--FN);
\coordinate (G) at (1.2,0.9);
\coordinate (Gp) at (intersection of XN--X and N--G);
\coordinate (GN) at (-1.2,0.9);
\coordinate (GNp) at (intersection of XN--X and N--GN);
\draw[-, color=black] (XN) - -(X);
\draw[->, color=gray] (S) - -(Sp);
\draw[->, color=gray] (Q) - -(Qp);
\draw[->, color=gray] (QN) - -(QNp);
\draw[->, color=gray] (W) - -(Wp);
\draw[->, color=gray] (WN) - -(WNp);
\draw[->, color=gray] (E) - -(Ep);
\draw[->, color=gray] (EN) - -(ENp);
\draw[->, color=gray] (R) - -(Rp);
\draw[->, color=gray] (RN) - -(RNp);
\draw[->, color=gray] (T) - -(Tp);
\draw[->, color=gray] (TN) - -(TNp);
\draw[->, color=gray] (Z) - -(Zp);
\draw[->, color=gray] (ZN) - -(ZNp);
\draw[->, color=gray] (U) - -(Up);
\draw[->, color=gray] (UN) - -(UNp);
\draw[->, color=gray] (A) - -(Ap);
\draw[->, color=gray] (AN) - -(ANp);
\draw[->, color=gray] (D) - -(Dp);
\draw[->, color=gray] (DN) - -(DNp);
\draw[->, color=gray] (F) - -(Fp);
\draw[->, color=gray] (FN) - -(FNp);
\draw[->, color=gray] (G) - -(Gp);
\draw[->, color=gray] (GN) - -(GNp);
\draw (O) circle (1.5);
\draw (N) node[anchor=south, above]{$\tilde{N}$};
\draw (S) node[anchor=north east, below]{$\tilde{S}$};
\filldraw[black] (N) circle (1pt);
\filldraw[black] (S) circle (1pt);
\filldraw[black] (O) circle (1pt);
\end{tikzpicture}
\subcaption{ }
\label{subfig:StereographicProjection2}
\end{subfigure}
\caption{The stereographic projection of a circle with projection origin~$\tilde{N}$. (a) Three different points on the circle ($\tilde{S}$, $\tilde{Q}$ and $\tilde{B}$) and their projections ($\hat{S}$, $\hat{Q}$ and $\hat{B}$). The South Pole~$\tilde{S}$ is projected onto the center of the circle, as we chose the projection origin to be the North Pole~$\tilde{N}$. The projected point~$\hat{Q}$ is located inside the circle, whereas the projected point~$\hat{B}$ lies outside of the circle. This behavior is again shown in (b): Points on the Southern hemisphere of the circle are projected onto points inside the circle, whereas points on the Northern hemisphere are projected onto points outside of the circle. Being the projection point, the North Pole~$\tilde{N}$ is the only point on the circle whose projection is not defined.}
\label{fig:StereographicProjection}
\end{figure}
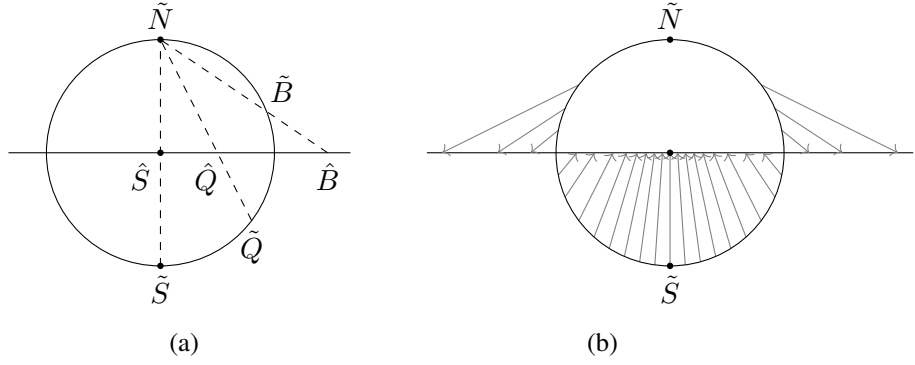

The geometry of a Universe with positive spatial curvature~$K$ can be described as a three-dimensional sphere with radius~$R_K=\frac{1}{\sqrt{K}}$ in 4D Euclidean space.
Therefore, we can perform a stereographic projection, which is a mapping that projects a sphere onto a plane.
In our case, the projection origin in tilde coordinates~$\xo=(\xo_1, \xo_2, \xo_3, \xo_4)$ is the North Pole~$\tilde{N}=(0,0,0,R_K)$, so the projected coordinate~$\hat{x}$ is determined through
\beeq
\hat{x}=\tilde{N} + \tau(\xo-\tilde{N})~,
\label{eqn:SP1}
\eneq
where $\tau$ is the projection parameter that measures the location of the projected point between the projection origin and the point in the sphere.
Mind that we project a 3D hypersphere in 4D onto a 3D plane, hence the projected coordinate satisfy $\hat{x}=(\hat{x}_1, \hat{x}_2, \hat{x}_3, 0)$.
Eq.~\eqref{eqn:SP1} is a system of four equations:
\beeq
\hat{x}_i=\tau\xo_i\qquad\text{for}~i=1,2,3~,\qquad\qquad
\hat{x}_4=0=R_K+\tau(\xo_4-R_K)~.
\label{eqn:SP2}
\eneq
Since the tilde coordinate satisfy
\beeq
\xo^2_1 + \xo^2_2 + \xo^2_3 + \xo^2_4 = \ro^2 + \xo^2_4 = R_K^2~, \qquad\qquad
\xo_4 = \pm \sqrt{R_K^2 - \ro^2}~,
\label{eqn:SP2b}
\eneq
the fourth equation in Eq.~\eqref{eqn:SP2} translates into
\beeq
\tau_{\text{\tiny{$\pm$}}}
=\frac{1\pm \sqrt{1 - K\ro^2}}{K\ro^2}~,
\eneq
where the plus and minus solutions correspond to $\xo_4>0$ (Northern hemisphere) and $\xo_4<0$ (Southern hemisphere), respectively.
Therefore, the radius of the projected coordinate is related to the radius~$\ro$ through
\beeq
\hat{r}^2=\hat{x}^2_1 + \hat{x}^2_2 + \hat{x}^2_3 = (\tau~\ro)^2~, \qquad\qquad
\hat{r}_{\text{\tiny{$\pm$}}}:=\tau_{\text{\tiny{$\pm$}}}\ro= \frac{1}{K\ro}\left(1\pm \sqrt{1 - K\ro^2}\right)~.
\label{eqn:SP3}
\eneq
Note that expressing the radius~$\ro$ in terms of the radius~$\hat{r}$ leads to the same expression for both the minus and plus solution in Eq.~\eqref{eqn:SP3}:
\beeq
\ro=\frac{2\hat{r}}{1+K\hat{r}^2}~.
\eneq
Evident from this relation, the untilde coordinate~$x$ introduced in Eqs.~\eqref{rrro} and \eqref{eqn:relationRr} is related to the projected coordinate~$\hat{x}$ through
\beeq
\rr_{\text{\tiny{$\pm$}}} = 2\hat{r}_{\text{\tiny{$\pm$}}}~.
\eneq
For better understanding of its geometrical meaning, we slice the hypersphere into three parts:
\begin{itemize}
\item Southern hemisphere ($\frac{1}{2}\leq\tau<1$):
\beeq
-R_K\leq\xo_4 <0~, \qquad \qquad
0\leq\ro<R_K~, \qquad\qquad
0\leq\hat{r}= \hat{r}_{\text{\tiny{$-$}}}<R_K~,
\eneq
Points in the Southern hemisphere are projected onto the area inside the intersection of the hypersphere with the $\xo_4=0-$plane.
The projection of the South Pole corresponds to the origin of the projected coordinate.
\item Equator ($\tau=1$):
\beeq
\xo_4 =0~, \qquad \qquad
\hat{r}= \ro=R_K~,
\eneq
The equator is projected onto itself.
\item Northern hemisphere ($\tau>1$):
\beeq
0<\xo_4\leq R_K~, \qquad \qquad
0\leq\ro<R_K~, \qquad\qquad
R_K<\hat{r}=\hat{r} _{\text{\tiny{$+$}}}<\infty~,
\eneq
Points in the Northern hemisphere are projected onto the area outside of the intersection of the hypersphere with the $\xo_4=0-$plane. 
The projection of the North Pole is mapped into infinity and hence ill-defined, as it serves as the projection origin.
\end{itemize}
This projection is illustrated in Fig.~\ref{fig:StereographicProjection}, where the stereographic projection of a circle is shown.
The points $\tilde{S}$, $\tilde{Q}$ and $\tilde{B}$ on the circle are projected onto $\hat{S}$, $\hat{Q}$ and $\hat{B}$ respectively. 
In our case there are two additional dimensions, but the principle is the same.
The analogy of the radius~$\ro$ in two dimensions would be the absolute value of the $x$-coordinate of any point on the circle, whereas the $y$-coordinate in the Figure corresponds to our $\xo_4$-coordinate.
The radius~$\hat{r}$ is therefore represented by the distance of the projected point to the origin ($\hat{S}$ in the Fig.~\ref{subfig:StereographicProjection1}).
It is evident in Fig.~\ref{subfig:StereographicProjection2} that the points on the Southern hemisphere are projected onto points inside the circle, whereas points on the Northern hemisphere are projected onto points outside of the circle.
This is equivalent to $\hat{r}<R$ for points on the Southern hemisphere and $\hat{r}>R$ for points on the Northern hemisphere, where $R$ is the radius of the circle.

Eq.~\eqref{eqn:SP2b} leads to a degeneracy in the sense that a specific choice of the spatial coordinates $(\ro, \ttt, \pp)$ corresponds to two different points, one on the Northern hemisphere (with $\xo_4>0$) and one on the Southern hemisphere (with $\xo_4<0$). 
The stereographic projection breaks this degeneracy.
This is also visible in the boundary conditions: While the radius in the tilde coordinate satisfies $0\leq\ro\leq R_K$, the radii~$\rr_{\text{\tiny{$\pm$}}}$ are constrained through $0\leq\rr_{\text{\tiny{$-$}}}\leq2 R_K$ and $2R_K\leq\rr_{\text{\tiny{$+$}}}<\infty$.

\subsection{Projection for $K<0$}
\label{app:ProjectionOpen}
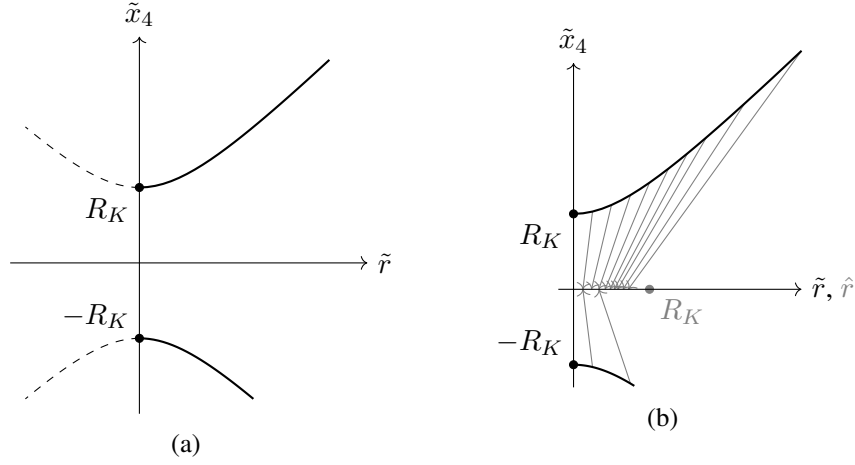
\begin{figure}
\centering
\begin{subfigure}{0.3\textwidth}
\begin{tikzpicture} [domain=0:2]
\coordinate (O) at (0,0);
\coordinate (X) at (2,0);
\coordinate (XN) at (-2,0);
\coordinate (RK) at (0,1);
\coordinate (RKN) at (0,-1);
\coordinate (RKx) at (1,0);
\draw (RK) node[anchor=nord east, below left]{$R_K$};
\draw (RKN) node[anchor=south east, above left]{$-R_K$};
\filldraw[black] (RK) circle (1.5pt);
\filldraw[black] (RKN) circle (1.5pt);
\draw[->] (-1.7,0) -- (3,0) node[right] {$\ro$}; 
\draw[->] (0,-2) -- (0,3) node[above] {$\xo_4$};
\draw[domain=0:2.5, thick]   plot (\x,{(1+(\x)*(\x))^(1/2)}); 
\draw[domain=0:1.5, thick]   plot (\x,{-(1+(\x)*(\x))^(1/2)}); 
\draw[domain=0:1.5, dashed]   plot (-\x,{(1+(\x)*(\x))^(1/2)}); 
\draw[domain=0:1.5,dashed]   plot (-\x,{-(1+(\x)*(\x))^(1/2)}); 
\end{tikzpicture}
\subcaption{ }
\label{subfig:ProjectionOpen1}
\end{subfigure}
\qquad\qquad
\begin{subfigure}{0.3\textwidth}
\begin{tikzpicture} [domain=0:2]
\coordinate (O) at (0,0);
\coordinate (X) at (2,0);
\coordinate (XN) at (-2,0);
\coordinate (RK) at (0,1);
\coordinate (RKN) at (0,-1);
\coordinate (RKx) at (1,0);
\coordinate (Q) at (1,1.414);
\coordinate (Qp) at (0.414,0);
\coordinate (QN) at (1,-1.414);
\coordinate (W) at (0.5,1.12);
\coordinate (Wp) at (0.236,0);
\coordinate (WN) at (0.5,-1.12);
\coordinate (E) at (0.25,1.03);
\coordinate (Ep) at (0.123,0);
\coordinate (EN) at (0.25,-1.03);
\coordinate (R) at (1.5,1.803);
\coordinate (Rp) at (0.535,0);
\coordinate (RN) at (1.5,-1.803);
\coordinate (T) at (1.75,2.016);
\coordinate (Tp) at (0.58,0);
\coordinate (TN) at (1.75,-2.016);
\coordinate (A) at (1.25,1.6);
\coordinate (Ap) at (0.48,0);
\coordinate (AN) at (1.25,-1.6);
\coordinate (S) at (0.75,1.25);
\coordinate (Sp) at (0.333,0);
\coordinate (SN) at (0.75,-1.25);
\coordinate (D) at (3,3.162);
\coordinate (Dp) at (0.721,0);
\coordinate (F) at (2.25,2.46);
\coordinate (Fp) at (0.65,0);
\draw[->, color=gray] (Q) - -(Qp);
\draw[->, color=gray] (W) - -(Wp);
\draw[->, color=gray] (E) - -(Ep);
\draw[->, color=gray] (R) - -(Rp);
\draw[->, color=gray] (T) - -(Tp);
\draw[->, color=gray] (A) - -(Ap);
\draw[->, color=gray] (S) - -(Sp);
\draw[->, color=gray] (D) - -(Dp);
\draw[->, color=gray] (F) - -(Fp);
\draw[->, color=gray] (EN) - -(Ep);
\draw[->, color=gray] (SN) - -(Sp);
\draw (RK) node[anchor=nord east, below left]{$R_K$};
\draw (RKN) node[anchor=south east, above left]{$-R_K$};
\draw (RKx) node[anchor=nord, below right]{\textcolor{gray}{$R_K$}};
\filldraw[black] (RK) circle (1.5pt);
\filldraw[black] (RKN) circle (1.5pt);
\filldraw[gray] (RKx) circle (1.5pt);
\draw[->] (-0.2,0) -- (3,0) node[right] {$\ro$,~\textcolor{gray}{$\hat{r}$}}; 
\draw[->] (0,-1.3) -- (0,3) node[above] {$\xo_4$};
\draw[domain=0:3, thick]   plot (\x,{(1+(\x)*(\x))^(1/2)}); 
\draw[domain=0:0.8, thick]   plot (\x,{-(1+(\x)*(\x))^(1/2)}); 
\end{tikzpicture}
\subcaption{ }
\label{subfig:ProjectionOpen2}
\end{subfigure}
\caption{(a) The geometric relation between the radius~$\ro$ and the coordinate~$\xo_4$ is a hyperbola. The solid line represents the physical solution, while the dashed line represents the unphysical solution of the hyperbola equation~\eqref{eqn:PO2} with $\ro<0$. (b) The gray arrows indicate the projections of points in the tilde coordinate system. The $x$-axis corresponds to the radius, $\ro$ in the tilde coordinate and $\hat{r}$ in the projected coordinate respectively. Points with opposite sign but equal absolute value in the $\xo_4$-coordinate that share the same value for $\ro$ are projected onto the same point.}
\label{fig:ProjectionOpen}
\end{figure}

A Universe with negative spatial curvature~$K$ can be described by a hyperbolic geometry.
In this case, Eq.~\eqref{eqn:SP2b} has to be slightly adapted to
\beeq
\xo^2_1 + \xo^2_2 + \xo^2_3 - \xo^2_4 = \ro^2 - \xo^2_4 = -R_K^2 = \frac{1}{K}~, \qquad\qquad
\xo_4 = \pm \sqrt{R_K^2 + \ro^2}~.
\label{eqn:PO1}
\eneq
Rearranging the terms makes the hyperbolic form evident:
\beeq
\frac{\xo^2_4}{R_K^2} - \frac{\ro^2}{R_K^2} = 1~.
\label{eqn:PO2}
\eneq
The points satisfying this equation are shown in Fig.~\ref{subfig:ProjectionOpen1}.
Both the solid and the dashed lines are solutions to Eq.~\eqref{eqn:PO2}, but the dashed line corresponds to negative $\ro$-values and is therefore unphysical.
It is evident from Eq.~\eqref{eqn:PO1} that $\xo_4\notin(-R_K, R_K)$ by definition.
The projection of the tilde coordinate~$\xo=(\xo_1, \xo_2, \xo_3, \xo_4)$ onto the coordinate~$\hat{x}=(\hat{x}_1, \hat{x}_2, \hat{x}_3, 0)$ through the projection origin~$\tilde{P}$ is
\beeq
\hat{x}=\tilde{P} + \tau(\xo-\tilde{P})~,
\label{eqn:PO3}
\eneq
as in Eq.~\eqref{eqn:SP1} for a closed Universe.
We slice the geometry into two parts:
\begin{itemize}
\item $\xo_4= \sqrt{R_K^2 + \ro^2}\geq R_K$:

The projection origin for any point with a positive $\xo_4$-value is $\tilde{P}=(0, 0, 0, -R_K)$, and Eq.~\eqref{eqn:PO3} leads to the following system of four equations:
\beeq
\hat{x}_i=\tau\xo_i\qquad\text{for}~i=1,2,3~,\qquad\qquad
\hat{x}_4=0=-R_K+\tau(\xo_4+R_K)~,
\eneq
which results in
\beeq
\hat{r}=\tau~\ro~, \qquad\qquad
\tau=\frac{1}{1+\sqrt{1-K\ro^2}}~.
\label{eqn:PO4}
\eneq

\item $\xo_4=- \sqrt{R_K^2 + \ro^2}\leq -R_K$:

The projection origin for any point with a negative $\xo_4$-value is $\tilde{P}=(0, 0, 0, R_K)$, so Eq.~\eqref{eqn:PO3} leads to the following system of four equations:
\beeq
\hat{x}_i=\tau\xo_i\qquad\text{for}~i=1,2,3~,\qquad\qquad
\hat{x}_4=0=R_K+\tau(\xo_4-R_K)~,
\eneq
which has the exact same solution for the projection parameter~$\tau$ as in Eq.~\eqref{eqn:PO4}.
\end{itemize}
Therefore, the radius~$\hat{r}$ is related to the radius~$\ro$ through
\beeq
\hat{r}=\frac{\ro}{1+\sqrt{1-K\ro^2}}=\frac{1}{K\ro}\left(1-\sqrt{1-K\ro^2}\right)~.
\label{eqn:PO5}
\eneq
Figure~\ref{subfig:ProjectionOpen2} shows the projection of the tilde coordinate~$\xo$.
Contrary to the closed Universe, there exists a degeneracy in the $\xo_4$-coordinate, as the points $(\xo_1^*, \xo_2^*, \xo_3^*, \xo_4^*)$ and $(\xo_1^*, \xo_2^*, \xo_3^*, -\xo_4^*)$ are projected onto the same point.
Evident in Eq.~\eqref{eqn:PO5} and clearly visible in the plot, the projected radius is constrained by $0\leq\hat{r}<R_K$ for $0\leq\ro<\infty$.
Comparing Eq.~\eqref{eqn:PO5} to Eq.~\eqref{eqn:relationRr} with $\hat{r}=\frac{1}{2}\rr$, we notice that we solely obtain the minus solution~$\rr_{\text{\tiny{$-$}}}$. 
However, as mentioned in Sec.~\ref{ssec:BGmetric}, the plus solution~$\rr_{\text{\tiny{$+$}}}$ is unphysical, as it is negative for $0\leq\ro<\infty$.

\section{Derivation of the tangential distortion}
\label{angdetail}

Here we present the detailed derivations of the angular solution 
$\TT_\al\dNn^\al$ in Eq.~\eqref{polardn} to the geodesic equation and 
the angular distortion~$\dtt$ in Eq.~\eqref{tangent} of the observed source 
position.

We start with the formal solution for $\TT_\al\dNn^\al_\cc$
in Eq.~\eqref{forsol2} 
\beeq
(\TT_\al\dNn^\al)_\cc-(\TT_\al\dNn^\al)_0
= \left[\frac{d}{d\cc}\left(\TT_\alpha\dNn^\al\right)\right]_{0} 
\int_{0}^\cc  d\cc'\ff_{\cc'}
+ \int_{0}^\cc  d\cc' \ff_{\cc'}\int_{0}^{\cc'} 
 d\cc''\frac{1}{\ff_{\cc''}}\frac{d}{d\cc''}
(\TT_\alpha\GGH^\al )_{\cc''}~,
\eneq
where the integral constants are obtained by using Eqs.~\eqref{boundary2} 
and~\eqref{const} as
\bear
\label{bcan}
(\TT_\al\dNn^\al)_0 &=&-\ttt^i\de_{i\al}\left(\VV^\alpha+\Nn^\beta
\mathcal{G}^{\al}{}_{,\be}+C^\al_\be \Nn^\beta \right)_{\bobs}+\pp^i
\Omega^i_{\bobs}~,\\
\label{bcan2}
\left[\frac{d}{d\cc} \left(\TT_\alpha\dNn^\alpha\right)\right]_0 
&=& \theta^i\de_{i\al} \left(\GGH^\alpha
+\frac{1}{2}K\delta\xx^\alpha\right)_{\bobs}~.
\enar 
The double integral term is first simplified by considering
\beeq
\int_{0}^{\cc'}
 d\cc''\frac{1}{\ff_{\cc''}}\frac{d}{d\cc''}
(\TT_\alpha\GGH^\al )_{\cc''} = \left.\frac{1}{\ff_{\cc''}}\TT_\alpha\GGH^\al 
 \right|_0^{\cc'}
- \frac{K}{2}\int_{0}^{\cc'} d\cc''\frac{\bar{\rr}_{\cc''}}{\ff_{\cc''}}
(\TT_\alpha\GGH^\al )_{\cc''}~,
\eneq
and the formal solution becomes
\beeq
\bigg[\TT_\al\dNn^\al\bigg]^\cc_0
=-\frac{K\bar{\rr}_\cc}{2}\left(\TT_\al\delta\xx^\alpha\right)_{\bobs}
+ \int_{0}^\cc  d\cc' (\TT_\alpha\GGH^\al)_{\cc'}
- \frac{K}{2} \int_{0}^\cc  d\cc' \ff_{\cc'}\int_{0}^{\cc'} d\cc''
\frac{\bar{\rr}_{\cc''}}{\ff_{\cc''}}(\TT_\alpha\GGH^\al )_{\cc''}~,
\eneq
where the boundary term from the double integration is cancelled by the first
term in Eq.~\eqref{bcan2} and we used the background relation
\beeq
\int_{0}^\cc  d\cc'\ff_{\cc'} = -\int_{0}^{\bar{\rr}_\cc}  d\bar{\rr} + 
\mathcal{O}(1)~,\Dquad d\cc=-{d\rbar\over \ff(\rbar)}+\OO(1)~.
\eneq
With the help of Eq.~\eqref{GGHCO}, we can determine $\TT^\alpha\GGH_\al$ to be
\beeq
\TT^\alpha\GGH_\al\equiv
~\TT^\alpha\frac{\partial}{\partial x^\alpha} \mathcal{T}
 + \frac{\ff(\bar{\rr})}{\bar{\rr}}\TT^\beta\mathcal{T}_\beta
-\TT^\alpha \frac{d}{d\lambda}\mathcal{T}_\alpha + 
\TT^\alpha\frac{1}{\ff}\frac{d^2}
{d\lambda^2}(\ff\mathcal{G}_\alpha) + K\TT^\alpha\mathcal{G}_\alpha~,
\eneq
where we defined
\beeq
\mathcal{T}:=\alpha_\chi-\varphi_\chi- \Psi_\parallel - C_\parallel~, \Dquad
\mathcal{T}_\alpha:=\Psi_\alpha + 2 \Nn^\beta C_{\alpha\beta}~.
\eneq

We will compute two integrals in the formal solution for each term in
$\TT^\al\GGH_\al$.
First, by noting that
\beeq
\TT^\alpha\frac{\partial}{\partial x^\alpha} =
\frac{\ff(\bar{\rr})}{\bar{\rr}} \frac{\partial}{\partial \theta}
+ \mathcal{O}(1)~,
\eneq
two integrals in the formal solution for the term with~$\mathcal{T}$
can be obtained as
\beeq
\TT^\al{\pa\mathcal{T}\over\pa x^\al}:~~
-\int_{0}^{\bar{\rr}_\cc}  \frac{d\bar{\rr}'}{\bar{\rr}'}
 \frac{\partial}{\partial \theta}\mathcal{T}
- \frac{K}{2} \int_{0}^{\bar{\rr}_\cc}  d\bar{\rr}' \int_{0}^{\bar{\rr}'} 
d\bar{\rr}''\frac{1}{\ff_{\cc''}} \frac{\partial}{\partial \theta}\mathcal{T}
=- \int_{0}^{\bar{\rr}_\cc}  {d\bar{\rr}'\over\rbar'}\left[1 + 
\frac{1}{2}K\frac{(\bar{\rr}_\cc-\bar{\rr}')\rbar'}{\ff(\bar{\rr}')}\right]
\frac{\partial}{\partial \theta}\mathcal{T}~.
\eneq
For two terms including $\mathcal{T}_\alpha$,  two integrals in the formal
solution contain
\bear
\frac{\ff}{\bar{\rr}}\TT^\beta\mathcal{T}_\beta
-\TT^\alpha \frac{d}{d\lambda}\mathcal{T}_\alpha :~~&&
-\int_{0}^{\bar{\rr}_\cc} \frac{d\bar{\rr}'}{\bar{\rr}'}
\TT^\beta\mathcal{T}_\beta
-\int_{0}^\cc  d\cc'\TT^\alpha \frac{d}{d\lambda'}\mathcal{T}_\alpha\nnn
&&
- \frac{K}{2} \int_{0}^{\bar{\rr}_\cc}  d\bar{\rr}'\int_{0}^{\bar{\rr}'} 
\frac{d\bar{\rr}''}{\ff(\bar{\rr}'')}\TT^\beta\mathcal{T}_\beta
+\frac{K}{2} \theta^i\delta_i^\alpha\int_{0}^\cc  d\cc' \ff_{\cc'}\int_{0}^{\cc'} d\cc''\bar{\rr}_{\cc''} \frac{d}{d\lambda''}\mathcal{T}_\alpha \nnn
&&=
-\left(\TT^\alpha \mathcal{T}_\alpha\right)_\cc
+\left(\TT^\alpha \mathcal{T}_\alpha\right)_{\bobs}
- \int_{0}^{\bar{\rr}_\cc} \frac{d\bar{\rr}'}{\bar{\rr}'} 
 \TT^\alpha \mathcal{T}_\alpha~,
\enar
where we used
\beeq
\frac{d}{d\cc}\left(\TT^\al\mathcal{T}_\alpha\right)
=\TT^\al\frac{d}{d\cc}\mathcal{T}_\alpha-\frac12K\rbar\TT^\al
\mathcal{T}_\alpha~.
\eneq
For two terms involving~$\CCG_\al$, we first perform the single integral
in the formal solution:
\beeq
\int_0^\cc d\cc'\left[
\TT^\alpha\frac{1}{\ff}\frac{d^2}
{d\lambda^2}(\ff\mathcal{G}_\alpha) + K\TT^\alpha\mathcal{G}_\alpha\right]
=
\left[ \TT^\al\frac{d}{d\lambda^{\prime}}\mathcal{G}_\alpha \right]_{0}^\cc
- \frac{K\bar{\rr}_\cc}{2} \TT^\al\mathcal{G}_\alpha
- K\int_{0}^{\bar{\rr}_\cc}  \frac{d\bar{\rr}'}{\ff(\bar{\rr}')}
\TT^\alpha\mathcal{G}_\alpha~,
\eneq
where we noted $\TT^\al/\ff=\ttt^i\de_{i\al}$ can be pulled out of the 
integration. Finally, the double integration in the formal solution
for two terms involving~$\CCG_\al$ are performed separately. First, we 
compute
\bear
&&
- \frac{K}{2}\int_{0}^\cc  d\cc' \ff_{\cc'}\int_{0}^{\cc'} d\cc''~
\frac{\bar{\rr}_{\cc''}}{\ff_{\cc''}} ~\left[{\TT^\al\over \ff}
\frac{d^2}{d\lambda^{\prime\prime2}}(\ff\mathcal{G}_\alpha)\right] \\
&&\qquad
=- \frac{K}{2} \theta^i\delta_i^\al\left[
\int_{0}^\cc  d\cc' \bar{\rr}_{\cc'}\frac{d}{d\lambda^{\prime}}
(\ff\mathcal{G}_\alpha)
-\int_{0}^\cc  d\cc' \ff_{\cc'}\int_{0}^{\cc'} d\cc''~
{d\over d\cc''}\left({\rbar\over \ff}\right)
\frac{d}{d\lambda^{\prime\prime}}(\ff\mathcal{G}_\alpha)
\right] \nnn
&&\qquad
=- \frac12K\left[\bar{\rr}_{\cc}  \TT^\al\mathcal{G}_\alpha
- \int_{0}^{\bar{\rr}_\cc}  d\bar{\rr}' \TT^\al\mathcal{G}_\alpha\right]\nnn
&&\Dquad
- \frac{1}{2}K\left[\int_{0}^{\bar{\rr}_\cc}  \frac{d\bar{\rr}'}{\ff(\bar{\rr}')}
\left( \frac{K}{4}\bar{\rr}_{\cc''}^2-1\right)\TT^\al\mathcal{G}_\alpha
+\bar{\rr}_{\cc} \bigg(\TT^\al\mathcal{G}_\alpha\bigg)_{\bobs}
-K\int_{0}^{\bar{\rr}_\cc}  d\bar{\rr}' ~
\frac{\bar{\rr}'(\bar{\rr}_\cc-\bar{\rr}')}{\ff^2(\bar{\rr}')}
~\TT^\al\mathcal{G}_\alpha\right] \nnn
&&\qquad =
- \frac12K\bar{\rr}_{\cc}\left[ \TT^\al\mathcal{G}_\alpha + 
\big(\TT^\al\mathcal{G}_\alpha\big)_{\bobs}\right]
+ K  \int_{0}^{\bar{\rr}_\cc} \frac{d\bar{\rr}'}{\ff(\bar{\rr}')} 
~\TT^\al\mathcal{G}_\alpha
+ \frac{K^2}{2} \int_{0}^{\bar{\rr}_\cc}  d\bar{\rr}'~
 \frac{\bar{\rr}'(\bar{\rr}_\cc-\bar{\rr}')}{\ff^2(\bar{\rr}')}
~\TT^\al\mathcal{G}_\alpha~, \nonumber
\enar
where we used
\beeq
{d\over d\cc}\left({\rbar\over \ff}\right)=-1+\frac1{2\ff}K\rbar^2=
-\frac1\ff+\frac1{4\ff}K\rbar^2~.
\eneq
Mind the dimension $[\CCG_\al]=[\Psi_\al]=L$.
Next, we compute the double integration for the remaining term:
\beeq
- \frac{1}{2}K \int_{0}^\cc  d\cc' \ff_{\cc'}\int_{0}^{\cc'} d\cc''
\frac{\bar{\rr}_{\cc''}}{\ff_{\cc''}} \bigg(K\TT^\alpha\mathcal{G}_\alpha \bigg)=
- \frac{1}{2}K^2 \int_{0}^{\bar{\rr}_\cc}  d\bar{\rr}'~ \frac{\bar{\rr}'
(\bar{\rr}_\cc-\bar{\rr}')}{\ff^2(\bar{\rr}')}~ \TT^\alpha\mathcal{G}_\alpha~.
\eneq
The sum of these integrals involving terms with~$\CCG_\al$ 
gives rise to several cancellations, leading to 
\beeq
(\TT_\al\dNn^\al)_\cc-(\TT_\al\dNn^\al)_0 ~~\ni~~
\left[ \frac{d}{d\lambda^{\prime}}\left(\TT_\al\mathcal{G}^\alpha\right)
 \right]_{0}^\cc
- \frac12K\bar{\rr}_{\cc}\left[\TT^\al\mathcal{G}_\alpha+
\bigg(\TT^\al\mathcal{G}_\alpha\bigg)_{\bobs}\right]~.
\eneq
Adding up all the computed terms, we arrive at the solution
for $\TT_\al\dNn^\al$ in Eq.~\eqref{polardn}:
\bear
\label{herepolar}
&&(\TT_\al\dNn^\al)_\cc-(\TT_\al\dNn^\al)_0=
-\frac12K\bar{\rr}_\cc\bigg[\left(\TT_\al\delta\xx^\alpha+\TT^\al\CCG_\alpha
\right)_{\bobs}+\TT^\al\CCG_\al\bigg]
- \int_{0}^{\bar{\rr}_\cc} \frac{d\bar{\rr}'}{\bar{\rr}'}  \TT^\alpha
\left(\Psi_\alpha + 2 \Nn^\beta C_{\alpha\beta} \right)\\
&&\qquad
-\left[\TT^\alpha \Psi_\alpha +2\TT^\alpha \Nn^\beta C_{\alpha\beta}
- \frac{d}{d\lambda^{\prime}}\left(\TT^\al\mathcal{G}_\alpha\right)
\right]_{0}^\cc
- \int_{0}^{\bar{\rr}_\cc}  {d\bar{\rr}'\over\rbar'}\left[1+\frac12K
\frac{(\bar{\rr}_\cc-\bar{\rr}')\rbar'}{\ff(\bar{\rr}')}\right]
\frac{\partial}{\partial \theta}\left( \alpha_\chi-\varphi_\chi-
\Psi_\parallel - C_\parallel \right)~.\nonumber
\enar

Using this solution, we derive the tangential 
distortion~$\rbar_z\dtt$ in Eq.~\eqref{tangent} of the observed source position,
which requires one more integration over the affine parameter as in 
Eq.~\eqref{dtt}:
\beeq
\rbar_z\dtt=(\TT_{\al}\de\xx^\al)_{\bobs}-\int_0^{\cc_s}d\cc~\ff\TT_\al\dNn^\al
~.
\eneq
As there are several terms for the expression $\TT_\al\dNn^\al$ in
Eq.~\eqref{herepolar}, we again compute the integral for these individual terms
separately. First, those terms that are evaluated at the observer position
can be integrated trivially:
\beeq
-\int_0^{\cc_s}d\cc~\ff_\cc 
\left[\TT^\alpha \Psi_\alpha +2\TT^\alpha \Nn^\beta C_{\alpha\beta} 
- \frac{d}{d\cc}\left(\TT_\al\mathcal{G}^\alpha\right)
  \right]_{0}
 = \bar{\rr}_z\left[\TT^\alpha \Psi_\alpha +2\TT^\alpha \Nn^\beta 
C_{\alpha\beta} - \frac{d}{d\cc}\left(\TT_\al\mathcal{G}^\alpha\right)
  \right]_{\bobs}~,
\eneq
and also the boundary term for~$\TT_\al\dNn^\al$ at the observer position:
\beeq
-\int_0^{\cc_s}d\cc~\ff_\cc(\TT_\al\dNn^\al)_{\bobs} 
= \bar{\rr}_z(\TT_\al\dNn^\al)_{\bobs}~,
\eneq
where the boundary term is already given in Eq.~\eqref{bcan}.
Finally, those terms at the observer position, but multiplied by~$\rbar_\cc$
can be integrated by using
\beeq
\int_0^{\cc_s}d\cc~\ff_\cc \bar{\rr}_\cc 
= - \frac{1}{2} \bar{\rr}_z^2+\OO(1)~,
\eneq
as
\beeq
-\int_0^{\cc_s}d\cc~\ff_\cc\left[-
\frac{1}{2}K\bar{\rr}_\cc
\left(\TT_\al\delta\xx^\alpha+
\TT^\al\mathcal{G}_\alpha\right)_{\bobs} \right]
= -\frac14K\bar{\rr}_z^2\left(\TT_\al\delta\xx^\alpha+
\TT^\al\mathcal{G}_\alpha\right)_{\bobs}~.
\eneq
Next, those terms that are at $\xx_\cc$ can be integrated as
\bear
&&
-\int_0^{\cc_s}d\cc~\ff_\cc \left[-\frac12K\rbar_\cc\TT^\al\CCG_\al
-\left(\TT^\al\Psi_\alpha 
+2 \TT^\al\Nn^\beta C_{\alpha\beta}-{d\over d\cc}\left(\TT^\al\CCG_\al\right)
\right)\right]~\nnn
&&\qquad
=-\int_0^{\bar{\rr}_z}d\bar{\rr}~ \TT^\al\left(\Psi_\alpha 
+2 \Nn^\beta C_{\alpha\beta} \right)
-\ff_\cc\TT^\al\mathcal{G}_\alpha+\bigg(\ff_\cc\TT^\al\mathcal{G}_\alpha\bigg)_
{\bobs}~, 
\enar
where we used Eq.~\eqref{trick} and the term with~$K$ is cancelled by the
integration by part of the term with derivative with respect to~$\cc$.
The integral term in Eq.~\eqref{herepolar} becomes the double integral,
which can be readily manipulated as
\beeq
\int_0^{\cc_s}d\cc~\ff_\cc \int_{0}^{\bar{\rr}_\cc} \frac{d\bar{\rr}'}
{\bar{\rr}'}  \TT^\alpha \left[ \Psi_\alpha + 2 \Nn^\beta C_{\alpha\beta}
 \right]= -\int_0^{\bar{\rr}_z}d\bar{\rr}\left(\frac{\bar{\rr}_z-\bar{\rr}}
{\bar{\rr}} \right) \TT^\alpha \left[ \Psi_\alpha
 + 2 \Nn^\beta C_{\alpha\beta} \right]~.
\eneq
Finally, the last term in Eq.~\eqref{herepolar} is integrated to yield
\bear
&&
-\int_0^{\bar{\rr}_z}d\bar{\rr}~ \int_{0}^{\bar{\rr}}  d\bar{\rr}'\left(\frac{1}{\bar{\rr}'} + \frac{K}{2}\frac{\bar{\rr}-\bar{\rr}'}{\ff(\bar{\rr}')}\right) \frac{\partial}{\partial \theta}\left[ \alpha_\chi-\varphi_\chi- \Psi_\parallel - C_\parallel \right] \nnn
&&\qquad
=-\left\{\int_0^{\bar{\rr}_z}d\bar{\rr}~ \int_{0}^{\bar{\rr}}  d\bar{\rr}'\frac{1 - \frac{K}{4}\bar{\rr}^{\prime 2}}{\bar{\rr}'\ff(\bar{\rr}')}
+ \frac{K}{2}\int_0^{\bar{\rr}_z}d\bar{\rr}~\bar{\rr} \int_{0}^{\bar{\rr}} \frac{ d\bar{\rr}'}{\ff(\bar{\rr}')} \right\}\frac{\partial}{\partial \theta}\left[ \alpha_\chi-\varphi_\chi- \Psi_\parallel - C_\parallel \right]\nnn
&&\qquad
=-\left\{\int_0^{\bar{\rr}_z}d\bar{\rr}~(\bar{\rr}_z-\bar{\rr})\frac{1 - \frac{K}{4}\bar{\rr}^{2}}{\bar{\rr}\ff(\bar{\rr})}
+ \frac{K}{2}\int_0^{\bar{\rr}_z}d\bar{\rr}~\bar{\rr}~\frac{\bar{\rr}_z^2-\bar{\rr}^2}{2\bar{\rr}\ff(\bar{\rr})} \right\}\frac{\partial}{\partial \theta}\left[ \alpha_\chi-\varphi_\chi- \Psi_\parallel - C_\parallel \right] \nnn
&&\qquad
=-\int_0^{\bar{\rr}_z}d\bar{\rr}\left(\frac{\bar{\rr}_z
  -\bar{\rr}}{\rbar \ff(\bar{\rr})}\right)
\left(1+ \frac{1}{4}K\bar{\rr}_z\rbar\right)\frac{\partial}{\partial \theta}\left[ \alpha_\chi-\varphi_\chi- \Psi_\parallel - C_\parallel \right]
~.
\enar
Adding up all the terms,
the tangential distortion $\rbar_z\dtt$ of the source position in
Eq.~\eqref{tangent} is therefore obtained as
\bear
\rbar_z\dtt
&=&\left(1 -  \frac{1}{4}K\bar{\rr}_z^2\right)
\bigg(\TT_{\al}\de\xx^\al +\TT^\al\mathcal{G}_\alpha \bigg)_{\bobs}
+\bar{\rr}_z\left[-\TT_\alpha V^\al+\TT_\al\Nn^\be C^\al_\be
  +\pp^i\Omega^i \right]_{\bobs}    -\ff(\bar{\rr}_z)\TT^\al\mathcal{G}_\alpha
\Dquad\\
&&
-  \int_{0}^{\bar{\rr}_z} d\bar{\rr}~\frac{\bar{\rr}_z}{\bar{\rr}}
~\TT^\alpha\left(\Psi_\alpha + 2\Nn^\beta C_{\alpha\beta}\right) 
- \int_{0}^{\bar{\rr}_z} d\bar{\rr}\left(
\frac{\bar{\rr}_z-\bar{\rr}}{\rbar \ff(\bar{\rr})}\right)
\left(1+\frac{1}{4}K\bar{\rr}_z\rbar\right)
\frac{\partial}{\partial\theta}\left[\alpha_\chi-\varphi_\chi
  - \Psi_\parallel - C_\parallel\right]~,\nonumber
\enar
where the expression was further simplified by noting that
\bear
&& \Psi_\alpha - \mathcal{U}_\alpha = \mathcal{G}'_\alpha - V_\alpha~,\\
&&
\frac{d}{d\cc}\left(\TT_\al\mathcal{G}^\alpha\right)
=\TT_\al\frac{d}{d\cc}\mathcal{G}^\alpha+\mathcal{G}^\alpha\frac{d}{d\cc}
\TT_\al
= \TT_\al\left(\mathcal{G}^{\alpha\prime} - \Nn^\beta
\mathcal{G}^\alpha_{~~,\beta}\right) + \frac{K\bar{\rr}}{2}\TT_\al
\mathcal{G}^\alpha~.
\enar

\bibliographystyle{unsrt}
\bibliography{paper.bbl}

\appendix

\clearpage

\clearpage

\end{document}